\documentclass[%
 reprint,
superscriptaddress,
 amsmath,amssymb,
 aps,
]{revtex4-2}

\usepackage{graphicx}
\usepackage{tikz}
\usepackage{dcolumn}
\usepackage{bm}
\allowdisplaybreaks[2]
\usepackage[colorlinks,linkcolor=blue, urlcolor=blue, anchorcolor=blue, citecolor=blue]{hyperref}

\begin{document}

\preprint{APS/123-QED}

\title{Single-photon triggered quantum entanglement between two qubits or at least 2,000 identical qubits}

\author{Wangjun Lu}
\email{wjlu1227@zju.edu.cn}
\affiliation{Institute of Engineering Education and Engineering Culture Innovation and Department of Maths and Physics, Hunan Institute of Engineering, Xiangtan 411104, China}
\affiliation{Key Laboratory of Low-Dimensional Quantum Structures and Quantum Control of Ministry of Education, Department of Physics
	and Synergetic Innovation Center for Quantum Effects and Applications, Hunan Normal University, Changsha 410081, China}

\author{Cuilu Zhai}
\affiliation{Institute of Engineering Education and Engineering Culture Innovation and Department of Maths and Physics, Hunan Institute of Engineering, Xiangtan 411104, China}

\author{Hong Tao}
\affiliation{Institute of Engineering Education and Engineering Culture Innovation and Department of Maths and Physics, Hunan Institute of Engineering, Xiangtan 411104, China}

\author{Yaju Song}
\affiliation{College of Physics and Electronic Engineering, Hengyang Normal University, Hengyang 421002, China}

\author{Shiqing Tang}
\affiliation{College of Physics and Electronic Engineering, Hengyang Normal University, Hengyang 421002, China}

\author{Lan Xu}
\affiliation{School of Physics and Chemistry, Hunan First Normal University, Changsha 410205, China}


\date{\today}

\begin{abstract}
In this paper, we study the effect of single-photon light fields on quantum entanglement between two qubits or multiple identical qubits that are initially in the direct state. Within the system where light interacts with two qubits, we first study the effect of the weight of the excited state of the other qubit in the initial state on the single-photon control of quantum entanglement between two qubits, when the initial state of one qubit is determined. We find that an excessively large weight of the excited state disrupts the single-photon triggering of quantum entanglement between two qubits. We then investigate the effect of the magnitude of the initial coherence on the single-photon control of the quantum entanglement of two qubits when the two qubits initially have the same coherence. We find that the single photon is able to trigger the maximum quantum entanglement of the two qubits when the initial coherence is maximum. In the system where light interacts with multiple identical qubits, we similarly investigate the effect of the weight of the excited state of each qubit in the initial state and the initial coherence on the single-photon control of quantum entanglement between any two qubits of the multi-qubits for different numbers of qubits. In the limit of large qubit number, we find that a single photon cannot trigger quantum entanglement between any two qubits of multi-qubits when the initial excited-state weights of each qubit are larger than the ground-state weights or all qubits are initially in the ground-state.  Interestingly, we find that a single photon is capable of triggering quantum entanglement between any two qubits out of at least 2000 qubits. Moreover, in this limit, the maximum quantum entanglement between any two qubits of multi-qubits triggered by a single photon varies with the initial state parameter of the qubits and almost no longer depends on the number of qubits.
\end{abstract}

\maketitle


\section{introduction}
Quantum entanglement is a core concept in quantum information science, which describes the extraordinary correlations between particles in a quantum system that cannot be explained by classical physics \cite{PhysRev.47.777,PhysRev.48.696,PhysicsPhysiqueFizika.1.195, RevModPhys.81.865}. Quantum entanglement has become a cornerstone in the practical applications of quantum systems. For instance, in quantum computing, quantum entangled states can greatly enhance computing power \cite{nielsen2010quantum, ladd2010quantum, berman1998introduction,preskill1998reliable,PhysRevX.10.041038,steane1998quantum}, allowing quantum computers to far surpass traditional computers in handling certain types of problems. In the field of quantum communication, quantum key distribution (QKD) uses quantum entanglement to ensure the security of communications \cite{gisin2007quantum, cozzolino2019high,chen2021integrated,yuan2010entangled,nauerth2013air,pirandola2017fundamental,long2007quantum,duan2001long, dai2008teleportation,peng2013construction,RevModPhys.94.035001}. In quantum measurement and sensing, quantum entanglement is used to improve the accuracy of measurements \cite{giovannetti2004quantum, xiang2011entanglement, wolfgramm2013entanglement, PhysRevA.97.042112, PhysRevLett.114.110506, PhysRevResearch.3.033114,PhysRevLett.118.233603, PhysRevLett.129.070502, PhysRevLett.132.210801,RevModPhys.90.035005, PhysRevA.105.023718, huang2024entanglementenhanced}.

Due to the importance of quantum entanglement in various fields, the preparation of quantum entangled states is a critical step. In two-qubit quantum systems, a large amount of theoretical and experimental work has already achieved the preparation of quantum entangled states \cite{shankar2013autonomously, PhysRevA.73.062306, PhysRevA.73.042319, PhysRevA.74.024304, PhysRevApplied.11.014017, PhysRevA.89.042328, PhysRevA.62.022311, PhysRevResearch.4.023010, PhysRevLett.109.240505, shen2014preparation, Jiang_2024,PhysRevLett.85.2392,PhysRevA.68.052312,PhysRevA.104.063701,PhysRevA.108.023728}. For example, quantum entanglement can arise between two qubits that interact with a common environment \cite{PhysRevA.73.062306}. The entanglement of the two-mode non-classical state field can almost be entirely transferred to two qubits \cite{PhysRevA.73.042319}. Using guided photons in a one-dimensional waveguide to generate, control, and measure the entanglement between two qubits \cite{PhysRevA.89.042328}. Using a bichromatic filed excitation scheme to deterministically produce entangled states in ions in thermal motion \cite{PhysRevA.62.022311}. Controlling the generation of two qubits quantum entangled states using different types of light fields \cite{Jiang_2024,PhysRevLett.85.2392,PhysRevA.68.052312,PhysRevA.104.063701}, and so on.

In terms of enhancing quantum computing capabilities and improving quantum measurement precision, multi-qubit quantum entangled states have stronger advantages compared to two-qubit quantum entangled states. There is also a significant amount of theoretical and experimental work in preparing multi-qubit quantum entangled states \cite{neeley2010generation, mcconnell2015entanglement, PhysRevA.62.062314, PhysRevA.71.060310, PhysRevA.75.052306, PhysRevA.78.032101, PhysRevA.85.032112, PhysRevA.90.042324,PhysRevA.94.012302, PhysRevLett.95.110503, PhysRevLett.98.130502,PhysRevLett.112.170501,PhysRevLett.129.063603,PhysRevLett.131.250801, zhong2021deterministic}. For example, using weak laser pulses can achieve quantum entanglement of nearly 3000 atoms \cite{mcconnell2015entanglement}. Multi-qubit entanglement can be achieved in a bidirectional chiral waveguide QED systems \cite{PhysRevA.94.012302}. Deterministic generation of entangled multi-qubit states using sequential coupling of an ancillary system to initially uncorrelated qubits \cite{PhysRevLett.95.110503}. Deterministic multi-qubit quantum entanglement can be achieved in a quantum network \cite{PhysRevLett.131.250801,zhong2021deterministic}.

Inspired by research that used a weak laser pulse containing about 210 photons to control close to 3,000 atoms to produce quantum entanglement \cite{ mcconnell2015entanglement}, in this paper, we propose the generation of quantum entanglement between two qubits or multiple identical qubits using a single-photon light filed. The single-photon light field plays a very important role in many quantum systems. For example, in a one-dimensional waveguide, single-photon control of single-photon transmission can be achieved \cite{PhysRevA.90.053807}. Single photons can induce quantum entanglement and quantum superposition states \cite{PhysRevA.97.033807}. In a hybrid quantum device combining cavity quantum electrodynamics and optomechanics, a single photon can manipulate quantum chaos \cite{PhysRevA.100.023825}. A single photon can trigger spin squeezing and reduce decoherence in open systems \cite{PhysRevA.104.053517}. A single photon can trigger a superradiant quantum phase transition from the equilibrium state \cite{PhysRevApplied.9.064006}. A single photon can trigger multi-molecular reactions \cite{PhysRevLett.119.136001}, and so on. Here, we investigate single-photon triggering of quantum entanglement between two or more qubits.

This paper is organized as follows. In Section \ref{sec2}, we solve the two qubits Tavis-Cummings (TC) model, obtaining the reduced density matrix of the two qubits at time $t$ under the control of a single-photon light field. In Section \ref{sec3}, we study the effects of the initial excitation state weight of the qubit and quantum coherence on the quantum entanglement of the two qubits under the control of a single-photon light field. In Section \ref{sec4}, we solve the multi-qubit TC model and discuss how the initial excitated-state weight of any qubit in identical qubits, quantum coherence, and the number of qubits influence the quantum entanglement between any two qubits under the control of a single-photon light field.

\section{two-qubit TC Model and solution}\label{sec2}

\begin{figure}[t]
\centering
\begin{tikzpicture}
\draw (0, 0) node[inner sep=0] {\includegraphics[width=7cm,height=2.5cm]{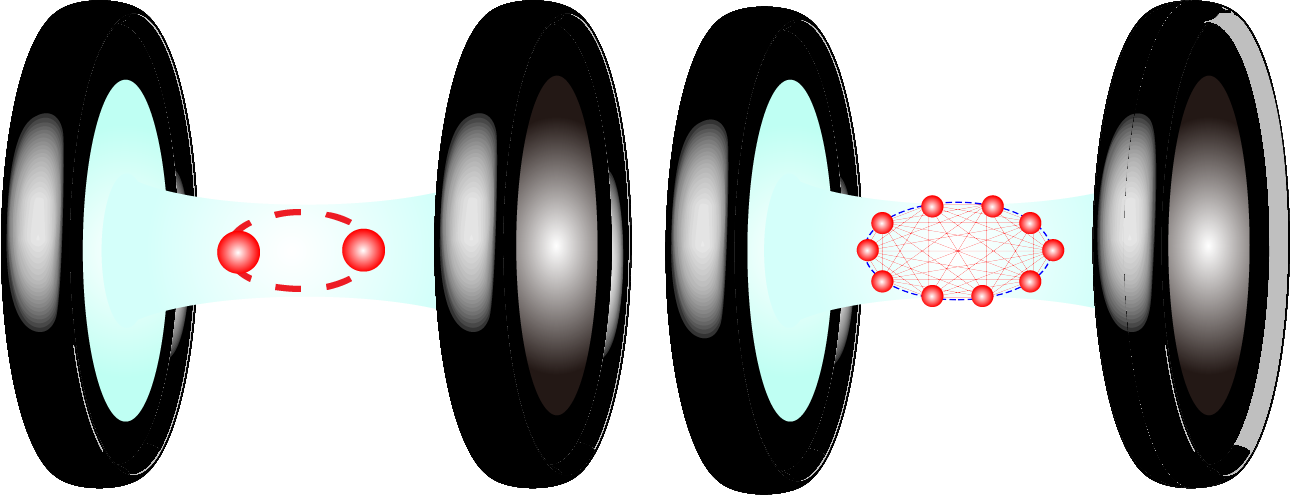}};
\draw (-2.5, 1.5) node {(a)};
\draw (1.1, 1.5) node {(b)};
\end{tikzpicture}	
\caption{\label{fig1}(a) and (b) are schematic diagrams of the two-qubits TC model and the multi-qubits TC model, respectively. Each red sphere represents each qubit, and the connecting line between each sphere denotes the quantum entanglement between qubits.}
\end{figure}

As shown in Fig.~(\ref{fig1}), we investigated the interaction of the single-photon light field with two qubits and $N$ identical qubits, respectively. We assume that there are no interactions between all qubits and only dipole interactions between the light field and each qubit. We analyze this system under conditions of near-resonance and relatively weak coupling, using the following Tavis-Cummings (TC) model to describe the Hamiltonian of the entire system ($\hbar=1$).
\begin{equation}
\hat{H}_{TC}=\text{\ensuremath{\omega_{a}}}\hat{a}^{\dagger}\hat{a}+\frac{1}{2}\omega_{0}\sum_{i=1}^{N}\hat{\sigma}_{z}^{i}+g\sum_{i=1}^{N}(\hat{a}^{\dagger}\hat{\sigma}_{-}^{i}+\hat{\sigma}_{+}^{i}\hat{a}),  \label{Eq1}
\end{equation}
where $\omega_{a}$ represents the eigenfrequency of the optical cavity, while $\hat{a}^{\dagger}$ and $\hat{a}$ are the creation and annihilation operators for bosons, respectively, adhering to the commutation relations $[\hat{a}, \hat{a}^{\dagger}]=1$ and $[\hat{a}, \hat{a}]=[\hat{a}^{\dagger}, \hat{a}^{\dagger}]=0$. The frequency $\omega_{0}$ corresponds to the transition between two energy levels of a single qubit within a system of $N$ identical qubits. The parameter $g$ denotes the coupling strength between the single-mode optical field and an individual qubit. The operators $\hat{\sigma}_{\pm}^{i}=\frac{1}{2}(\hat{\sigma}_{x}^{i}\pm i\hat{\sigma}_{y}^{i})$ represent the raising and lowering operators for the $i$th qubit, where $\hat{\sigma}_{x}^{i}$, $\hat{\sigma}_{y}^{i}$, and $\hat{\sigma}_{z}^{i}$ are the standard Pauli operators.

We initially investigate the scenario where a single-mode optical field interacts with two qubits ($N=2$). Assuming resonance such that $\omega_{a}=\omega_{0}$ and working within the interaction picture, the unitary evolution operator of the system is represented as \cite{fujii2004explicit}
\begin{eqnarray}
\ensuremath{\hat{U}(t)=\exp(-i\hat{H}_{int}t)=\left[\begin{array}{cccc}
\hat{U}_{11} & \hat{U}_{12} & \hat{U}_{13} & \hat{U}_{14}\\
\hat{U}_{21} & \hat{U}_{22} & \hat{U}_{23} & \hat{U}_{24}\\
\hat{U}_{31} & \hat{U}_{32} & \hat{U}_{33} & \hat{U}_{34}\\
\hat{U}_{41} & \hat{U}_{42} & \hat{U}_{43} & \hat{U}_{44}
\end{array}\right]}, \label{Eq2}
\end{eqnarray}
where $\hat{H}_{int}=g\sum_{i=1}^{2}(\hat{a}^{\dagger}\hat{\sigma}_{-}^{i}+\hat{\sigma}_{+}^{i}\hat{a})$ and all the matrix elements in Eq.~(\ref{Eq2}) are provided in Appendix \ref{Appendix A}.

In the following, we focus on the effect of the initial state parameters of the two qubits, such as the initial excited-state weights of the qubits and the initial coherence, on the quantum entanglement between the two qubits under the action of a single-photon light field.  Firstly, we prepare the light field in a single-photon number state $|1\rangle$ with the two qubits in a superposition of the ground state $|g\rangle$ and excited state $|e\rangle$, respectively. Then, the initial state of the total system is
\begin{eqnarray}
|\Psi(0)\rangle &=& (\alpha_{1}|e\rangle + \beta_{1}|g\rangle) \otimes (\alpha_{2}|e\rangle + \beta_{2}|g\rangle) \otimes |1\rangle \nonumber\\
&=&\left[\begin{array}{c}
\alpha_{1}\alpha_{2}|1\rangle \\
\alpha_{1}\beta_{2}|1\rangle \\
\beta_{1}\alpha_{2}|1\rangle \\
\beta_{1}\beta_{2}|1\rangle
\end{array}\right], \label{Eq3}
\end{eqnarray}
where $\left|\alpha_{1}\right|^{2}+\left|\beta_{1}\right|^{2}=1$, and $\left|\alpha_{2}\right|^{2}+\left|\beta_{2}\right|^{2}=1$.Then, the state of the system at time $t$ is
\begin{eqnarray}
|\Psi(t)\rangle&=&\hat{U}(t)|\Psi(0)\rangle\nonumber\\
&=&\left[\begin{array}{cccc}
\hat{U}_{11} & \hat{U}_{12} & \hat{U}_{13} & \hat{U}_{14}\\
\hat{U}_{21} & \hat{U}_{22} & \hat{U}_{23} & \hat{U}_{24}\\
\hat{U}_{31} & \hat{U}_{32} & \hat{U}_{33} & \hat{U}_{34}\\
\hat{U}_{41} & \hat{U}_{42} & \hat{U}_{43} & \hat{U}_{44}
\end{array}\right]\left[\begin{array}{c}
\alpha_{1}\alpha_{2}|1\rangle \\
\alpha_{1}\beta_{2}|1\rangle \\
\beta_{1}\alpha_{2}|1\rangle \\
\beta_{1}\beta_{2}|1\rangle
\end{array}\right]\nonumber\\
&=&\left[\begin{array}{c}
\Psi_{1}\\
\Psi_{2}\\
\Psi_{3}\\
\Psi_{4}
\end{array}\right],  \label{Eq4}
\end{eqnarray}
where
\begin{eqnarray}
\Psi_{1}&=&\hat{U}_{11}|1\rangle\alpha_{1}\alpha_{2} + \hat{U}_{12}|1\rangle\alpha_{1}\beta_{2} + \hat{U}_{13}|1\rangle\beta_{1}\alpha_{2} \nonumber\\
&&+ \hat{U}_{14}|1\rangle\beta_{1}\beta_{2},   \label{Eq5}\\
\Psi_{2}&=&\hat{U}_{21}|1\rangle\alpha_{1}\alpha_{2} + \hat{U}_{22}|1\rangle\alpha_{1}\beta_{2} + \hat{U}_{23}|1\rangle\beta_{1}\alpha_{2} \nonumber\\
&&+ \hat{U}_{24}|1\rangle\beta_{1}\beta_{2},  \label{Eq6}\\
\Psi_{3}&=&\hat{U}_{31}|1\rangle\alpha_{1}\alpha_{2} + \hat{U}_{32}|1\rangle\alpha_{1}\beta_{2} + \hat{U}_{33}|1\rangle\beta_{1}\alpha_{2} \nonumber\\
&&+ \hat{U}_{34}|1\rangle\beta_{1}\beta_{2},   \label{Eq7}\\
\Psi_{4}&=&\hat{U}_{41}|1\rangle\alpha_{1}\alpha_{2} + \hat{U}_{42}|1\rangle\alpha_{1}\beta_{2} + \hat{U}_{43}|1\rangle\beta_{1}\alpha_{2} \nonumber\\
&&+ \hat{U}_{44}|1\rangle\beta_{1}\beta_{2}.  \label{Eq8}
\end{eqnarray}
Meanwhile, by tracing out the states of the light field, we can obtain the reduced density matrix of the two qubits at time $t$ as follows
\begin{eqnarray}
\hat{\rho}_{q}(t)&=&\textbf{Tr}_{a}[|\Psi(t)\rangle\langle\Psi(t)|] \nonumber\\
&=&\sum_{n=0}^{\infty}\langle n|\Psi(t)\rangle\langle\Psi(t)|n\rangle \nonumber \\
&=&\left[\begin{array}{cccc}
q_{11} & q_{12} & q_{13} & q_{14}\\
q_{12}^{*} & q_{22} & q_{23} & q_{24}\\
q_{13}^{*} & q_{23}^{*} & q_{33} & q_{34}\\
q_{14}^{*} & q_{24}^{*} & q_{34}^{*} & q_{44}
\end{array}\right],  \label{Eq9}
\end{eqnarray}
where $\textbf{Tr}_{a}[\cdot]$ denotes tracing out the light field state of the total system, $|n\rangle$ denotes the number state of the light field, and
\begin{eqnarray}
q_{11}&=&|\alpha_{1}|^{2} |\alpha_{2}|^{2} f_{1}^{2}(gt) + \Big( |\alpha_{1}|^{2} |\beta_{2}|^{2} + 2\textbf{Re}(\beta_{1} \alpha_{2} \alpha_{1}^{*} \beta_{2}^{*}) \nonumber\\
&&+ |\beta_{1}|^{2} |\alpha_{2}|^{2} \Big) g^{2}(6, gt),  \label{Eq10}\\
q_{12}&=&|\alpha_{1}|^{2} \alpha_{2} \beta_{2}^{*} f_{1}(gt)f_{2}(gt) + \alpha_{1} \beta_{1}^{*} |\alpha_{2}|^{2} f_{1}(gt)f_{3}(gt) \nonumber\\
&&+ \left( \alpha_{1} \beta_{1}^{*} |\beta_{2}|^{2} + \alpha_{2} \beta_{2}^{*} |\beta_{1}|^{2} \right) g(6, gt)g(2, gt) ,    \label{Eq11}\\
q_{13}&=&|\alpha_{1}|^{2} \alpha_{2} \beta_{2}^{*} f_{1}(gt)f_{3}(gt) + \alpha_{1} \beta_{1}^{*} |\alpha_{2}|^{2} f_{1}(gt)f_{2}(gt) \nonumber\\
&&+ \left( \alpha_{1} \beta_{1}^{*} |\beta_{2}|^{2} + \alpha_{2} \beta_{2}^{*} |\beta_{1}|^{2} \right) g(6, gt)g(2, gt),\label{Eq12} \\
q_{14}&=&\alpha_{1} \alpha_{2} \beta_{1}^{*} \beta_{2}^{*} f_{1}(gt) \cos(gt \sqrt{2}),\label{Eq13}\\
q_{22}&=&2 |\alpha_{1}|^{2} |\alpha_{2}|^{2} g^{2}(10, gt) + |\alpha_{1}|^{2} |\beta_{2}|^{2} f_{2}^{2}(gt) \nonumber\\
&&+ 2\textbf{Re}(\beta_{1} \alpha_{2} \alpha_{1}^{*} \beta_{2}^{*})f_{2}(gt)f_{3}(gt) + |\beta_{1}|^{2} |\alpha_{2}|^{2} f_{3}^{2}(gt) \nonumber\\
&&+ |\beta_{1}|^{2} |\beta_{2}|^{2} g^{2}(2, gt),\label{Eq14} \\
q_{23}&=&2 |\alpha_{1}|^{2} |\alpha_{2}|^{2} g^{2}(10, gt) + \beta_{1} \alpha_{2} \alpha_{1}^{*} \beta_{2}^{*} f_{3}^{2}(gt) + \alpha_{1} \beta_{2} \beta_{1}^{*} \nonumber\\
&&\times\alpha_{2}^{*} f_{2}^{2}(gt) + |\beta_{1}|^{2} |\beta_{2}|^{2} g  ^{2}(2, gt) + (|\alpha_{1}|^{2} |\beta_{2}|^{2} \nonumber\\
&&+ |\beta_{1}|^{2} |\alpha_{2}|^{2})f_{2}(gt)f_{3}(gt),\label{Eq15}\\
q_{24}&=&2(|\alpha_{1}|^{2} \alpha_{2} \beta_{2}^{*} + \alpha_{1} \beta_{1}^{*} |\alpha_{2}|^{2})g(6, gt)g(10, gt) + [\alpha_{1} \nonumber\\
&&\times\beta_{1}^{*} |\beta_{2}|^{2} f_{2}(gt) + \alpha_{2} \beta_{2}^{*} |\beta_{1}|^{2} f_{3}(gt)] \cos(gt \sqrt{2}),\label{Eq16}\\
q_{33}&=& 2 |\alpha_{1}|^{2} |\alpha_{2}|^{2} g^{2}(10, gt) + |\alpha_{1}|^{2} |\beta_{2}|^{2} f_{3}^{2}(gt) + (\beta_{1} \alpha_{2} \nonumber\\
&&\times\alpha_{1}^{*} \beta_{2}^{*} + \alpha_{1} \beta_{2} \beta_{1}^{*} \alpha_{2}^{*})f_{2}(gt)f_{3}(gt) + |\beta_{1}|^{2} |\alpha_{2}|^{2} \nonumber\\
&&\times f_{2}^{2}(gt)+ |\beta_{1}|^{2} |\beta_{2}|^{2} g^{2}(2, gt),\label{Eq17}\\
q_{34}&=&2(\alpha_{2} \beta_{2}^{*} |\alpha_{1}|^{2} + \alpha_{1} \beta_{1}^{*} |\alpha_{2}|^{2})g(6, gt)g(10, gt) + [\alpha_{1} \beta_{1}^{*} \nonumber\\
&&\times|\beta_{2}|^{2} f_{3}(gt) + \alpha_{2} \beta_{2}^{*} |\beta_{1}|^{2} f_{2}(gt)] \cos(gt \sqrt{2}),\label{Eq18}\\
q_{44} &=& 2 \left( |\alpha_{1}|^{2} |\beta_{2}|^{2} + 2 \textbf{Re}(\beta_{1} \alpha_{2} \alpha_{1}^{*} \beta_{2}^{*}) + |\beta_{1}|^{2} |\alpha_{2}|^{2} \right) \nonumber\\
&&\times g^{2}(6, gt) + |\beta_{1}|^{2} |\beta_{2}|^{2} \cos^{2}\left(gt \sqrt{2}\right)
,\label{Eq19} 
\end{eqnarray}
where 
\begin{eqnarray}
f_{1}(gt)&=&\frac{3+2\cos(gt\sqrt{10})}{5},f_{2}(gt)=\frac{\cos(gt\sqrt{6})+1}{2} ,\label{Eq20}\\
f_{3}(gt)&=&\frac{\cos(gt\sqrt{6})-1}{2},g(x,gt)=\frac{\sin(gt\sqrt{x})}{\sqrt{x}}. \label{Eq21} 
\end{eqnarray}

\section{The effect of single photon on quantum entanglement of two qubits}\label{sec3}

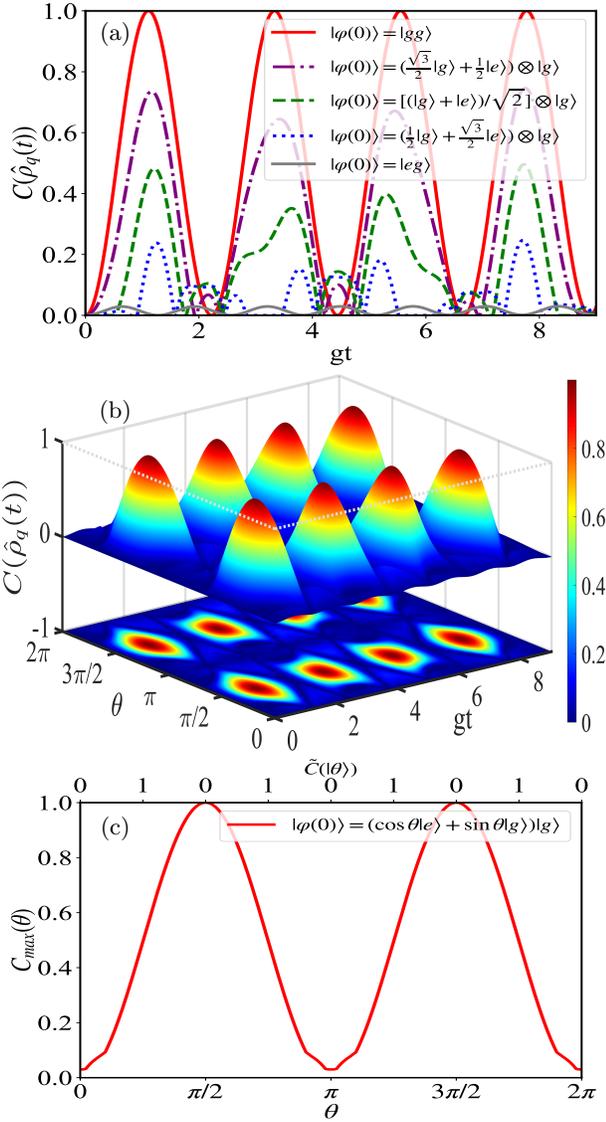
\begin{figure}[t]
	\centering
	\begin{tikzpicture}
	\draw (0, 0) node[inner sep=0] {\includegraphics[width=8cm,height=5cm]{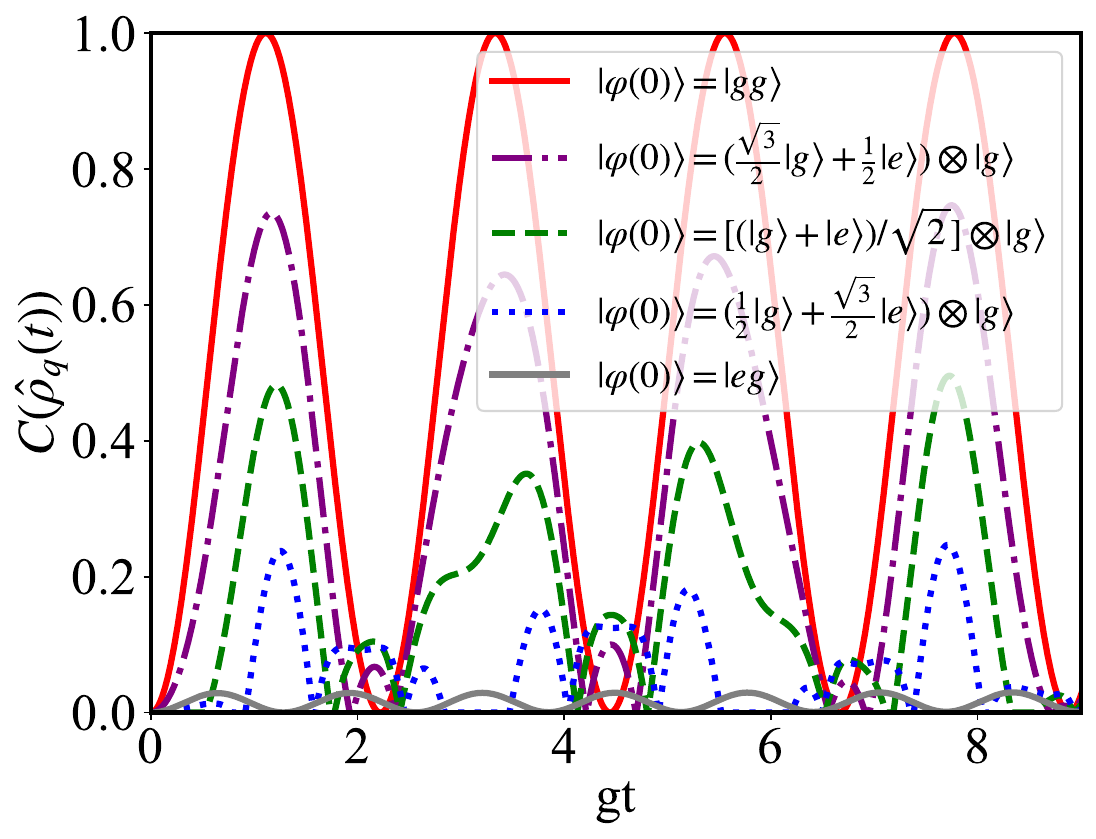}};
	\draw (-2.5, 2) node {(a)};
	\end{tikzpicture}
	\begin{tikzpicture}
	\draw (0, 0) node[inner sep=0] {\includegraphics[width=8cm,height=5cm]{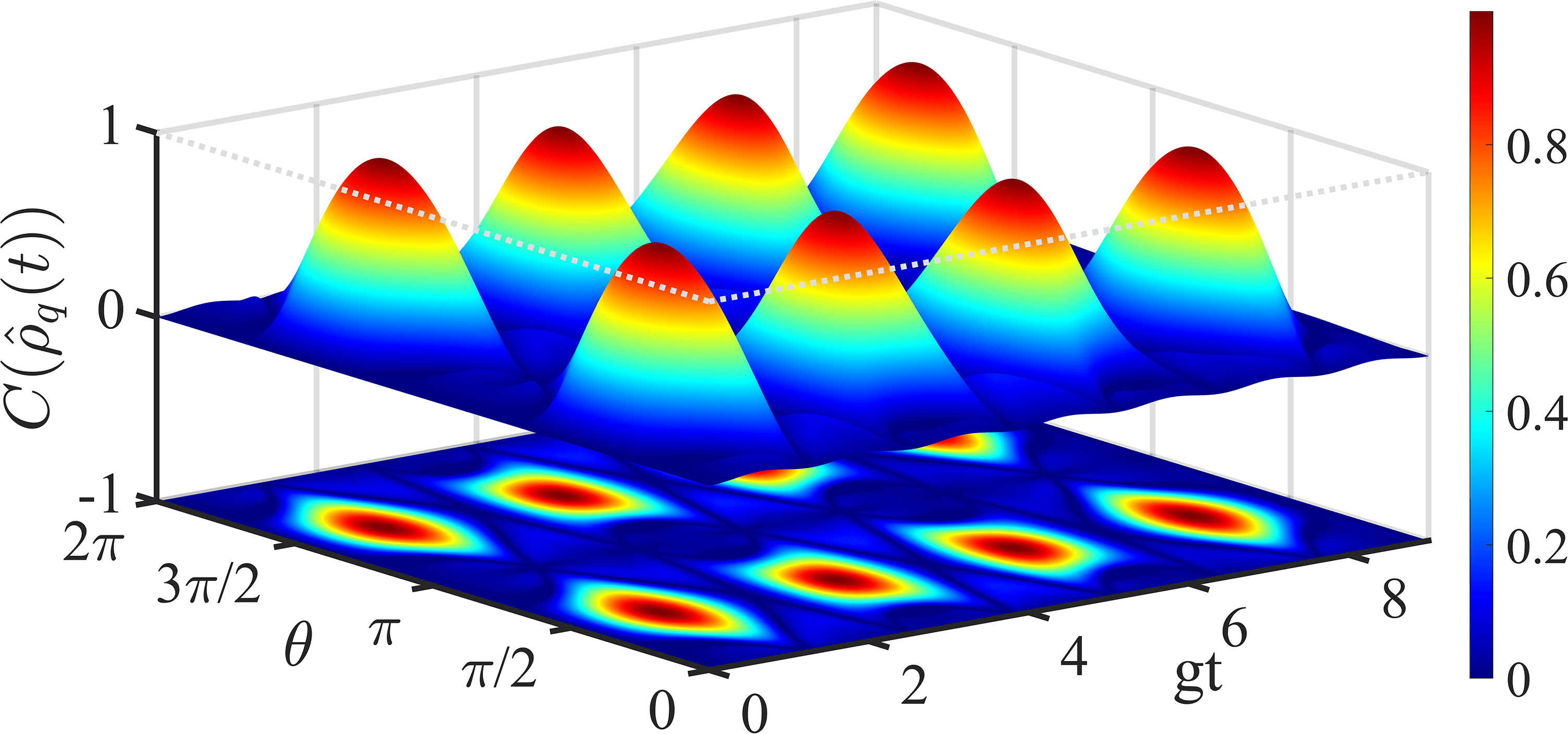}};
	\draw (-2.5, 2) node {(b)};
	\end{tikzpicture}
	\begin{tikzpicture}
	\draw (0, 0) node[inner sep=0] {\includegraphics[width=8cm,height=5cm]{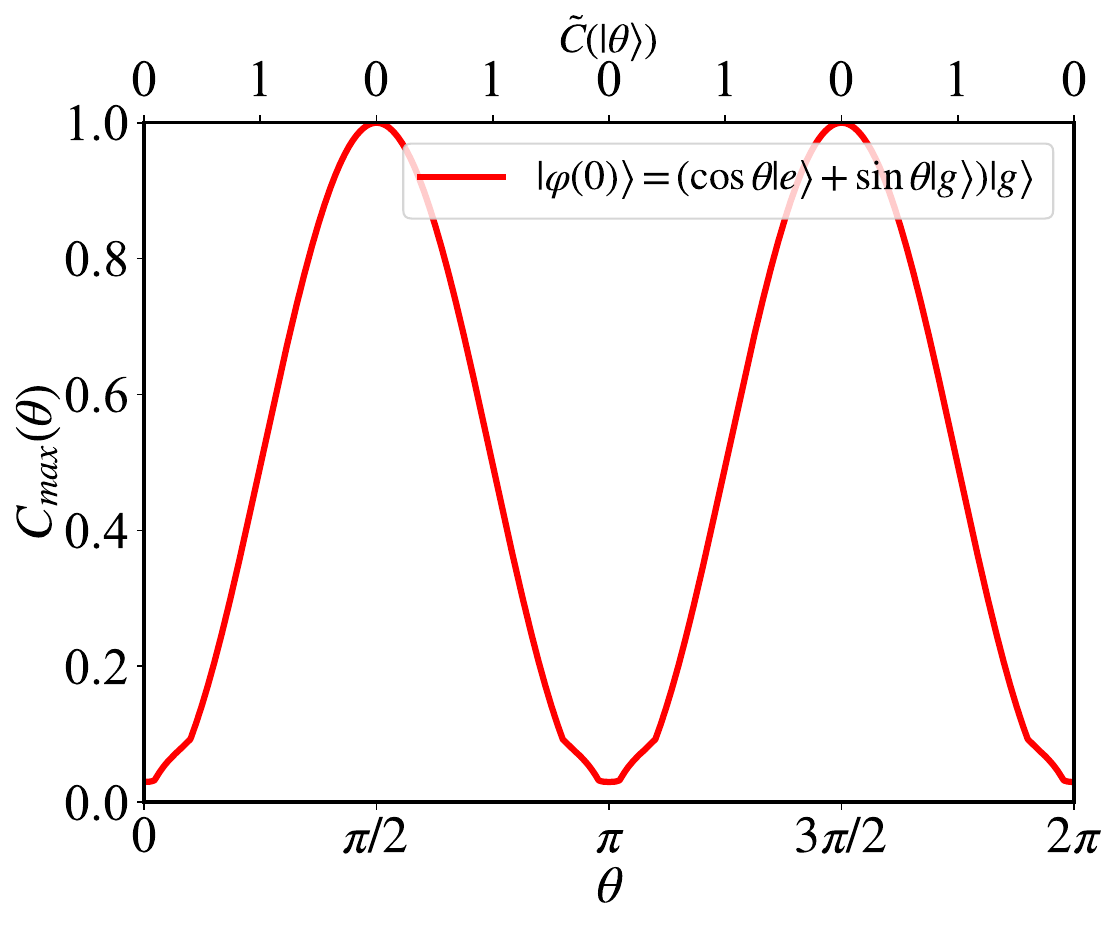}};
	\draw (-2.5, 1.5) node {(c)};
	\end{tikzpicture}
	\caption{When the light field is initially in a single-photon number state and one of the qubits is initially in the ground state, (a) the figure represents the variation of the quantum Concurrence between the two qubits with time $gt$ when the other qubit is in a different initial state, (b) the figure represents the variation of the quantum Concurrence between the two qubits with the parameter $\theta$ of the initial state of the other qubit and time $gt$, and (c) the figure represents the variation of the maximum quantum Concurrence between the two qubits with the parameter $\theta$ of the initial state of the other qubit.\label{fig2}}
\end{figure}

In the following, in a single-photon light field interacting system with two qubits, we investigate the effects of the initial excited state weights and the initial coherence of the qubits on the quantum entanglement between the two qubits, respectively. We choose the initial state of the two qubits to be 
\begin{equation}
|\varphi(0)\rangle=\left(\cos \theta|e\rangle+\sin \theta|g\rangle\right)\otimes\left|g\right\rangle=\left|\theta\right\rangle\otimes\left|g\right\rangle, \label{Eq22}
\end{equation}
that is, $\alpha_{1}=\cos\theta$, $\beta_{1}=\sin\theta$, $\alpha_{2}=0$, and $\beta_{2}=1$ in Eq.~(\ref{Eq3}). $\left|\theta\right\rangle=\left(\cos \theta|e\rangle+\sin \theta|g\rangle\right)$, $\theta\in[0, 2\pi)$.   The matrix elements of the reduced density matrix of the two qubits at time $t$ is then as follows
\begin{eqnarray}
q_{11}&=& g^{2}(6,gt)\cos^{2}\theta,  \label{Eq23}\\
q_{12}&=&\sin\theta\cos\theta g(6,gt)g(2,gt),     \label{Eq24}\\
q_{13}&=&\sin\theta\cos\theta g(6,gt)g(2,gt), \label{Eq25}\\
q_{14}&=&0 ,  \label{Eq26}\\
q_{22}&=&\cos^{2}\theta f_{2}^{2}(gt)+\sin^{2}\theta g^{2}(2,gt ), \label{Eq27}\\
q_{23}&=&\sin^{2}\theta g^{2}(2,gt)+\cos^{2}\theta f_{2}(gt)f_{3}(gt), \label{Eq28}\\
q_{24}&=&\sin\theta\cos\theta f_{2}(gt)\cos(gt\sqrt{2}), \label{Eq29}\\
q_{33}&=&\cos^{2}\theta f_{3}^{2}(gt)+\sin^{2}\theta g^{2}(2,gt), \label{Eq30}\\
q_{34}&=&\sin\theta\cos\theta f_{3}(gt)\cos(gt\sqrt{2}), \label{Eq31}\\
q_{44}&=&2\cos^{2}\theta g^{2}(6,gt)+\sin^{2}\theta \cos^{2}(gt\sqrt{2}) . \label{Eq32}
\end{eqnarray}

Now, we have obtained all the matrix elements of the reduced density matrices of the two qubits at the time $t$, so we can easily study the quantum entanglement between them in relation to the excited-state weights and coherence of the qubits initial state. To facilitate the subsequent discussion, we will now describe the measures of coherence and the measures of quantum entanglement between two qubits, according to Ref.~\cite{PhysRevLett.113.140401,RevModPhys.89.041003}  and Ref.~\cite{PhysRevLett.80.2245}, respectively. According to Ref.~\cite{PhysRevLett.113.140401}, for a quantum state $\hat{\rho}$, the magnitude of the coherence is
\begin{equation}
\tilde{C}(\hat{\rho})=\sum_{i,j,i\neq j}|\hat{\rho}_{ij}|, \label{Eq33}
\end{equation}
which sums the absolute values of all off-diagonal elements of the density matrix $\hat{\rho}$.

For a quantum state $\hat{\rho}$, the degree of quantum entanglement is quantified using the function \cite{PhysRevLett.80.2245}
\begin{eqnarray}
E(C(\hat{\rho}))&=&h(\frac{1+\sqrt{1-C^{2}(\hat{\rho})}}{2}),  \label{Eq34}
\end{eqnarray}
where $h(x)=-x\log_{2}x-(1-x)\log_{2}(1-x)$, and $E(C(\hat{\rho}))$ is the entanglement of formation, and 
\begin{equation}
C(\hat{\rho})=\max\{0, \lambda_{1}-\lambda_{2}-\lambda_{3}-\lambda_{4}\} \label{Eq35}
\end{equation}
is defined as the quantum concurrence. $\lambda_i$ (where $i=1,2,3,4$) are the square roots of the eigenvalues of the matrix $\hat{\rho}\hat{\varrho}$, arranged in descending order such that $\lambda_{1}>\lambda_{2}>\lambda_{3}>\lambda_{4}$. Here, $\hat{\varrho}$ is defined as $(\hat{\sigma}_{y}\otimes\hat{\sigma}_{y})\hat{\rho}^{*}(\hat{\sigma}_{y}\otimes\hat{\sigma}_{y})$,where $\hat{\sigma}_{y}$ represents the Pauli-Y matrix, and $\hat{\rho}^{*}$ is the complex conjugate of $\hat{\rho}$. Given that the quantum concurrence $C(\hat{\rho})$ ranges monotonically from $0$ to $1$, mirroring the behavior of the function $E(C(\hat{\rho}))$ from $0$ to $1$, the quantum concurrence $C(\hat{\rho})$ is frequently used as a direct measure of the degree of quantum entanglement between two qubits. In this study, we employ the quantum concurrence $C(\hat{\rho})$ to quantify the quantum entanglement between two qubits \cite{PhysRevLett.80.2245}.

Having introduced the measures of quantum coherence and quantum entanglement for quantum states, we now proceed to investigate how the initial excited state weight and coherence of qubits affect the quantum entanglement between two qubits under the control of a single-photon light field. As shown in Eq.~(\ref{Eq22}), at the initial moment, when one qubit is in the ground state and the other qubit is in the superposition of the ground and excited states, we can obtain the quantum entanglement of the two qubits with respect to the initial state parameter, $\theta$, and time $t$ based on all the matrix elements shown in Eq.~(\ref{Eq23})-Eq.~(\ref{Eq32}) and the entanglement metric shown in Eq.~(\ref{Eq35}). In Fig.~\ref{fig2}a, at the initial moment, when one qubit is in the ground state and the other qubit is in the superposition state of the ground and excited states, we plot the quantum entanglement between the two qubits over time $t$ under different superposition coefficients. We find that the smaller the initial excited state weight of the qubit in the superposition state, the larger the maximum quantum entanglement between the two qubits under the action of a single-photon light field. This conclusion can also be obtained from Fig.~\ref{fig2}b, where the single-photon light field is able to trigger the two ground-state qubits to reach the maximally entangled state only when $\theta=l\pi/2$ ($l=1,3$). According to the above analysis, when the initial state of the two qubits is $|\varphi(0)\rangle=\pm|gg\rangle$, that is, $\theta=l\pi/2$ ($l=1,3$). Substituting these conditions into Eq.~(\ref{Eq23})-Eq.~(\ref{Eq32}), we can get the density matrix of the two qubits at time $t$ as follows
\begin{eqnarray}
\hat{\rho}_{q}(t)=\left[\begin{array}{cccc}
0 & 0 & 0 & 0\\
0 & \frac{\sin^{2}(gt\sqrt{2})}{2} & \frac{\sin^{2}(gt\sqrt{2})}{2} & 0\\
0 & \frac{\sin^{2}(gt\sqrt{2})}{2} & \frac{\sin^{2}(gt\sqrt{2})}{2} & 0\\
0 & 0 & 0 & \cos^{2}(gt\sqrt{2})
\end{array}\right]. \label{Eq36}
\end{eqnarray}
Obviously, if $gt\sqrt{2}=\frac{k\pi}{2}$ ($k$ is a odd number), i.e., $gt=\frac{k\pi}{2\sqrt{2}}$, then $\frac{\sin^{2}(gt\sqrt{2})}{2}=\frac{1}{2}$ and $\cos^{2}(gt\sqrt{2})=0$. We can obtain a maximum entangled state between the two qubits
\begin{eqnarray}
\hat{\rho}_{q}(t)=\left[\begin{array}{cccc}
0 & 0 & 0 & 0\\
0 & \frac{1}{2} & \frac{1}{2} & 0\\
0 & \frac{1}{2} & \frac{1}{2} & 0\\
0 & 0 & 0 & 0
\end{array}\right]. \label{Eq37}
\end{eqnarray}
It is the density matrix form of the Bell state $|B\rangle=(|eg\rangle+|ge\rangle)/\sqrt{2}$ and it is generated by the single photon-mediated interaction of two qubits that are both initially in the ground state. The matrix elements image of the density matrix when the two qubits are maximally entangled from the initial both in the ground state has been given in Ref.~\cite{Jiang_2024} and we will not give it here. We focus here on the effect of the initial excited state weights of the qubit under the action of a single-photon light field on the quantum entanglement between two qubits. In order to show more intuitively the effect of the initial excited state weights on the quantum entanglement between two qubits, in Fig.~\ref{fig2}c we plot the variation of the maximum quantum entanglement between two qubits with respect to the initial state parameter $\theta$ or with respect to the initial coherence $\tilde{C}(|\theta\rangle)$. The maximum quantum entanglement between two qubits with respect to the initial state parameter $\theta$ is shown in Fig.~\ref{fig2}c. From the figure, we can intuitively see that the maximum quantum entanglement between two qubits is larger when $\theta$ converges more to $l\pi/2$ ($l=1,3$), and smaller when $\theta$ converges more to $k\pi$ ($k=0,1,2$). Therefore, under the action of single-photon light field, too large initial excited state weights will reduce the maximum quantum entanglement between two qubits. In addition, we can also see from Fig.~\ref{fig2}c that when the qubit in the superposition state initially has maximum coherence, the maximum quantum entanglement between two qubits induced by it is not 1.

In summary, when two qubits are initially in $|\varphi(0)\rangle=\left(\cos \theta|e\rangle+\sin \theta|g\rangle\right)\otimes\left|g\right\rangle$, under the action of single-photon light field, the larger the value of $\sin \theta$, the larger the quantum entanglement generated between the two qubits and able to reach 1, and the larger the value of $\cos \theta$, the smaller the maximum quantum entanglement generated between the two qubits. Therefore, under the action of single-photon light field, the excited state weights reduce the value of maximum entanglement between two qubits.

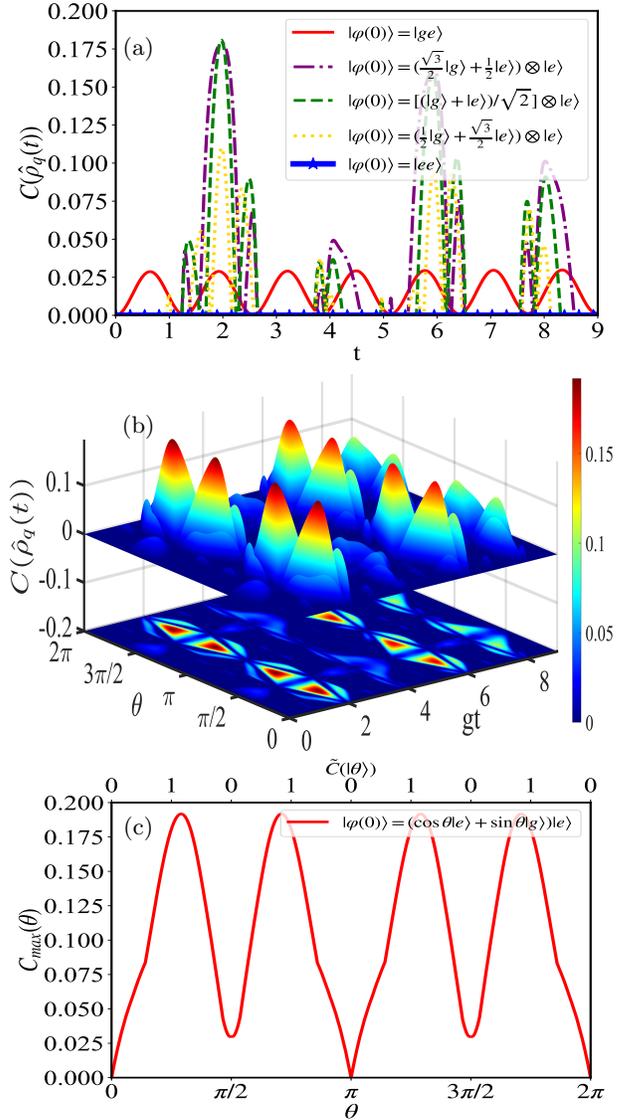
\begin{figure}[t]
	\centering
	\begin{tikzpicture}
	\draw (0, 0) node[inner sep=0] {\includegraphics[width=8cm,height=5cm]{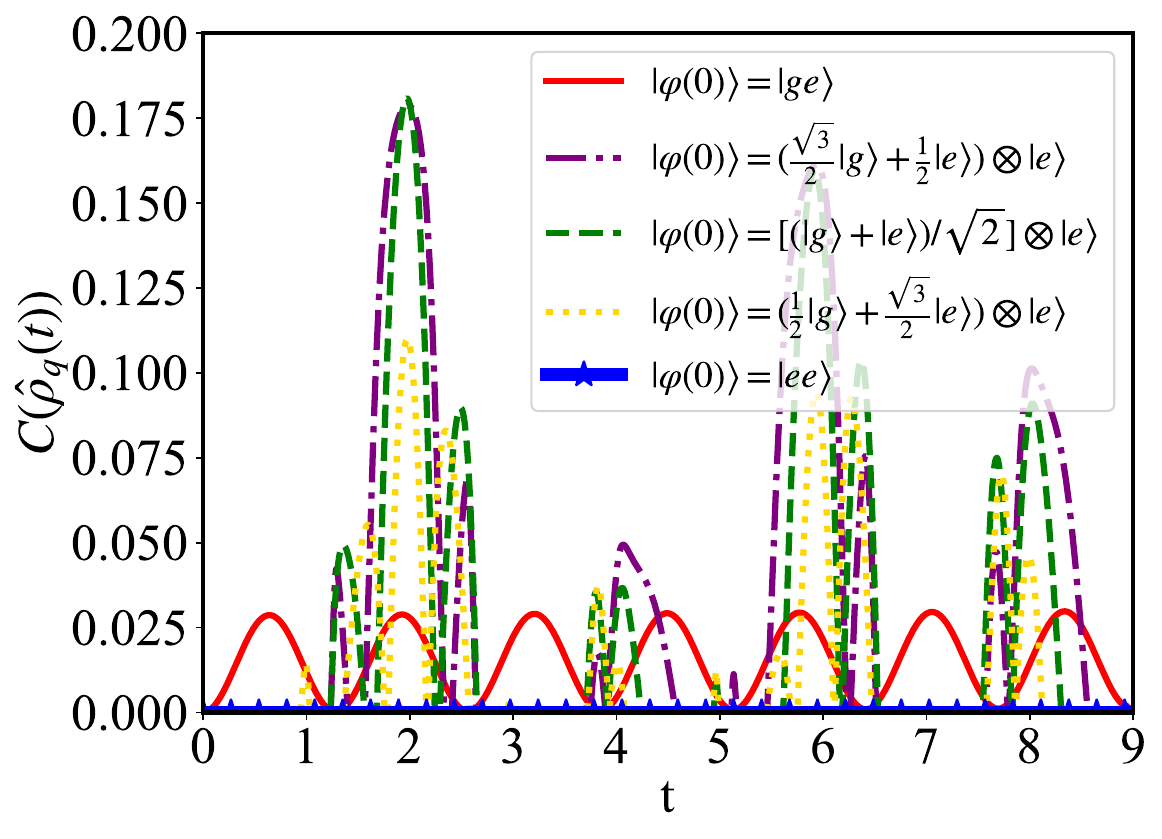}};
	\draw (-2.3, 1.8) node {(a)};
	\end{tikzpicture}
	\begin{tikzpicture}
	\draw (0, 0) node[inner sep=0] {\includegraphics[width=8cm,height=5cm]{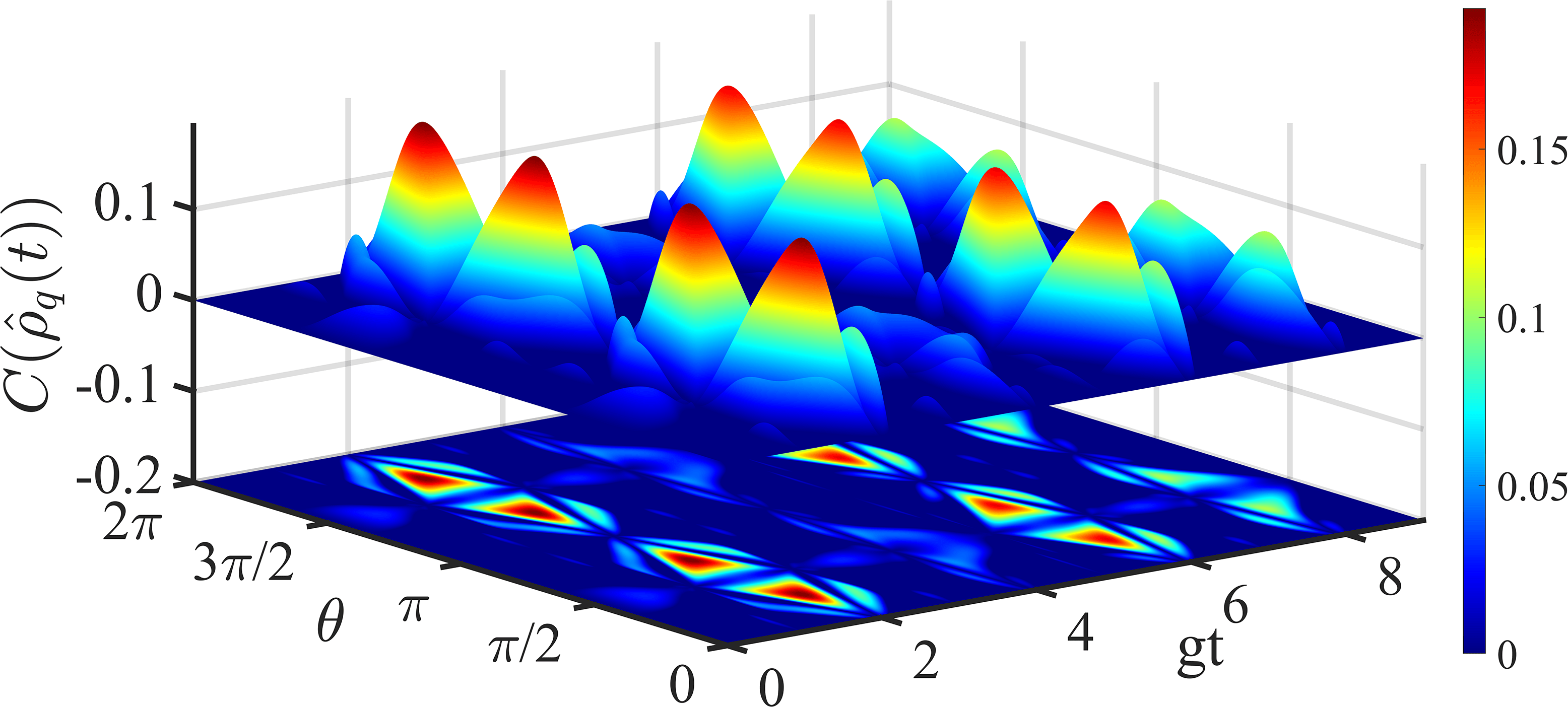}};
	\draw (-2.3, 1.8) node {(b)};
	\end{tikzpicture}
	\begin{tikzpicture}
	\draw (0, 0) node[inner sep=0] {\includegraphics[width=8cm,height=5cm]{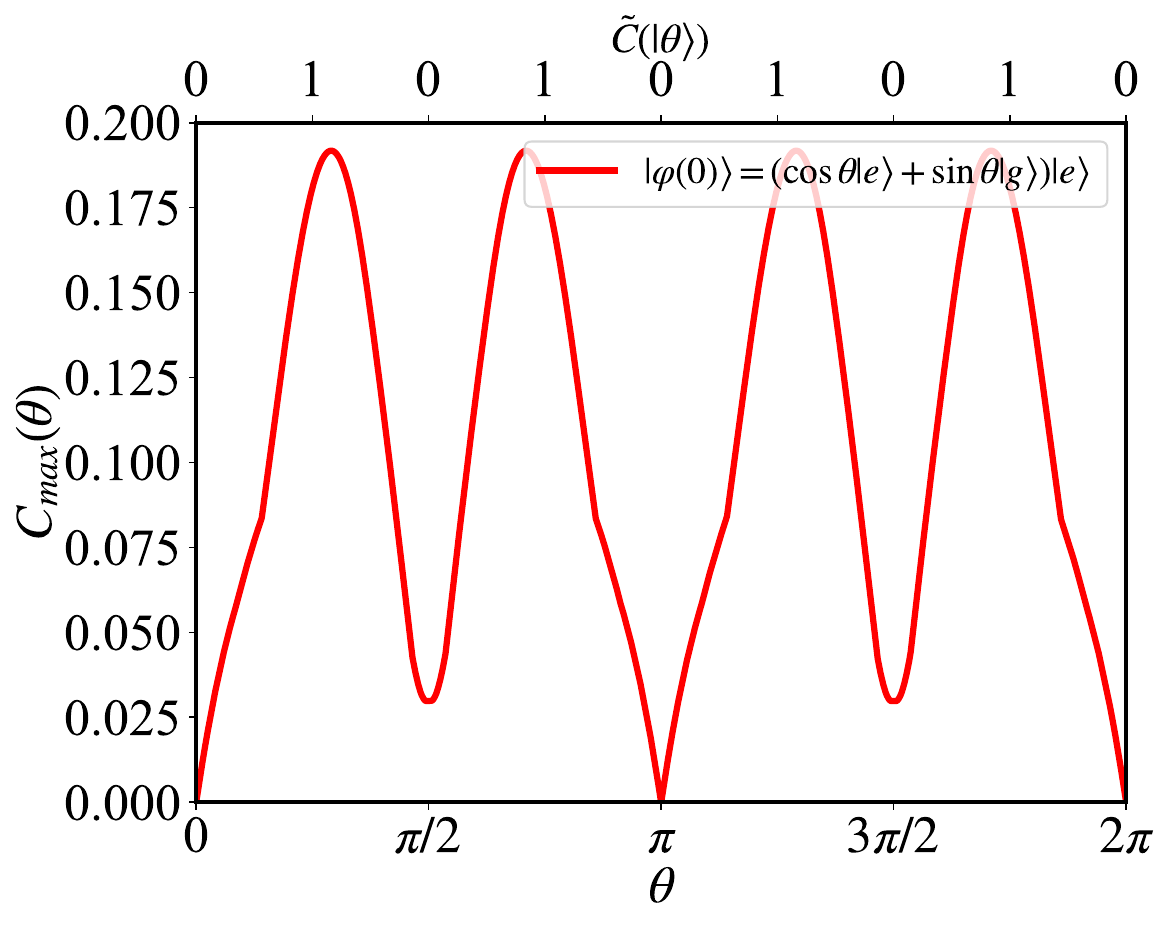}};
	\draw (-2.3, 1.5) node {(c)};
	\end{tikzpicture}
	\caption{When the light field is initially in a single-photon number state and one of the qubits is initially in the excited state, (a) the graph shows the variation of quantum Concurrence between two qubits over time $gt$ with the other qubit in different initial states, (b) the graph shows the change in quantum Concurrence between two qubits with the initial state parameter $\theta$ of the other qubit and time $gt$, (c) the graph shows the change in maximum quantum Concurrence between two qubits with the initial state parameter $\theta$ or with the initial coherence $\tilde{C}(|\theta\rangle)$ of the other qubit.\label{fig3} }
\end{figure}

Similarly to the above, when the initial state of the two-qubit system is
\begin{equation}
|\varphi(0)\rangle=\left(\cos \theta|e\rangle+\sin \theta|g\rangle\right)\otimes\left|e\right\rangle=|\theta\rangle\otimes\left|e\right\rangle. \label{Eq38}
\end{equation} 
That is, when one of the two-qubit is in an excited state and the another qubit is in a superposition of excited and ground states (i.e., $\alpha_{1}=\cos\theta$, $\beta_{1}=\sin\theta$, $\alpha_{2}=1$, and $\beta_{2}=0$ in Eq.~(\ref{Eq3})), we study the effect of excited state weights of qubits on quantum entanglement between two qubits under the action of a single-photon light field. From the initial conditions, we can get all matrix elements of the density matrix of the two qubits at time $t$ are
\begin{eqnarray}
q_{11}&=&\cos^{2}\theta f_{1}^{2}(gt)+\sin^{2}\theta g^{2}(6,gt), \label{Eq39}\\
q_{12}&=&\sin\theta\cos\theta f_{1}(gt)f_{3}(gt),     \label{Eq40}\\
q_{13}&=&\sin\theta\cos\theta f_{1}(gt)f_{2}(gt),\label{Eq41} \\
q_{14}&=&0,\label{Eq42}\\
q_{22}&=&2\cos^{2}\theta g^{2}(10,gt)+\sin^{2}\theta f_{3}^{2}(gt),\label{Eq43} \\
q_{23}&=&2\cos^{2}\theta g^{2}(10,gt)+\sin^{2}\theta f_{2}(gt)f_{3}(gt),\label{Eq44} \\
q_{24}&=&2\sin\theta\cos\theta g(6,gt)g(10,gt),\label{Eq45} \\
q_{33}&=&2\cos^{2}\theta g^{2}(10,gt)+\sin^{2}\theta f_{2}^{2}(gt),\label{Eq46} \\
q_{34}&=&2\sin\theta\cos\theta g(6,gt)g(10,gt),\label{Eq47} \\
q_{44}&=&2\sin^{2}\theta g(6,gt).\label{Eq48}
\end{eqnarray}

\begin{figure}[t]
\centering
\begin{tikzpicture}
\draw (0, 0) node[inner sep=0] {\includegraphics[width=8cm,height=5cm]{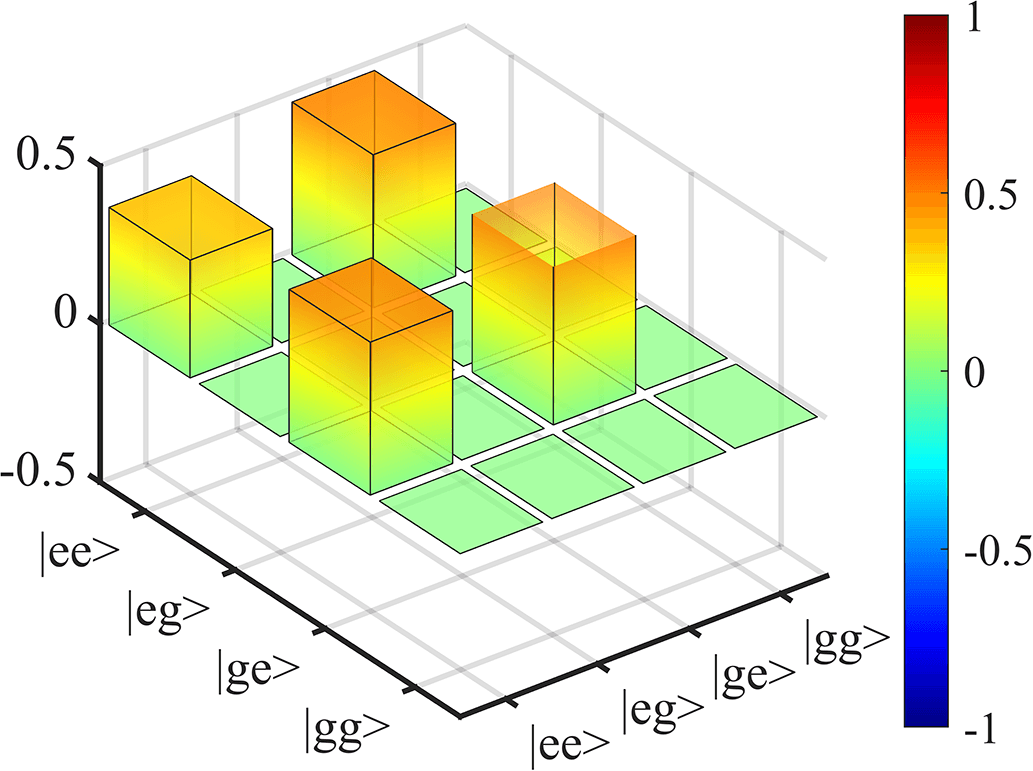}};
\draw (-2.5, 2) node {(a)};
\end{tikzpicture}
\begin{tikzpicture}
\draw (0, 0) node[inner sep=0] {\includegraphics[width=8cm,height=5cm]{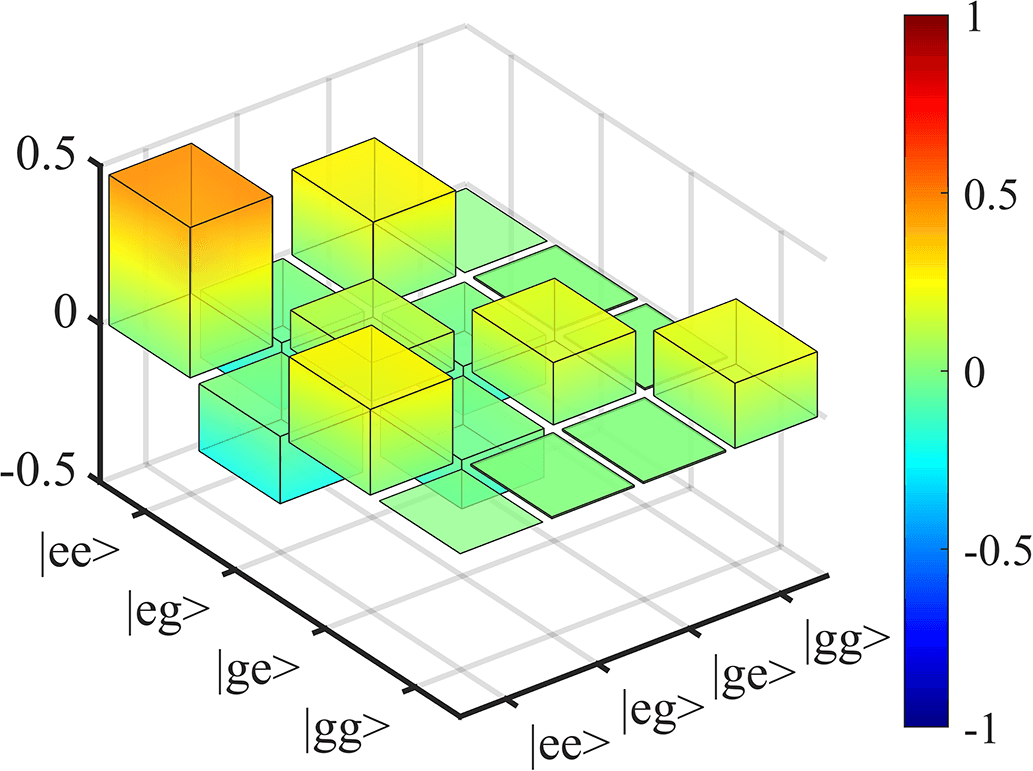}};
\draw (-2.5, 2) node {(b)};
\end{tikzpicture}
\caption{\label{fig4}From Fig.~\ref{fig3}c, we obtain an optimal initial state parameter $\cos\theta$. Based on this optimal parameter, when a qubit is in the excited state $|e\rangle$, we can obtain the initial state of the other qubit that makes the two qubits reach the maximum entangled state as $|\theta\rangle=\sqrt{0.37}|e\rangle+\sqrt{0.63}|g\rangle$. (a) shows the matrix element images of the density matrix for the optimal initial state of the two qubits.(b) displays the matrix element images of the density matrix for the maximally entangled state evolved from this optimal initial state.}
\end{figure}

When the initial state of the system is $|\varphi(0)\rangle=\left(\cos \theta|e\rangle+\sin \theta|g\rangle\right)\left|e\right\rangle$, based on all matrix elements of the density matrix of the two qubits above, we plot in Fig.~\ref{fig3}a the variation of the quantum concurrence of the two qubits over time $t$ for different superposition weights. We find that when both qubits are initially in the excited state, the single-photon light field cannot trigger these two qubits to produce quantum entanglement. In addition, when one qubit is in an excited state and the other qubit is in a superposition of excited and ground states, single photons can trigger the two qubits to produce quantum entanglement. However, it is worth noting that from Figs.~\ref{fig3}a and \ref{fig3}b, we can get that when the two qubits are introduced as Eq.~(\ref{Eq38}), the single photon is unable to trigger the two qubits to produce maximum entanglement with value 1 regardless of the value of $\theta$.  We draw the images of all density matrix elements of the initial state density matrix and the density matrix when the maximum entanglement is reached for the two qubits in Figs.~\ref{fig4}a and \ref{fig4}b, respectively. 

In order to study the effect of the initial coherence of the qubit on the quantum entanglement between two qubits under the action of a single-photon light field, in Fig.~\ref{fig3}c we plot the variation of the quantum concurrence of the two qubits with respect to the initial state parameter $\theta$ or with respect to the initial coherence $\tilde{C}(|\theta\rangle)$. From Fig.~\ref{fig3}c, we obtain the optimal initial state parameter $\cos\theta=\pm\sqrt{0.37}$. Based on this optimal parameter, we can obtain the optimal initial state of two qubits as $|\varphi(0)\rangle=\left(\sqrt{0.37}|e\rangle+\sqrt{0.63}|g\rangle\right)\otimes\left|e\right\rangle$. Following the measure of coherence magnitude given by Eq.~(\ref{Eq33}), we know that the quantum coherence of the two-qubits initial state is $\tilde{C}(|\theta\rangle)=0.96$. We can get a rough conclusion that when one qubit is in the excited state and the other qubit is in the superposition of the excited and the ground state, the larger initial coherence of the qubit in the superposition helps the two qubits to produce larger quantum entanglement. In addition, when the initial excited state weight of the qubit in the superposition state reaches the maximum, i.e., when the initial state of the two qubits is $|\varphi(0)\rangle=\left|ee\right\rangle$, the single-photon light field cannot trigger the generation of quantum entanglement between the two qubits.

In summary, when the initial state of the two qubits is $|\varphi(0)\rangle=|ee\rangle$, the two qubits cannot be triggered to produce quantum entanglement in a single-photon light field. When the initial state of the two qubits is $|\varphi(0)\rangle=\left(\sqrt{0.37}|e\rangle+\sqrt{0.63}|g\rangle\right)\otimes\left|e\right\rangle$, i.e., their initial coherence is $\tilde{C}(|\theta\rangle)=0.96$, which is a value converging to the maximum coherence 1, the quantum entanglement between the two qubits triggered by the single-photon light field reaches a larger value.

\begin{figure}[t]
\centering
\begin{tikzpicture}
\draw (0, 0) node[inner sep=0] {\includegraphics[width=8cm,height=5cm]{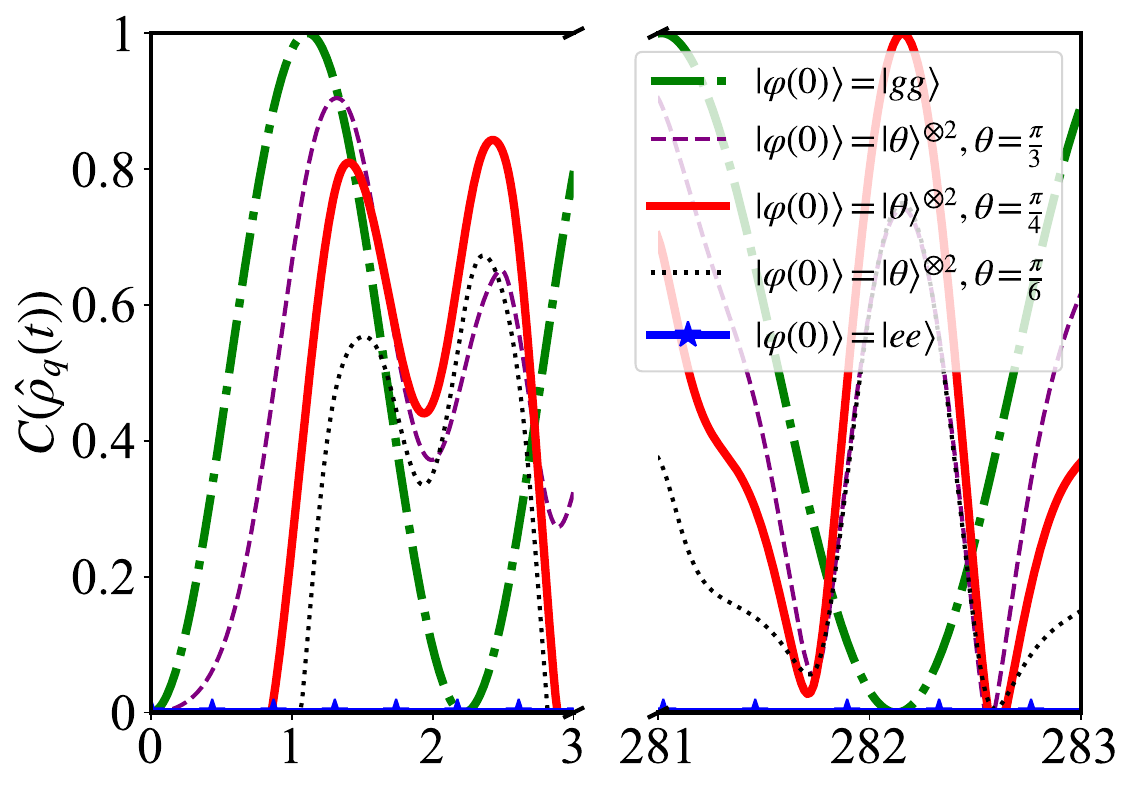}};
\draw (-2.7, 1.8) node {(a)};
\draw (0.4, -2.6) node {gt};
\end{tikzpicture}
\begin{tikzpicture}
\draw (0, 0) node[inner sep=0] {\includegraphics[width=8cm,height=5cm]{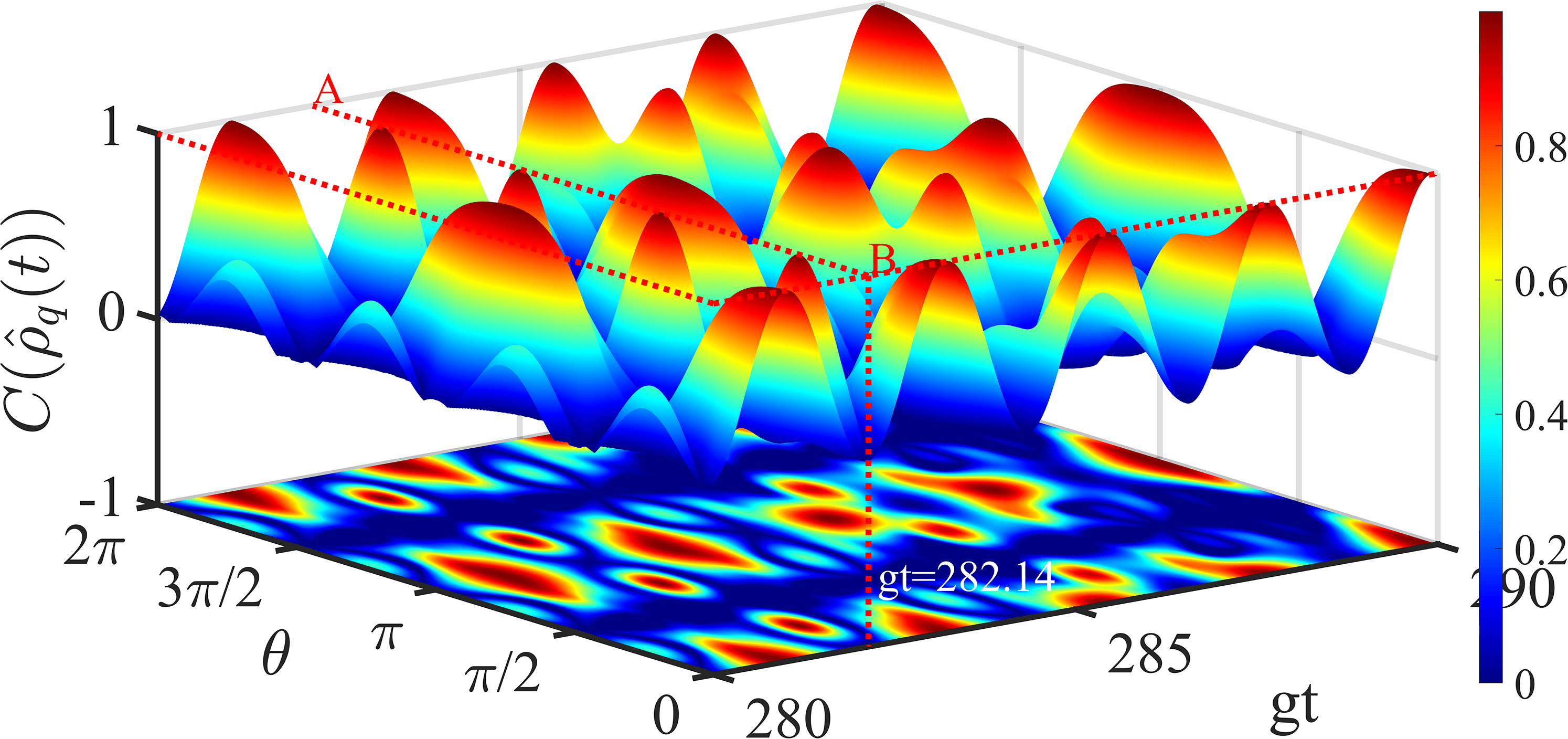}};
\draw (-2.7, 1.8) node {(b)};
\end{tikzpicture}
\begin{tikzpicture}
\draw (0, 0) node[inner sep=0] {\includegraphics[width=8cm,height=5cm]{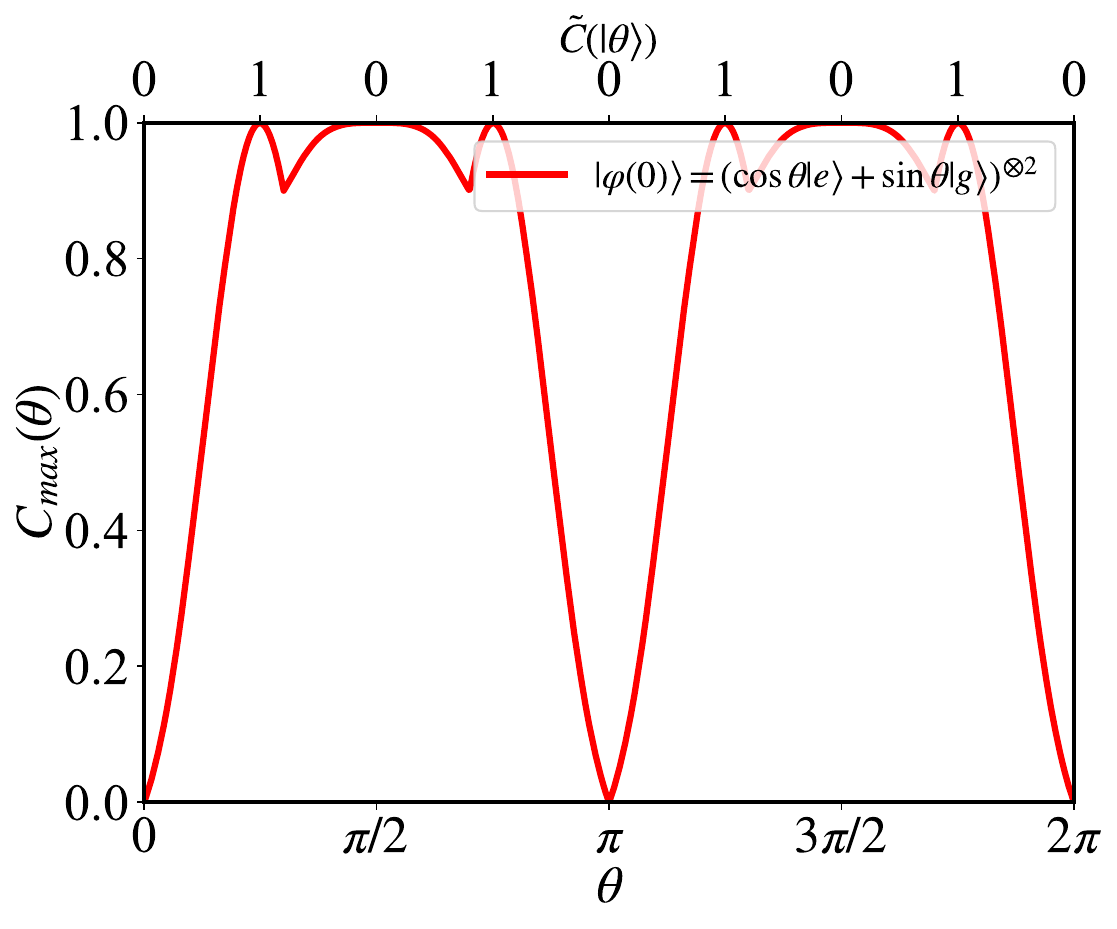}};
\draw (-2.7, 1.5) node {(c)};
\end{tikzpicture}
\caption{When the light field is initially in a single-photon number state, (a) the graph shows the change in quantum concurrence between two identical qubits over time $gt$ with different initial states, (b) the graph shows the change in quantum concurrence between two qubits with the initial state parameter $\theta$ and time $gt$, (c) the graph shows the change in maximum quantum concurrence between two qubits with the initial state parameter $\theta$. \label{fig5} }
\end{figure}
         
When two qubits are all identical qubits and are in a superposition of excited and ground states as follows
\begin{equation}
|\varphi(0)\rangle=|\theta\rangle\otimes|\theta\rangle=|\theta\rangle^{\otimes 2},\label{Eq49}
\end{equation}
where $|\theta\rangle=\left(\cos\theta|e\rangle+\sin\theta|g\rangle\right)$. Substituting the initial state parameters in Eq.~(\ref{Eq49}) into Eq.~(\ref{Eq10})-Eq.~(\ref{Eq19}) yields all matrix elements of the reduced density matrix of the two qubits at the time $t$ as 
\begin{eqnarray}
q_{11}&=&\cos^{4}\theta f_{1}^{2}(gt) +\sin^{2}(2\theta)   g^{2}(6,gt), \label{Eq50}  \\
q_{12}&=&\cos^{3}\theta\sin\theta f_{1}(gt)[f_{2}(gt)+f_{3}(gt)]+\sin (2\theta)\sin^{2}\theta  \nonumber\\
&&\times g(2,gt)g(6,gt),  \label{Eq51}   \\
q_{13}&=&q_{12} ,\label{Eq52}     \\
q_{14}&=&\sin^{2}\theta \cos^{2}\theta f_{1}(gt)\cos(gt\sqrt{2}) ,\label{Eq53}    \\
q_{22}&=&2\cos^{4}\theta g^{2}(10,gt)+\sin^{2}\theta \cos^{2}\theta [f_{2}(gt)+f_{3}(gt)]^{2} \nonumber\\
&&+ \sin^{4}\theta g^{2}(2,gt)  ,\label{Eq54}  \\
q_{23}&=& q_{22}  ,\label{Eq55}   \\
q_{24}&=&2 \cos^{2}\theta \sin(2\theta)g(6,gt)g(10,gt)+\cos\theta\sin^{3}\theta  \nonumber\\
&&\times[f_{2}(gt)+f_{3}(gt)]\cos(gt\sqrt{2})    ,\label{Eq56}\\
q_{33}&=&q_{22}    ,\label{Eq57}   \\
q_{34}&=&  q_{24}  ,\label{Eq58}   \\
q_{44}&=&2\sin^{2}(2\theta)g^{2}(6,gt)+\sin^{4}\theta\cos^{2}(gt\sqrt{2})   .\label{Eq59}    
\end{eqnarray}

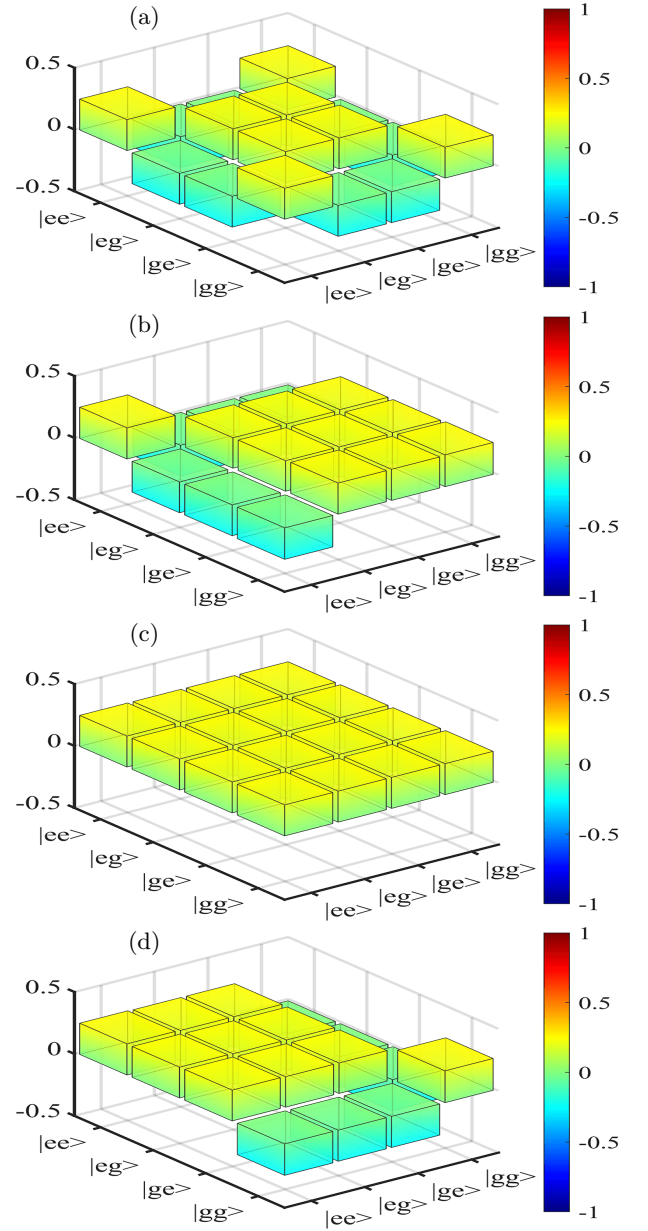
\begin{figure}[t]
\centering
\begin{tikzpicture}
\draw (0, 0) node[inner sep=0] {\includegraphics[width=8cm,height=4cm]{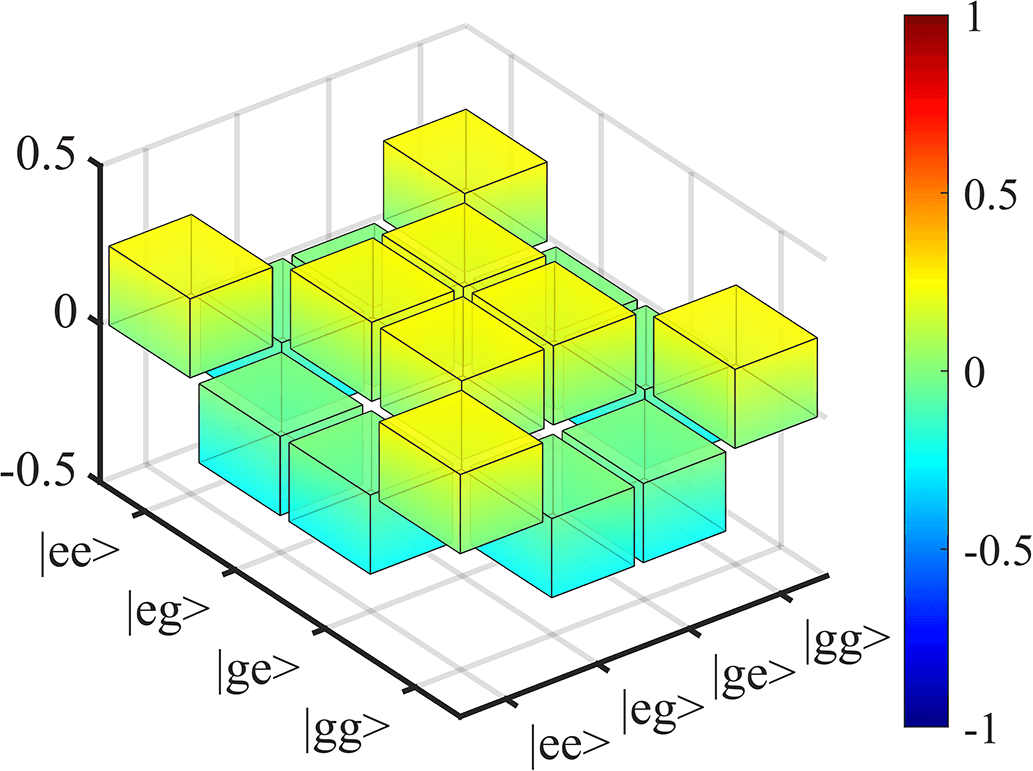}};
\draw (-2.3, 1.8) node {(a)};
\end{tikzpicture}
\begin{tikzpicture}
\draw (0, 0) node[inner sep=0] {\includegraphics[width=8cm,height=4cm]{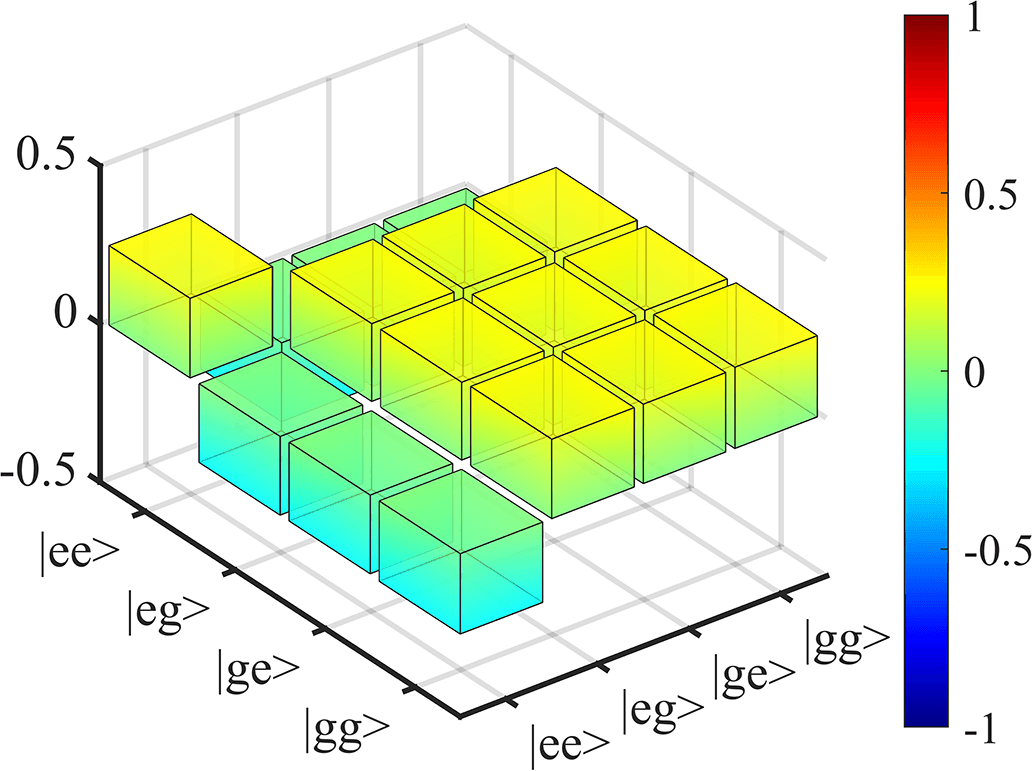}};
\draw (-2.3, 1.8) node {(b)};
\end{tikzpicture}
\begin{tikzpicture}
\draw (0, 0) node[inner sep=0] {\includegraphics[width=8cm,height=4cm]{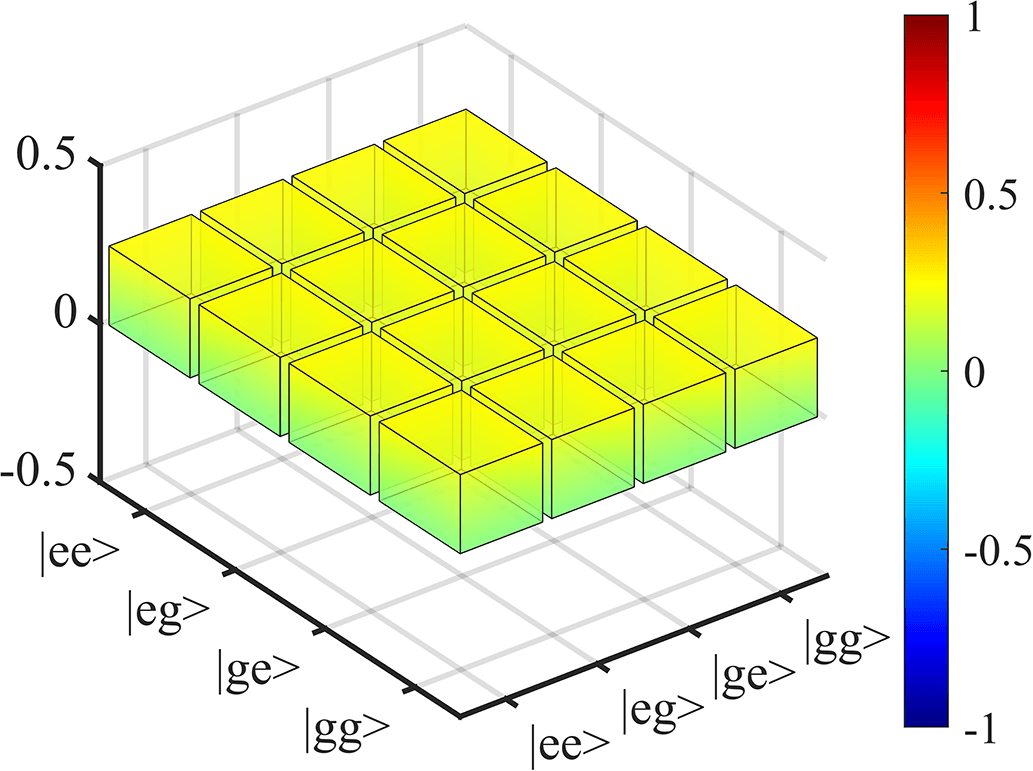}};
\draw (-2.3, 1.8) node {(c)};
\end{tikzpicture}
\begin{tikzpicture}
\draw (0, 0) node[inner sep=0] {\includegraphics[width=8cm,height=4cm]{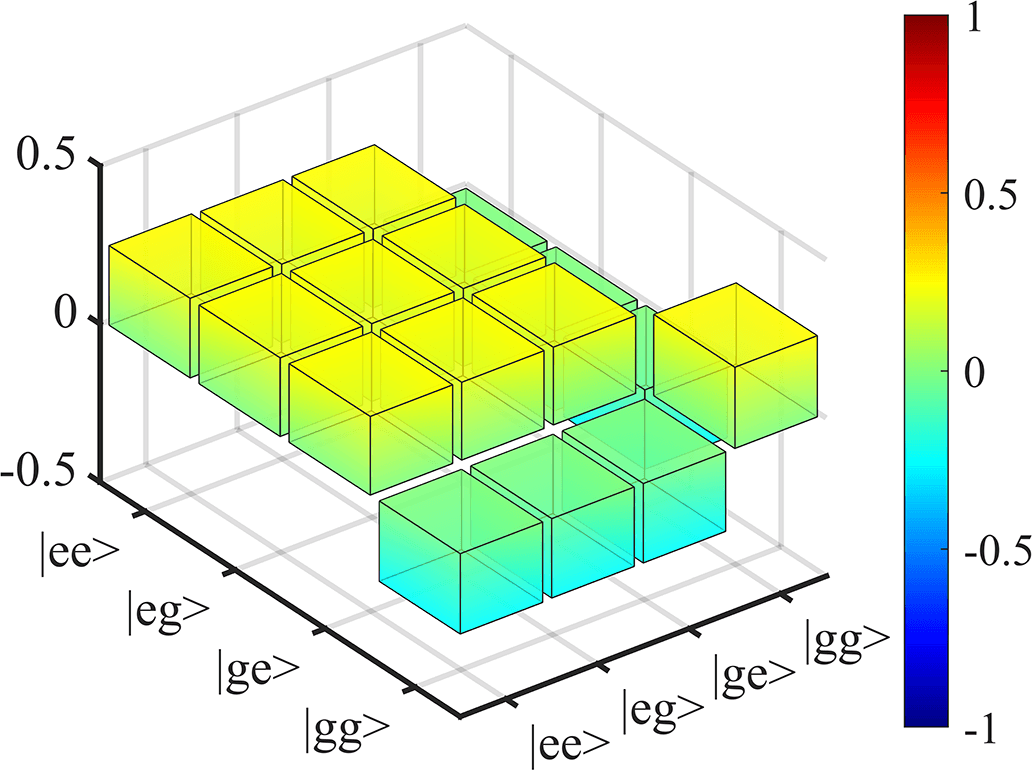}};
\draw (-2.3, 1.8) node {(d)};
\end{tikzpicture}
\caption{(a) and (c) respectively represent the matrix element images of the density matrices of the initial states of two qubits as shown in Eq.~(\ref{Eq60}) and Eq.~(\ref{Eq61}). (b) and (c) respectively represent the matrix element images of the density matrices when the two qubits, initially in the states shown by Eq.~(\ref{Eq60}) and Eq.~(\ref{Eq61}), reach maximum quantum entanglement under the control of a single-photon number state light field.\label{fig6}}
\end{figure}

For the initial state of the two identical qubits shown in Eq.~(\ref{Eq49}), all matrix elements of the reduced density matrix that evolves under the control of a single-photon light field up to time $t$ are shown above. We plot in Fig.~\ref{fig5}(a) the variation of the quantum concurrence of the two qubits with time for different superposition weights of the excited and ground states. We find that, in addition to the single-photon light field studied above, which is capable of triggering two qubits that are both initially in the ground state to reach a maximal entangled state with a value of 1, the single-photon light field is also capable of triggering two identical qubits that are initially maximally coherent to reach a maximal entangled state with a value of 1. In Fig.~\ref{fig5}(b), we plot the variation of the quantum concurrence of the two qubits with respect to the parameter $\theta$ and the time $t$. Fig.~\ref{fig5}(b) also confirms that the single-photon light field can produce maximum quantum entanglement not only for two qubits that are both initially in the ground state, but also for two qubits that are initially in the state with maximum coherence at time $gt=282.14$. In Fig.~\ref{fig5}(c), we plot the variation of quantum entanglement of two qubits with the initial state parameter $\theta$ and the initial coherence $\tilde{C}(|\theta\rangle)$. It is clear that when the initial coherence $\tilde{C}(|\theta\rangle)=1$ for each qubit, a single photon is able to trigger the two identical qubits to produce a maximally entangled state with a value of 1.

Above we have investigated the ability of a single photon to trigger two identical qubits initially with maximal coherence to produce a maximally entangled state with a value of 1. In the following, we begin to investigate the specific density matrix form of such entangled states. When each qubit has maximum coherence ($\tilde{C}(|\theta\rangle)=1$), i.e., $\theta=k\pi/4$ ($k=1,3,5,7$). When $\theta=\pi/4$ or $\theta=5\pi/4$, the initial states shown in Eq.~(\ref{Eq49}) is
\begin{equation}
|\varphi(0)\rangle=\frac{|e\rangle-|g\rangle}{\sqrt{2}}\frac{|e\rangle-|g\rangle}{\sqrt{2}} ,\label{Eq60} 
\end{equation} 
and when $\theta=3\pi/4$ or $\theta=7\pi/4$, the initial states shown in Eq.~(\ref{Eq49}) is
\begin{equation}
|\varphi(0)\rangle=\frac{|e\rangle+|g\rangle}{\sqrt{2}}\frac{|e\rangle+|g\rangle}{\sqrt{2}}.\label{Eq61} 
\end{equation}
The matrix element images of the density matrices for the above two initial states are shown in Figs.~\ref{fig6}(a) and \ref{fig6}(c), respectively. It is clear that each qubit in both initial states is a superposition state with maximum coherence of value 1, and both initial states are direct product states. From Figs.~\ref{fig5}(a) and  \ref{fig5}b we know that for the two initial states above, the two qubits reach the maximum entangled state with value 1 at time gt=282.14. Substituting the initial state information of Eq.~(\ref{Eq60}) and Eq.~(\ref{Eq61}), which is the $\theta$ in these two different initial states, and the time $gt=282.14$ into Eq.~(\ref{Eq50})-Eq.~(\ref{Eq59}), we can obtain the following density matrices
\begin{eqnarray}
\hat{\rho}_{q}(t)=\left[\begin{array}{cccc}
0.25 & -0.25 & -0.25 & -0.25\\
-0.25 & 0.25 & 0.25 & 0.25\\
-0.25 & 0.25 & 0.25 & 0.25\\
-0.25 & 0.25 & 0.25 & 0.25
\end{array}\right]\label{Eq62} 
\end{eqnarray}
and 
\begin{eqnarray}
\hat{\rho}_{q}(t)=\left[\begin{array}{cccc}
0.25 & 0.25 & 0.25 & -0.25\\
0.25 & 0.25 & 0.25 & -0.25\\
0.25 & 0.25 & 0.25 & -0.25\\
-0.25 & -0.25 & -0.25 & 0.25
\end{array}\right],\label{Eq63} 
\end{eqnarray}
respectively, when the two qubits reach the maximum entangled state with value 1 under the action of single-photon light field. Obviously, these two density matrices are the density matrice forms of the maximum entangled pure states
\begin{eqnarray}
|\varphi_{1}\rangle&=&\frac{1}{\sqrt{2}}[|e\rangle\frac{-|e\rangle+|g\rangle}{\sqrt{2}}+|g\rangle\frac{|e\rangle+|g\rangle}{\sqrt{2}}] \nonumber \\
&=&\frac{1}{2}(-|ee\rangle+|eg\rangle+|ge\rangle+|gg\rangle),\label{Eq64} 
\end{eqnarray}
and 
\begin{eqnarray}
|\varphi_{4}\rangle&=&\frac{1}{\sqrt{2}}[|e\rangle\frac{|e\rangle+|g\rangle}{\sqrt{2}}+|g\rangle\frac{|e\rangle-|g\rangle}{\sqrt{2}}] \nonumber \\
&=&\frac{1}{2}(|ee\rangle+|eg\rangle+|ge\rangle-|gg\rangle),\label{Eq65} 
\end{eqnarray}
respectively. Since the coefficients of the first term in Eq.~(\ref{Eq64}) and the fourth term in Eq.~(\ref{Eq65}) are negative, we name their states $|\varphi_{1}\rangle$ and $|\varphi_{4}\rangle$, respectively. We plot all matrix element images of the density matrices of these two maximally entangled states in Figs.~\ref{fig6}(b) and \ref{fig6}(d), respectively.

In summary, when the initial states of the two identical qubits are those shown in Eq.~(\ref{Eq49}), we find that the single-photon light field is capable of triggering not only the two qubits that are both in the ground state to reach the maximum entanglement with a value of 1, but also triggering the two qubits that have the maximum coherence to produce the maximum quantum entanglement with a value of 1.

\begin{figure}[t]
\centering
\begin{tikzpicture}
\draw (0, 0) node[inner sep=0] {\includegraphics[width=8cm,height=5cm]{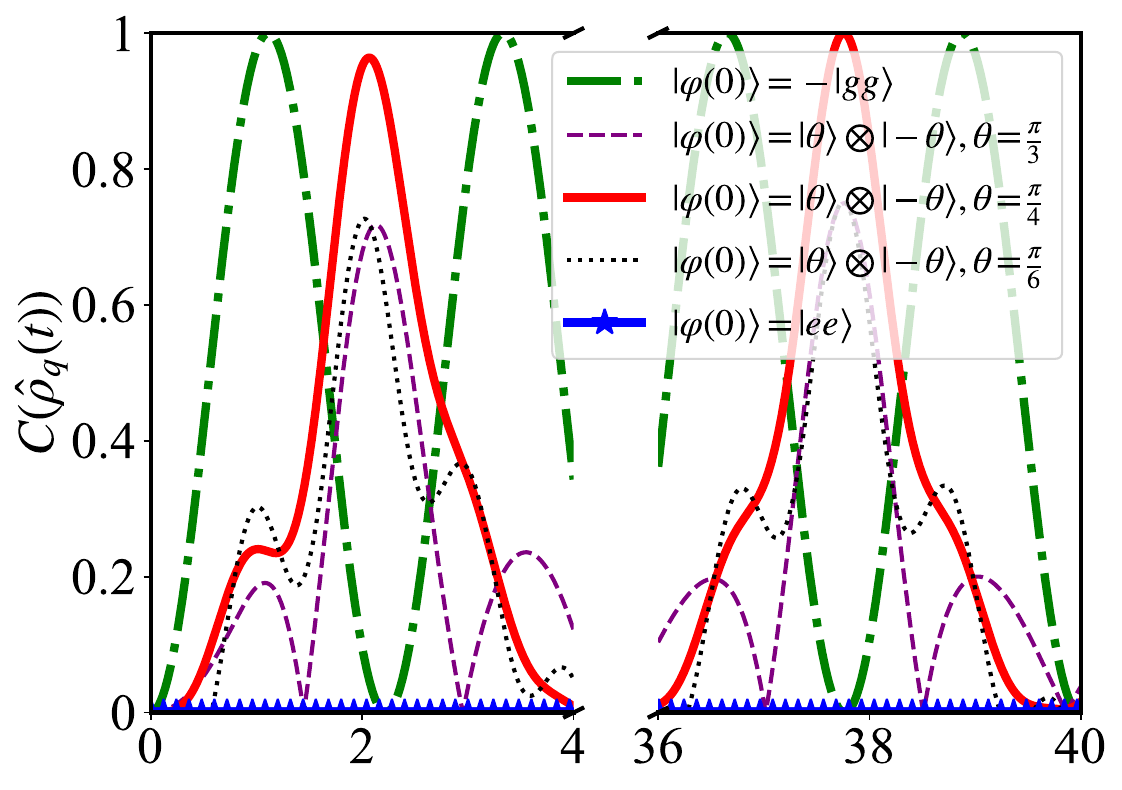}};
\draw (-2.5, 1.9) node {(a)};
\draw (0.4, -2.6) node {gt};
\end{tikzpicture}
\begin{tikzpicture}
\draw (0, 0) node[inner sep=0] {\includegraphics[width=8cm,height=5cm]{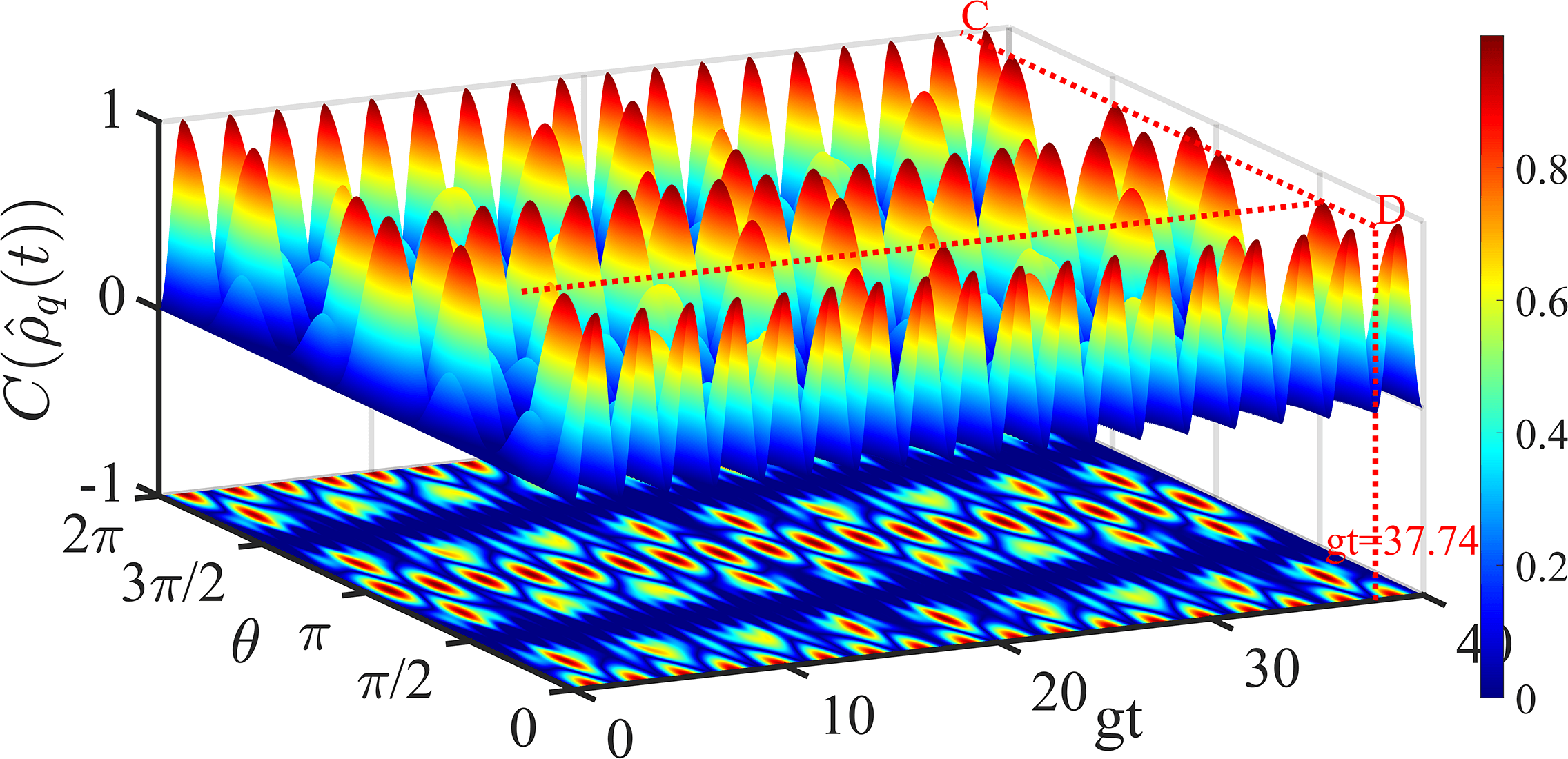}};
\draw (-2.5, 2) node {(b)};
\end{tikzpicture}
\begin{tikzpicture}
\draw (0, 0) node[inner sep=0] {\includegraphics[width=8cm,height=5cm]{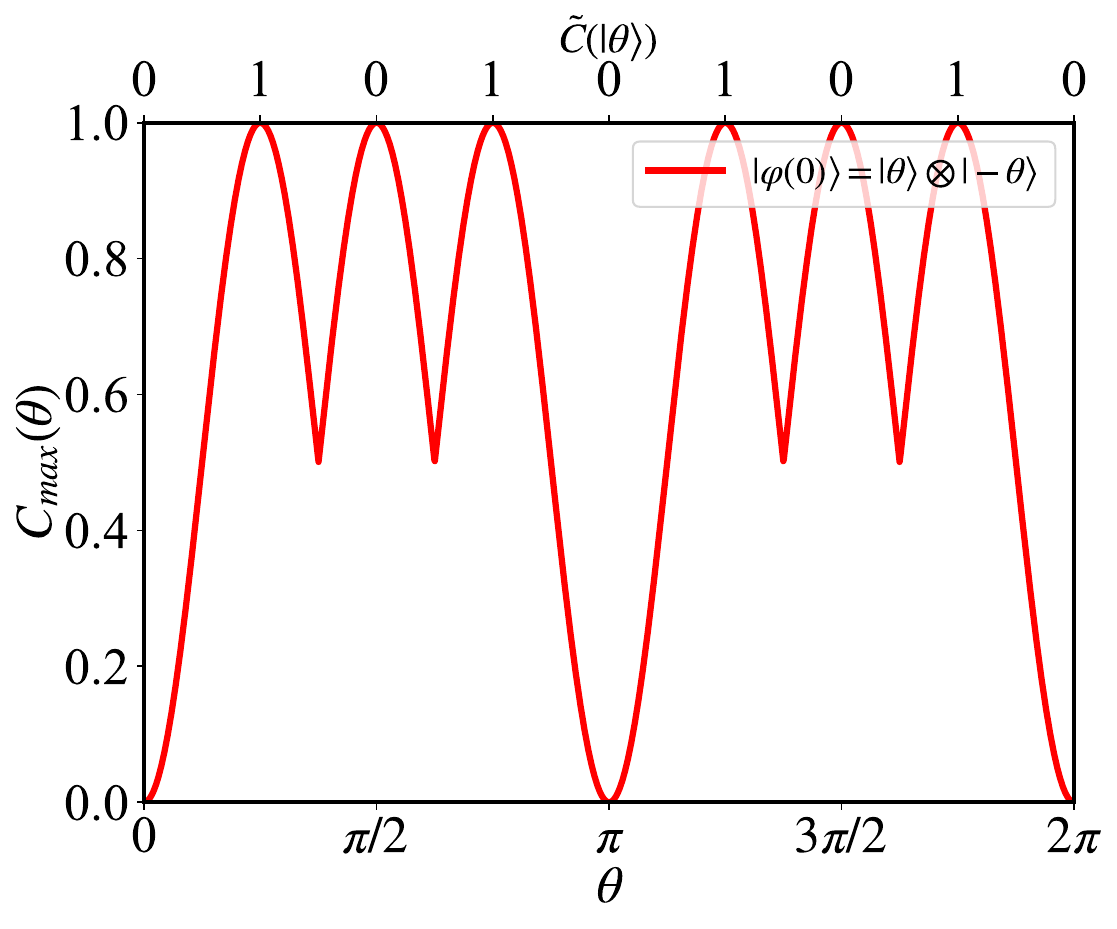}};
\draw (-2.5, 1.5   ) node {(c)};
\end{tikzpicture}
\caption{When the light field is initially in a single-photon number state, (a) the graph shows the change in quantum Concurrence over time $gt$ between two non-identical qubits with the same coherence in different initial states, (b) the graph shows the change in quantum Concurrence between the two qubits with the initial state parameter $\theta$ and time $gt$, (c) the graph shows the change in maximum quantum Concurrence between the two qubits with the initial state parameter $\theta$.\label{fig7}}
\end{figure}
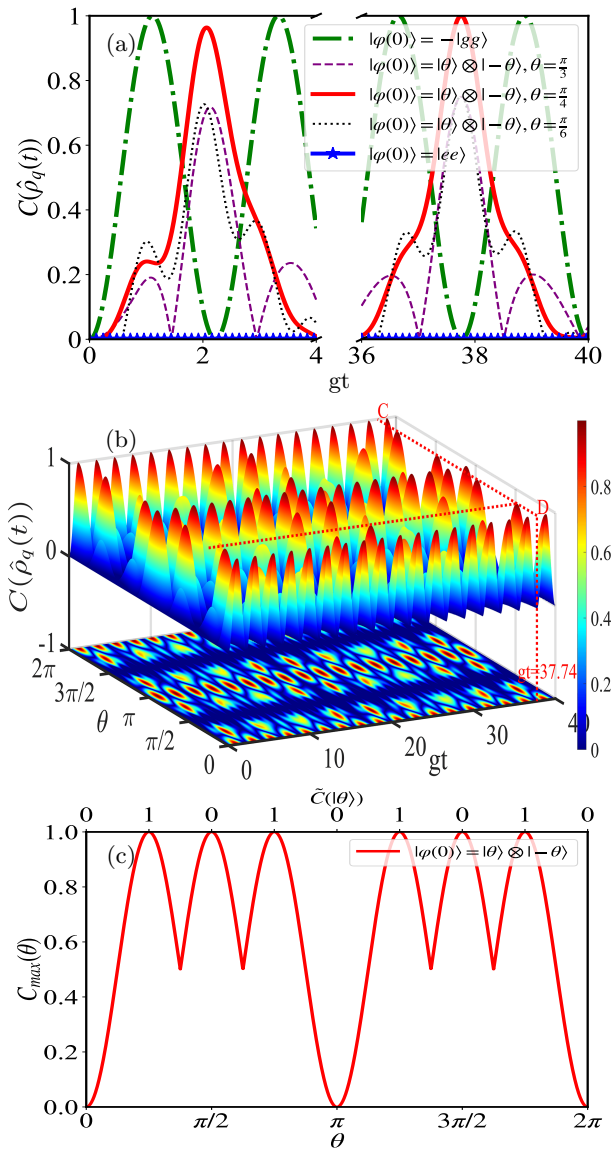

Finally, we study the effect of the single-photon light field on quantum entanglement between two qubits that are not identical but have the same coherence. The initial state of the two non-identical qubits with the same coherence is
\begin{equation}
|\varphi(0)\rangle=|\theta\rangle\otimes |-\theta\rangle,\label{Eq66} 
\end{equation}
where $|-\theta\rangle=\left(\cos\theta|e\rangle-\sin\theta|g\rangle\right)$. Substituting the initial parameter of this initial state into Eq.~(\ref{Eq10})-Eq.~(\ref{Eq19}), we can obtain all matrix elements of the density matrix of the two qubits at time $t$ as follows
\begin{eqnarray}
q_{11}&=&\cos^{4}\theta f_{1}^{2}(gt),\label{Eq67}   \\
q_{12}&=&\cos^{3}\theta\sin\theta f_{1}(gt)[f_{2}(gt)-f_{3}(gt)] ,\label{Eq68} \\
q_{13}&=&-q_{12}     ,\label{Eq69}  \\
q_{14}&=&-\sin^{2}\theta \cos^{2}\theta f_{1}(gt)\cos(gt\sqrt{2})     ,\label{Eq70} \\
q_{22}&=&2\cos^{4}\theta g^{2}(10,gt)+\sin^{2}\theta \cos^{2}\theta[f_{2}(gt)-f_{3}(gt)]^{2} \nonumber\\
&&+ \sin^{4}\theta g^{2}(2,gt)  ,\label{Eq71}   \\
q_{23}&=& 2\cos^{4}\theta g^{2}(10,gt)-\sin^{2}\theta \cos^{2}\theta[f_{3}(gt)-f_{2}(gt)]^{2} \nonumber\\
&&+ \sin^{4}\theta g^{2}(2,gt)     ,\label{Eq72}  \\
q_{24}&=&\cos\theta\sin^{3}\theta[f_{2}(gt)-f_{3}(gt)] \cos(gt\sqrt{2}) ,\label{Eq73}   \\
q_{33}&=&q_{22}    ,\label{Eq74}   \\
q_{34}&=&-q_{24}   ,\label{Eq75}  \\
q_{44}&=&\sin^{4}\theta\cos^{2}(gt\sqrt{2})    .\label{Eq76}   
\end{eqnarray}

For the initial state of the two different qubits that have the same coherence as shown in Eq.~(\ref{Eq66}), we have provided all the matrix elements of their density matrix at time $t$ from Eq.~(\ref{Eq67}) to Eq.~(\ref{Eq76}). Using these matrix elements, we plotted the quantum concurrence between the two qubits over time under different initial states in Fig.~\ref{fig7}(a). From the figure, we can see that when both qubits are initially in the ground state or in a state with maximum coherence, a single photon can manipulate the two qubits in these two initial states to reach the maximum entangled state. In Fig.~\ref{fig7}(b), we also plotted the quantum concurrence of two qubits as a function of the initial state parameter $\theta$ and time $gt$. From the red dashed line $CD$ in the Fig.~\ref{fig7}(b), it can be seen that when the initial state parameter $\theta$ equals $k\pi/4$ ($k$ is an odd number), i.e., when both qubits are initially in the state with maximum coherence, a single photon can manipulate them to reach the maximum entangled state at time $gt=37.74$. To more convincingly demonstrate the impact of the initial coherence of a single qubit on the entanglement of two qubits controlled by a single photon, we plotted the maximum quantum concurrence of two qubits as a function of the initial state parameter $\theta$ or the initial coherence $\tilde{C}(|\theta\rangle)$ of each qubit in Fig.~\ref{fig7}(c). We found that a single-photon light field can not only control two qubits, both initially in the ground state, to achieve a maximally entangled state with a concurrence value of 1, but also control two non-identical qubits, both having maximum coherence, to reach a maximally entangled state with a concurrence value of 1.

\begin{figure}[t]
\centering
\begin{tikzpicture}
\draw (0, 0) node[inner sep=0] {\includegraphics[width=8cm,height=4cm]{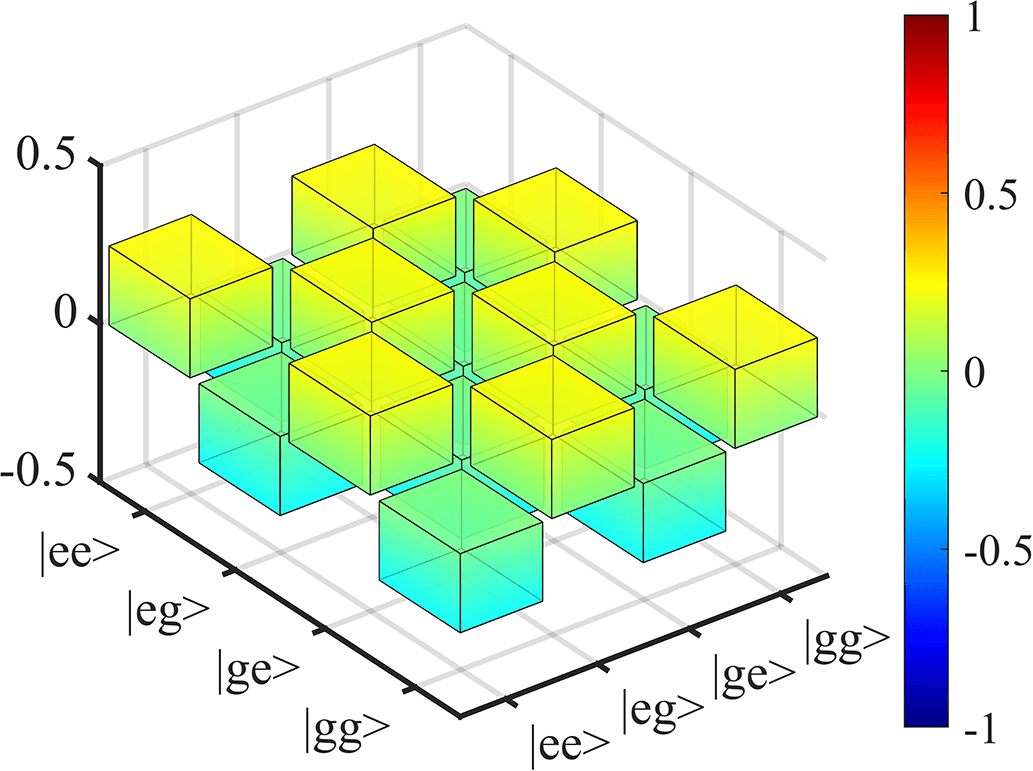}};
\draw (-2.3, 1.8) node {(a)};
\end{tikzpicture}
\begin{tikzpicture}
\draw (0, 0) node[inner sep=0] {\includegraphics[width=8cm,height=4cm]{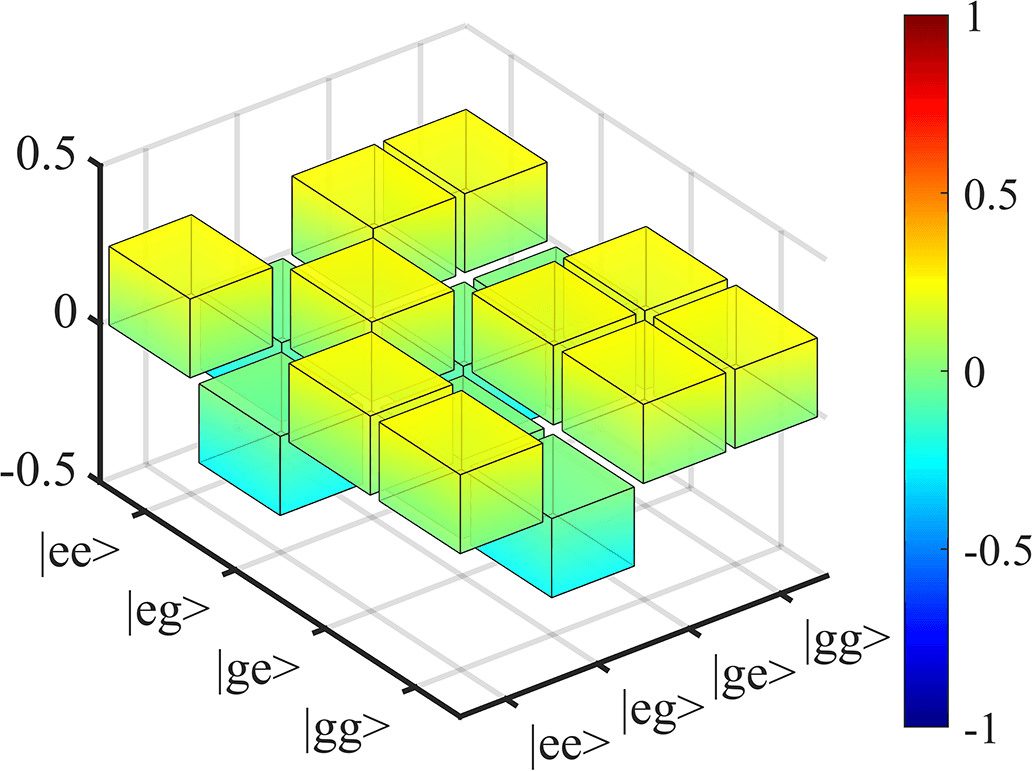}};
\draw (-2.3, 1.8) node {(b)};
\end{tikzpicture}
\begin{tikzpicture}
\draw (0, 0) node[inner sep=0] {\includegraphics[width=8cm,height=4cm]{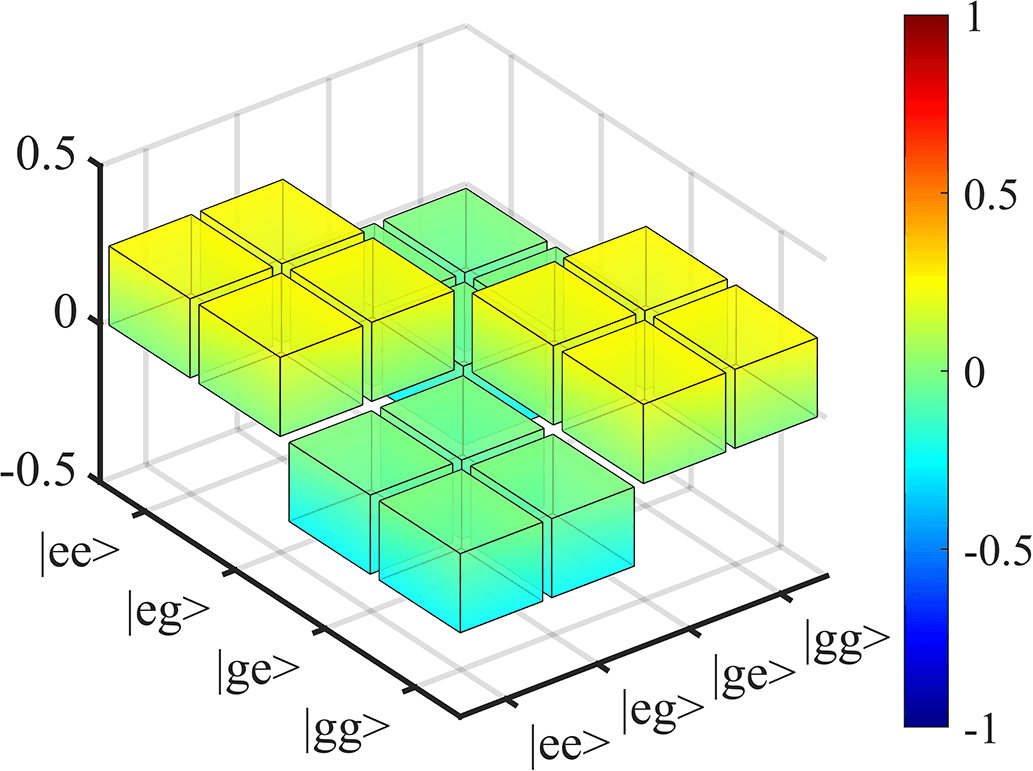}};
\draw (-2.3, 1.8) node {(c)};
\end{tikzpicture}
\begin{tikzpicture}
\draw (0, 0) node[inner sep=0] {\includegraphics[width=8cm,height=4cm]{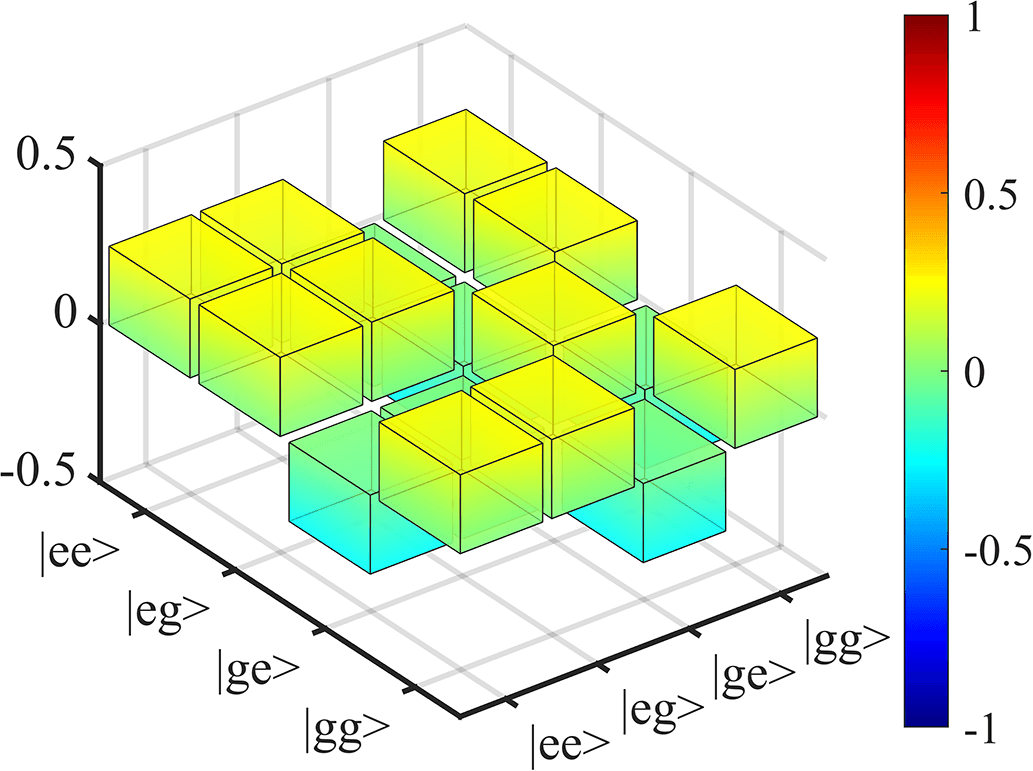}};
\draw (-2.3, 1.8) node {(d)};
\end{tikzpicture}
\caption{(a) and (c) respectively represent the matrix element images of the density matrices of the initial states of two qubits as shown in equations 85 and 86. (b) and (c) respectively represent the matrix element images of the density matrices when the two qubits, initially in the states shown by equations 85 and 86, reach maximum quantum entanglement under the control of a single-photon number state light field.\label{fig8}}
\end{figure}
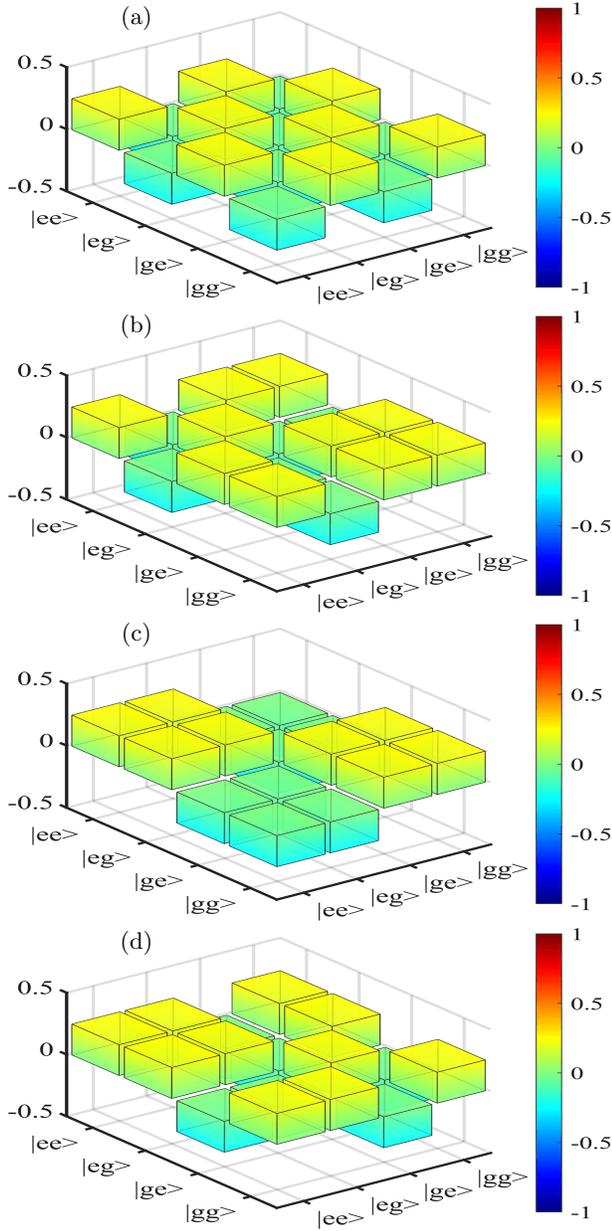

In the above study, we found that when two non-identical qubits both initially have maximum coherence, they can reach a maximally entangled state with a concurrence value of 1 under the action of a single-photon light field. Next, we begin to examine the specific form of the density matrix when the two qubits achieve maximal entanglement under these conditions. From the previous discussion, we know that the initial states of the two non-identical qubits with the same coherence are
\begin{equation}
|\varphi(0)\rangle=\frac{|e\rangle+|g\rangle}{\sqrt{2}}\frac{|e\rangle-|g\rangle}{\sqrt{2}} \label{Eq77} 
\end{equation} 
and 
\begin{equation}
|\varphi(0)\rangle=\frac{|e\rangle-|g\rangle}{\sqrt{2}}\frac{|e\rangle+|g\rangle}{\sqrt{2}},\label{Eq78} 
\end{equation} 
respectively. The initial states shown in Eq.~(\ref{Eq77}) and Eq.~(\ref{Eq78}) correspond to $\theta=k\pi/4$ ($k=1,5$)  and $\theta=l\pi/4$ ($l=3,7$) in Eq.~(\ref{Eq66}), respectively. Evidently, both initial states are direct product states of two non-identical qubits with maximum coherence. For the initial states shown in Eq.~(\ref{Eq77}) and Eq.~(\ref{Eq78}), we plotted the matrix elements of their density matrices in Figs.~\ref{fig8}(a) and \ref{fig8}(c), respectively. From Fig.~\ref{fig7}(b), we already know that under the action of a single-photon light field, these two initial states allow the two qubits to reach a maximally entangled state with a concurrence value of 1 at the time $gt=37.74$. By substituting the initial state parameters from Eq.~(\ref{Eq77}) and Eq.~(\ref{Eq78}) into Eq.~(\ref{Eq67}) to Eq.~(\ref{Eq76}), and then substituting the time $gt=37.74$, we can obtain the specific forms of the density matrices for the two qubits when they reach maximal entanglement under these conditions, which are respectively denoted as 
\begin{eqnarray}
\hat{\rho}_{q}(t)=\left[\begin{array}{cccc}
0.25 & -0.25 & 0.25 & 0.25\\
-0.25 & 0.25 & -0.25 & 0.25\\
0.25 & -0.25 & 0.25 & 0.25\\
0.25 & 0.25 & 0.25 & 0.25
\end{array}\right]\label{Eq79} 
\end{eqnarray} 
and 
\begin{eqnarray}
\hat{\rho}_{q}(t)=\left[\begin{array}{cccc}
0.25 & 0.25 & -0.25 & 0.25\\
0.25 & 0.25 & -0.25 & 0.25\\
-0.25 & -0.25 & 0.25 & -0.25\\
0.25 & 0.25 & -0.25 & 0.25
\end{array}\right],  \label{Eq80} 
\end{eqnarray}
Through these two forms of density matrices, we can easily obtain their maximum entangled pure state forms 
\begin{eqnarray}
|\varphi_{2}\rangle&=&\frac{1}{\sqrt{2}}[|e\rangle\frac{|e\rangle-|g\rangle}{\sqrt{2}}+|g\rangle\frac{|e\rangle+|g\rangle}{\sqrt{2}}] \nonumber\\
&=&\frac{1}{2}(|ee\rangle-|eg\rangle+|ge\rangle+|gg\rangle),\label{Eq81} 
\end{eqnarray}
and 
\begin{eqnarray}
|\varphi_{3}\rangle&=&\frac{1}{\sqrt{2}}[|e\rangle\frac{|e\rangle+|g\rangle}{\sqrt{2}}+|g\rangle\frac{-|e\rangle+|g\rangle}{\sqrt{2}}] \nonumber \\
&=&\frac{1}{2}(|ee\rangle+|eg\rangle-|ge\rangle+|gg\rangle).\label{Eq82} 
\end{eqnarray}
Since the coefficients of the second term in Eq.~(\ref{Eq81}) and the third term in Eq.~(\ref{Eq82}) are negative, we name their states $|\varphi_{2}\rangle$ and $|\varphi_{3}\rangle$, respectively. In Figs.~\ref{fig8}(b) and \ref{fig8}(d), when the initial states of the two qubits are as shown in Eq.~(\ref{Eq77}) and Eq.~(\ref{Eq78}), respectively, we plotted the matrix elements images  of the density matrices for the maximally entangled states shown in Eq.~(\ref{Eq79}) and Eq.~(\ref{Eq80}).

In summary, when two non-identical qubits both initially have maximum coherence, a single-photon light field can trigger these two qubits to produce maximal quantum entanglement with a concurrence value of 1.

\section{The effect of single photons on quantum entanglement of multiple qubits}\label{sec4}

In the above study, we investigated the effects of the initial excited state weight and initial coherence of the qubits on the quantum entanglement between two qubits triggered by a single-photon light field. Next, we will study the impact of a single-photon light field on the quantum entanglement between any two qubits among multiple identical qubits. The interaction between the light field and multiple qubits is described by the following multi-qubit TC model \cite{PhysRev.93.99,HEPP1973360,PhysRev.170.379} ($\hbar=1$)
\begin{equation}
\hat{H}_{TC}=\omega_{c}\hat{a}^{\dagger}\hat{a}+\omega_{a}\hat{J}_{z}+
g(\hat{a}^{\dagger}\hat{J}_{-}+\hat{J}_{+}\hat{a}),\label{Eq83}   
\end{equation}
In this context, $\hat{J}_{\alpha}=\Sigma_{i=1}^{N}\hat{\sigma}_{\alpha}^{i}/2$  $(\alpha=x,y,z)$ represents the collective angular momentum operator for a spin ensemble composed of $N$ identical qubits, where $\hat{\sigma}_{\alpha}^{i}$ are the Pauli matrices for the $i$-th qubit. These operators, denoted as $\{\hat{J}_{x}, \hat{J}_{y}, \hat{J}_{z}\}$, adhere to the commutation relations of the SU(2) algebra, and $\hat{J}_{\pm}=\hat{J}_{x}\pm i\hat{J}_{y}$.

Assume that the system is initially in a direct product state of a single-photon number state and a spin coherent state of $N$ identical qubits. \cite{JMRadcliffe_1971,MA201189}
\begin{eqnarray}
|\Psi(0)\rangle&=&|\tilde{\theta},\phi=0\rangle\otimes |1\rangle \nonumber \\
&=&(\cos\frac{\tilde{\theta}}{2}|e\rangle+\sin\frac{\tilde{\theta}}{2}|g\rangle)^{\otimes N}\otimes |1\rangle,\label{Eq84} 
\end{eqnarray}
where $\tilde{\theta}=2\theta$. The coherent spin state can be expanded on the Dicke state $|J,m\rangle$ in the following manner
\begin{equation}
|\tilde{\theta},\phi=0\rangle=\sum_{m=-J}^{J}d_{m}|J,m\rangle,\label{Eq85} 
\end{equation}
where $J=N/2$, 
\begin{eqnarray}
d_{m}&=&\langle J,m|\tilde{\theta},\phi=0\rangle  \nonumber\\
&=&(1+|\eta|^{2})^{-J}\sqrt{\frac{(2J)!}{(J+m)!(J-m)!}}\eta^{J+m},\label{Eq86} 
\end{eqnarray}
and $\eta=-\tan\frac{\tilde{\theta}}{2}$ \cite{JMRadcliffe_1971,MA201189}.
The reduced density matrix of any two qubits at any time is as follows \cite{wang2002pairwise, PhysRevA.68.012101}
\begin{eqnarray}
\hat{\rho}_{q}(t)=\left[\begin{array}{cccc}
v_{+}  &  h_{+}^{*}  &  h_{+}^{*}  &  \mu^{*}\\
h_{+}  &  w  &  p  &  h_{-}^{*}\\
h_{+}  &  p  &  w  &  h_{-}^{*}\\
\mu\   &  h_{-}  &  h_{-}  &  v_{-} 
\end{array}\right].\label{Eq87} 
\end{eqnarray}
Since $N$ identical qubits possess exchange symmetry, all matrix elements are \cite{wang2002pairwise, PhysRevA.68.012101}
\begin{eqnarray}
v_{\pm}&=&\frac{1}{4}\pm\frac{\langle\hat{J}_{z}\rangle}{N}+\frac{4\langle\hat{J}_{z}^{2}\rangle/N^{2}-1/N}{4(1-1/N)} ,\label{Eq88} \\
h_{\pm}&=&\frac{1}{2}\frac{\langle\hat{J}_{+}\rangle}{N}\pm\frac{\langle\hat{J}_{+}\hat{J}_{z}+\hat{J}_{z}\hat{J}_{+}\rangle/N^{2}}{2(1-1/N)} ,\label{Eq89} \\
w&=&p=\frac{1}{4}-\frac{4\langle\hat{J}_{z}^{2}\rangle/N^{2}-1/N}{4(1-1/N)} ,\label{Eq90} \\
\mu&=&\frac{\langle\hat{J}_{+}^{2}\rangle/N^{2}}{(1-1/N)} .\label{Eq91} 
\end{eqnarray}
In the following, in order to obtain the expectation values of these operators, we solve for the state of the system at time $t$ under the Hamiltonian shown in Eq.~(\ref{Eq83}) and the initial state shown in Eq.~(\ref{Eq84}). The state at time $t$ of the $N$ identical qubits is
\begin{equation}
|\Psi(t)\rangle=\exp(-i\hat{H}_{TC}t)|\Psi(0)\rangle,\label{Eq92} 
\end{equation}
We insert a unit operator in front of the initial state as follows \cite{bogoliubov1996exact,bogoliubov2013exactly,bogoliubov2017time,PhysRevA.104.043706}
\begin{equation}
\hat{I}_{M}=\sum_{\sigma}^{K}\frac{\left|\Phi_{J,M}\left(\{\lambda^{\sigma}\}\right)\right\rangle \left\langle \Phi_{J,M}\left(\{\lambda^{\sigma}\}\right)\right|}{N_{\sigma}^{2}},\label{Eq93} 
\end{equation}
where $M$ is an integer in the interval $[0, +\infty)$, $K=\min(2J,M)+1$, $\lambda^{\sigma}$ are complex solutions obtained via the Behte ansaz equation, and $N_{\sigma}$ is the normalization coefficient of the $\sigma$-th eigenstate $\left|\Phi_{J,M}\left(\{\lambda^{\sigma}\}\right)\right\rangle $. Then, for each $M$, we can obtain the state of the system at time $t$ as
\begin{eqnarray}
\left|\Phi_{M}(t)\right\rangle
&=&d_{M-J-1}\sum_{\sigma}^{K} \frac{C(M, \{\lambda^{\sigma}\})}{N_{\sigma}^{2}}  e^{-iE_{J,M}^{\sigma}t}\nonumber\\
&&\times\left|\Phi_{J,M}\left(\{\lambda^{\sigma}\}\right)\right\rangle  ,\label{Eq94} 
\end{eqnarray}
The detailed calculations of the above equations are shown in Appendix \ref{Appendix B}. For any operator $\hat{O}$ of the system, its average value at time $t$ is \cite{bogoliubov1996exact,bogoliubov2013exactly,bogoliubov2017time,PhysRevA.104.043706}
\begin{eqnarray}
\langle\hat{O}\rangle&=&\sum_{M=0}^{\infty}\sum_{E=0}^{\infty}\left\langle\Phi_{M}(t)\right| \hat{O} \left|\Phi_{E}(t)\right\rangle ,\label{Eq95}
\end{eqnarray}
Then, we can get the average values of all the operators in Eq.~(\ref{Eq88})-Eq.~(\ref{Eq91}) (Due to the lengthy and complex nature of the calculations, we have placed the detailed computation process for calculating the expectation values of all the operators below in Appendix \ref{Appendix C} )

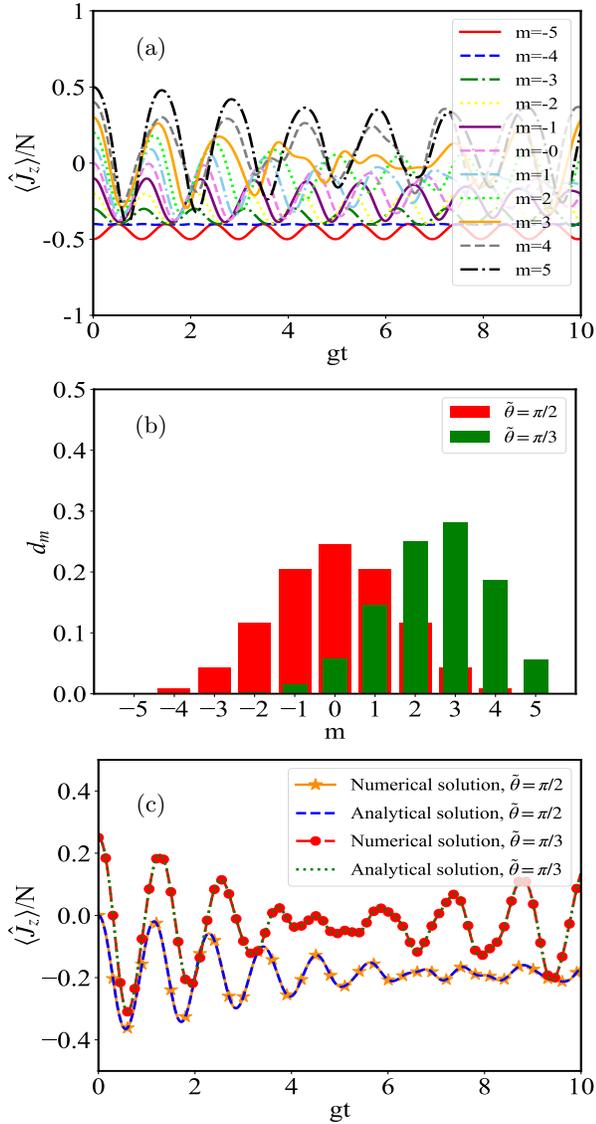
\begin{figure}[t]
\centering
\begin{tikzpicture}
\draw (0, 0) node[inner sep=0] {\includegraphics[width=8cm,height=5cm]{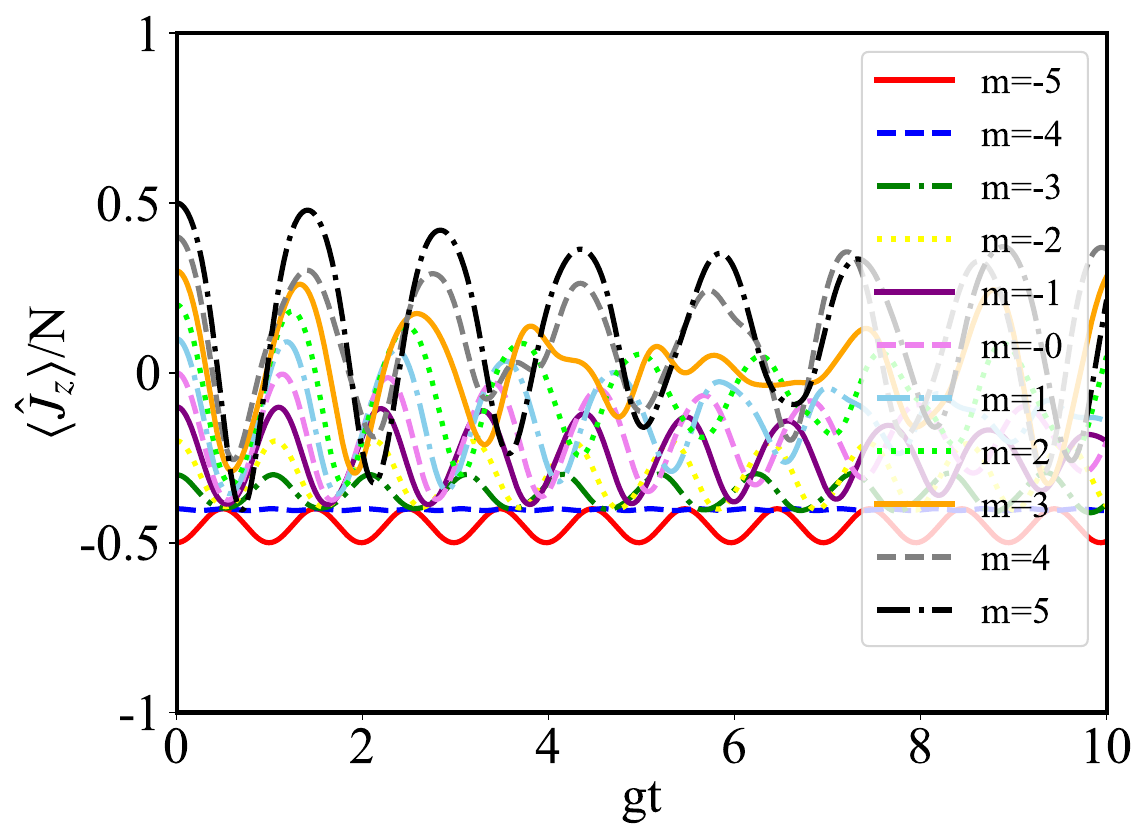}};
\draw (-2, 1.8) node {(a)};
\end{tikzpicture}
\begin{tikzpicture}
\draw (0, 0) node[inner sep=0] {\includegraphics[width=7.5cm,height=5cm]{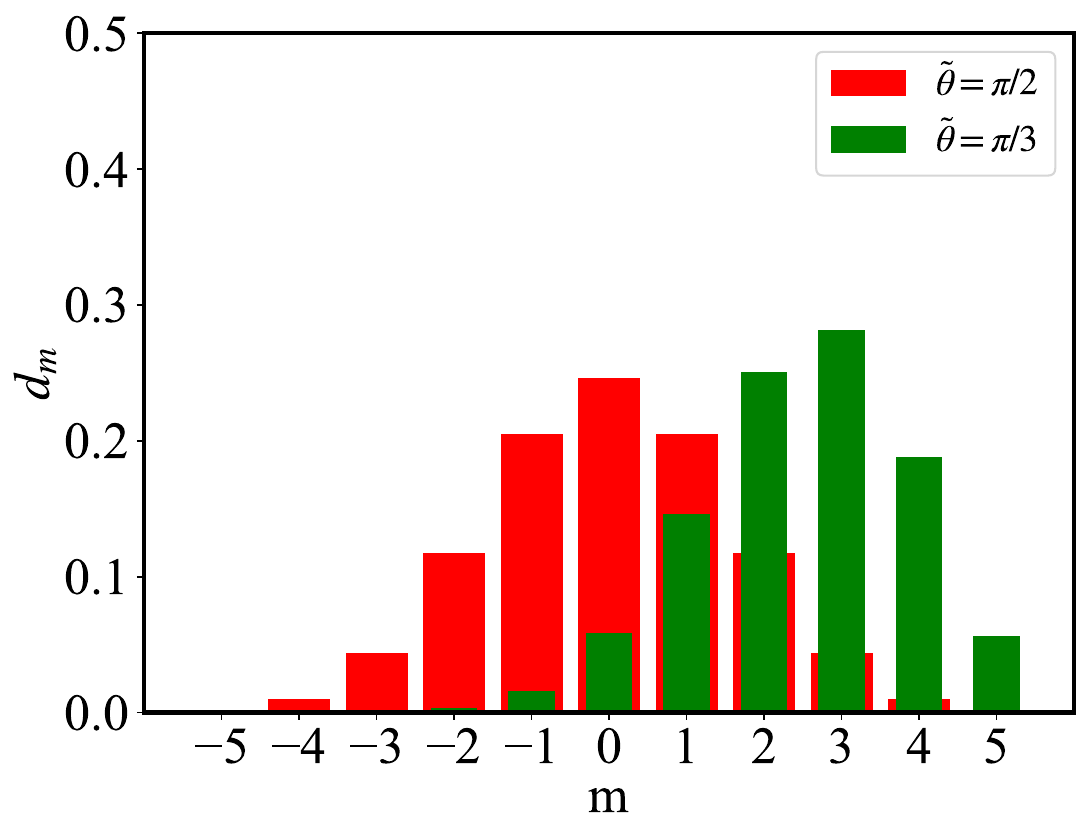}};
\draw (-2, 1.8) node {(b)};
\end{tikzpicture}
\begin{tikzpicture}
\draw (0, 0) node[inner sep=0] {\includegraphics[width=8cm,height=5cm]{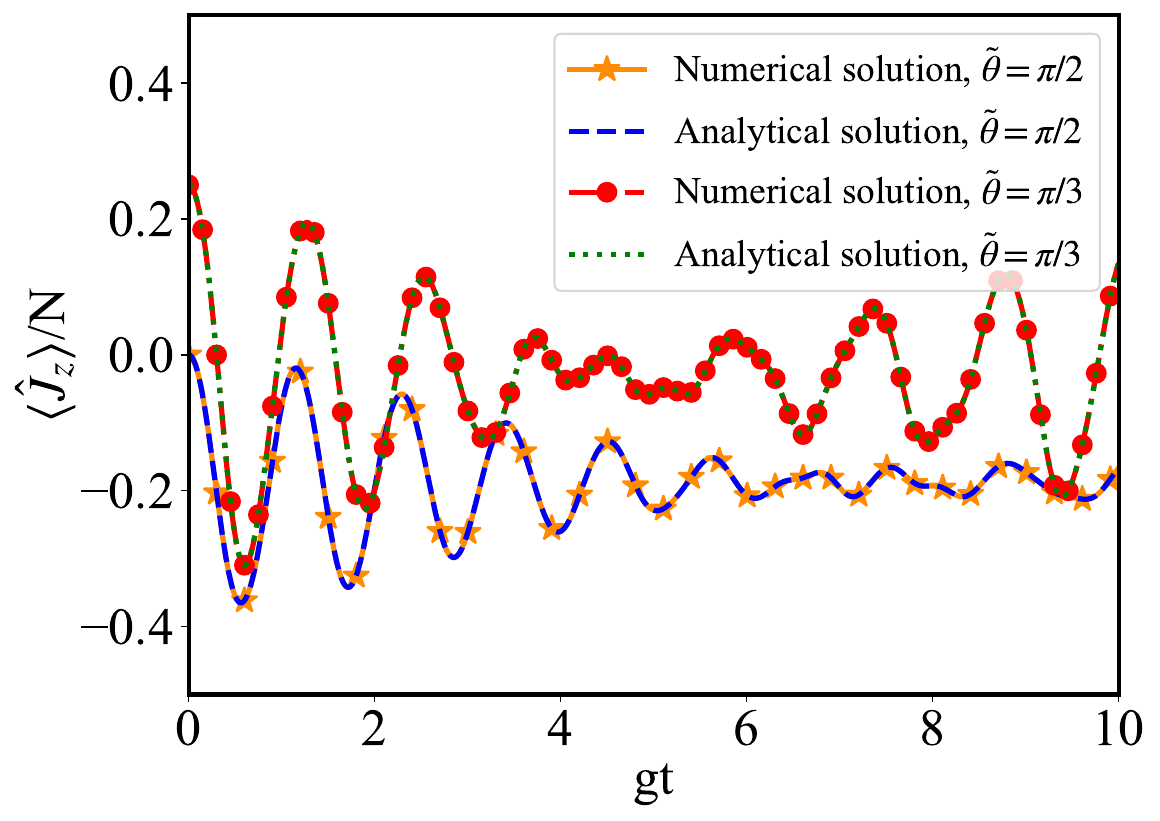}};
\draw (-2, 1.8) node {(c)};
\end{tikzpicture}
\caption{\label{fig9}When the light field is initially in a single-photon number state and the number of particles $N=10$, (a) the variation over time $gt$ of the average value of the operator $\hat{J}_{z}/N$ under different states $|J, m\rangle$, (b) the probability distribution of different initial states of multiple qubits in the Dicke state space, (c) analytical and numerical results of the variation over time $gt$ of the average value of the operator $\hat{J}_{z}/N$ under different initial states.}
\end{figure}

\begin{eqnarray}
\frac{\langle\hat{J}_{z}\rangle}{N}&=&\sum_{M=0}^{2J}|d_{M-J}|^{2}F(\hat{J}_{z}, M,  t),\label{Eq96}\\
\frac{\langle\hat{J}_{z}^{2}\rangle}{N^{2}}&=&\sum_{M=0}^{2J}|d_{M-J}|^{2}F(\hat{J}_{z}^{2}, M,  t),\label{Eq97} \\
\frac{\langle\hat{J}_{+}\rangle}{N}&=&\sum_{M=0}^{2J-1}d_{M-J+1}^{*}d_{M-J}F(\hat{J}_{+},M,t),\label{Eq98}  \\
&&\frac{\langle[\hat{J}_{+},\hat{J}_{z}]_{+}\rangle}{N^{2}}  \nonumber\\
&=&\sum_{M=0}^{2J-1}d_{M-J+1}^{*}d_{M-J}F([\hat{J}_{+},\hat{J}_{z}]_{+},M,t),\label{Eq99}\\
\frac{\langle\hat{J}_{+}^{2}\rangle}{N^{2}}&=&\sum_{M=0}^{2J-2}d_{M-J+2}^{*}d_{M-J}F(\hat{J}_{+}^{2},M,t),\label{Eq100}
\end{eqnarray}
where $[\hat{J}_{+},\hat{J}_{z}]_{+}=\hat{J}_{+}\hat{J}_{z}+\hat{J}_{z}\hat{J}_{+}$.
Due to the lengthy and complex nature of the specific expressions of the functions $F(\hat{O}, M, t)$ ($\hat{O}=\hat{J}_{z}, \hat{J}_{z}^{2}, \hat{J}_{+}, (\hat{J}_{+}\hat{J}_{z}+\hat{J}_{z}\hat{J}_{+}), \hat{J}_{+}^{2}$), we put them in the Appendix \ref{Appendix C}. When the number of qubits of the system is determined, then these functions are determined, so we can control the probability distribution $d_{m}$ of the initial states of mul-qubits over the Dicke states to control the average value of these operators. To verify the above conclusion, when the number of qubits $N = 10$, we plot in Fig.~\ref{fig9}(a) the variation of the function $F(\hat{J}_{z}, M,  t)$ with time $gt$ when $m$ takes different values, where $M = m + N/2$ ($N = 10$). In Fig.~\ref{fig9}(b) we plot the probability distribution $d_{m}$ of the initial states of multiple qubits shown in Eq.~(\ref{Eq84}) in the Dicke state space for different initial parameters $\tilde{\theta}$. From the results shown in Figs.~\ref{fig9}(a) and \ref{fig9}(b), we can obtain the mean value of the operator $\hat{J}_{z}/N$ over time $gt$ for these two different initial states. In Fig.~\ref{fig9}(c), we have plotted the variation over time $gt$ of the average value of the operator $\hat{J}_{z}/N$ under different initial states using numerical calculations and analytical methods based on Fig.~\ref{fig9}(a) and \ref{fig9}(b), respectively. We find that the numerical results are consistent with the analytical results. This indicates that our previous calculations and analysis are correct.

\begin{figure}[t]
\centering
\begin{tikzpicture}
\draw (0, 0) node[inner sep=0] {\includegraphics[width=8cm,height=5cm]{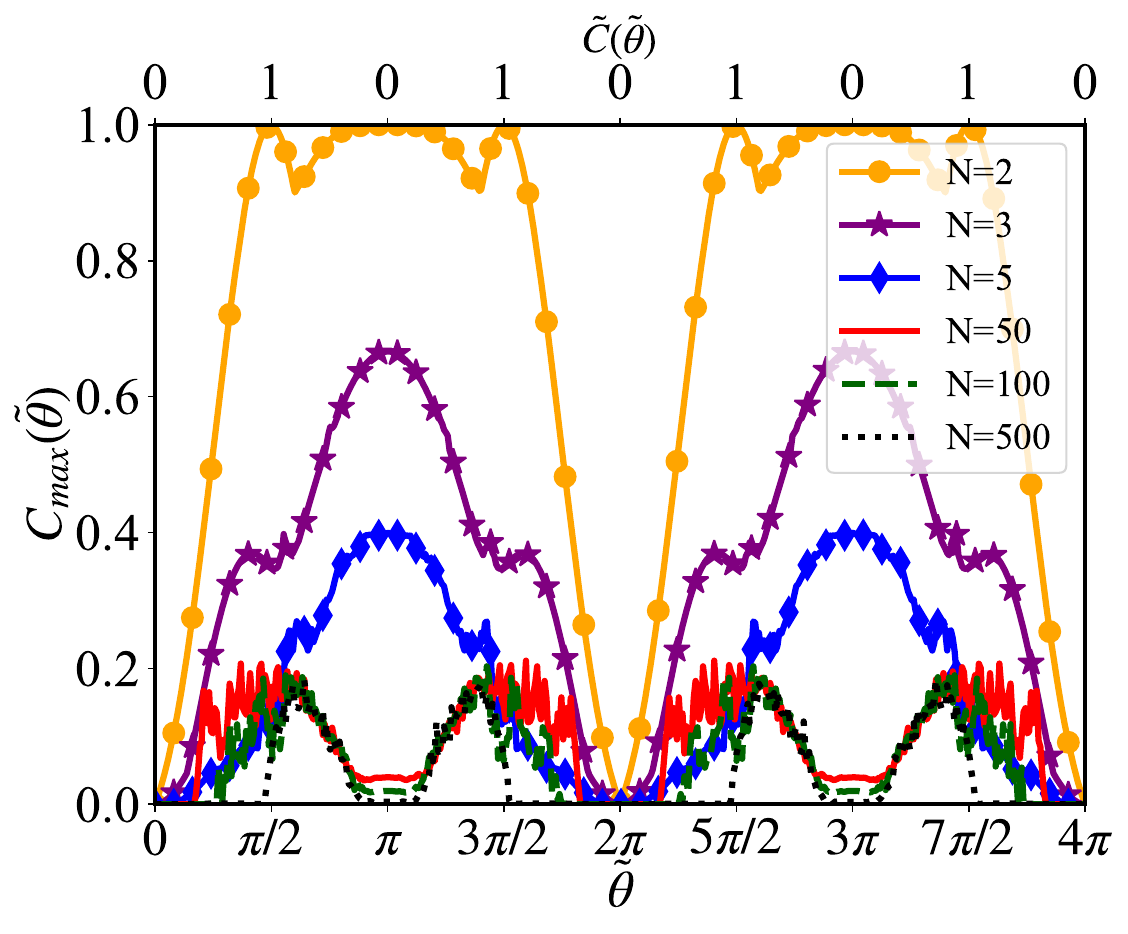}};
\draw (-2.6, 1.6) node {(a)};
\end{tikzpicture}
\begin{tikzpicture}
\draw (0, 0) node[inner sep=0] {\includegraphics[width=8cm,height=5cm]{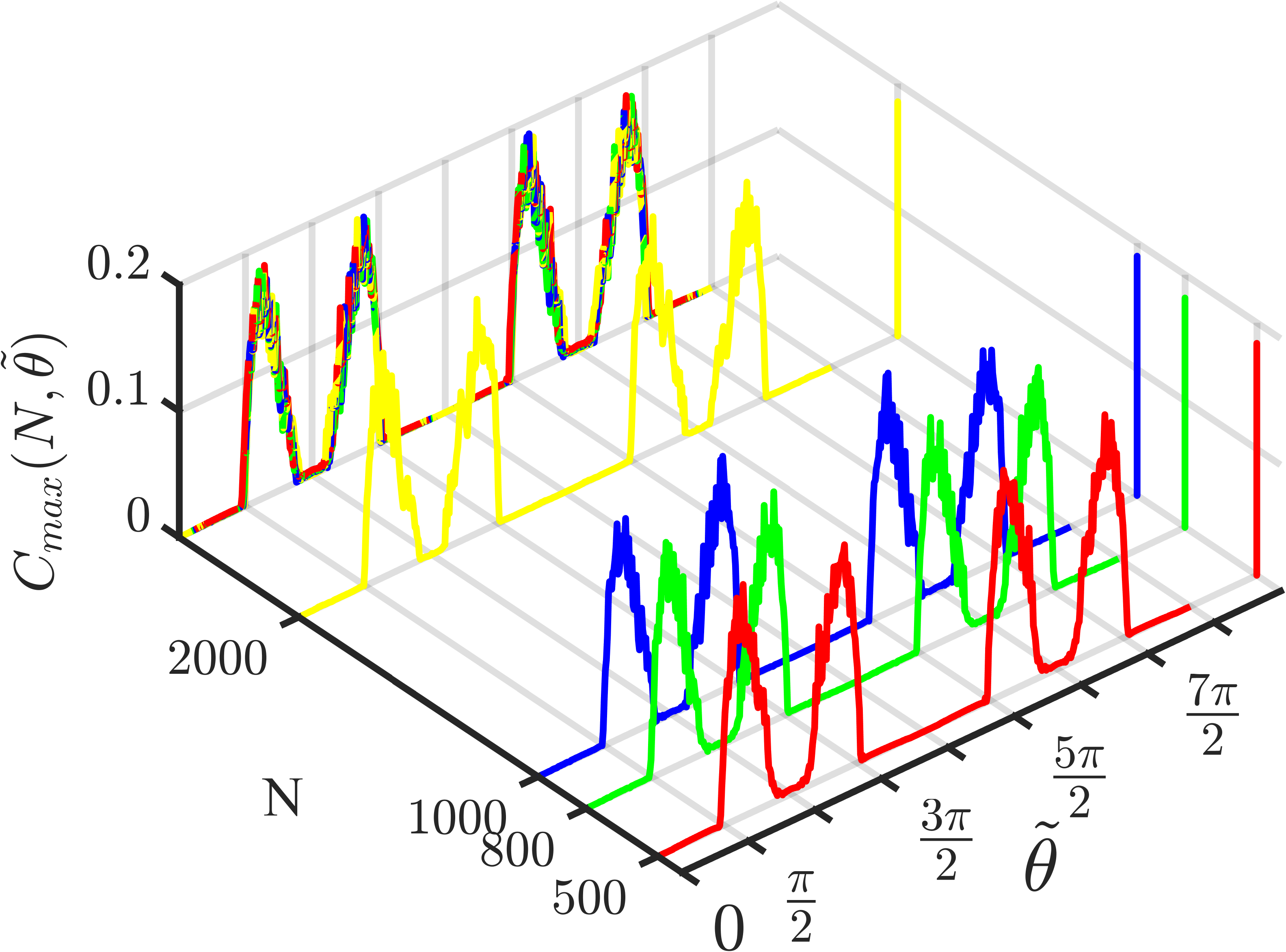}};
\draw (-2.6, 1.8) node {(b)};
\end{tikzpicture}
\caption{(a) Variation of the maximum quantum entanglement of any two qubits in multiple qubits with respect to the initial state parameter $\tilde{\theta}$ for different numbers of qubits in the multi-qubits TC model. (b) Variation of the maximum quantum entanglement of any two qubits in multiple qubits with respect to the initial state parameter $\tilde{\theta}$ in the large particle number limit. \label{fig10}}
\end{figure}

In the above, we have obtained all the matrix elements of the reduced density matrix for any two qubits in the TC model with $N$ identical qubits. In Fig.~\ref{fig10}(a) and \ref{fig10}(b), we plotted the variation of the maximum quantum entanglement that can be triggered between any two qubits among different numbers of identical qubits under the action of a single-photon light field, as a function of the initial state parameter $\theta$. We found that within a certain range of qubit numbers, the more qubits there are, the smaller the quantum entanglement that can be generated between any two qubits by a single photon. However, as the number of qubits increases, for example, when the number of qubits $N>500$, the maximum value of quantum entanglement between any two qubits among multiple identical qubits remains almost constant. Here, the multi-qubits TC model has some conclusions similar to the two-qubits TC model, e.g., from Figs.~\ref{fig10}(a) and \ref{fig10}(b) we can find that in the large particle number limit, when $\tilde{\theta}/2 \in [k\pi, k\pi+\pi/4]\cup[k\pi+3\pi/4, (k+1)\pi]$ ($k=0, 1$), i.e., when the probability of each of the $N$ identical qubits to initially be in the excited state is greater than the probability of the ground state, the single-photon light field cannot control the generation of quantum entanglement between any two qubits among multiple identical qubits. In the previous two-qubit TC model, we also found that the qubit in the excited state can weaken the ability of the single-photon light field to control the maximum quantum entanglement of these two qubits. However, unlike the two-qubits TC model, in the multi-qubit TC model, the larger the initial coherence of each qubit in the $N$ identical qubits, the quantum entanglement produced by the single-photon light field controlling any two qubits is not maximal. From Figs.~\ref{fig10}(a) and \ref{fig10}(b), we find that when the value of $\tilde{\theta}$ is near $1.9+k\pi$ or $4.4+k\pi$ ($k=0,2$), the quantum entanglement of any two qubits reaches the maximum. It is worth noting here that in the large particle number limit, when the value of $\tilde{\theta}$ is taken near $1.9+k\pi$ or $4.4+k\pi$ ($k=0,2$), a single photon is able to control the value of quantum entanglement of any two qubits out of $N$ identical qubits up to approximately $0.19$, and the value of this maximum quantum entanglement degree is independent of $N$. Furthermore, we find that the value of the maximum quantum entanglement degree tends to converge with the variation of the initial state parameter $\tilde{\theta}$ in the limit of a large number of particles, which means that this variation becomes independent of the particle number $N$ when the number of particles is large.

\begin{figure}[t]
\centering
\includegraphics[width=8cm,height=5cm]{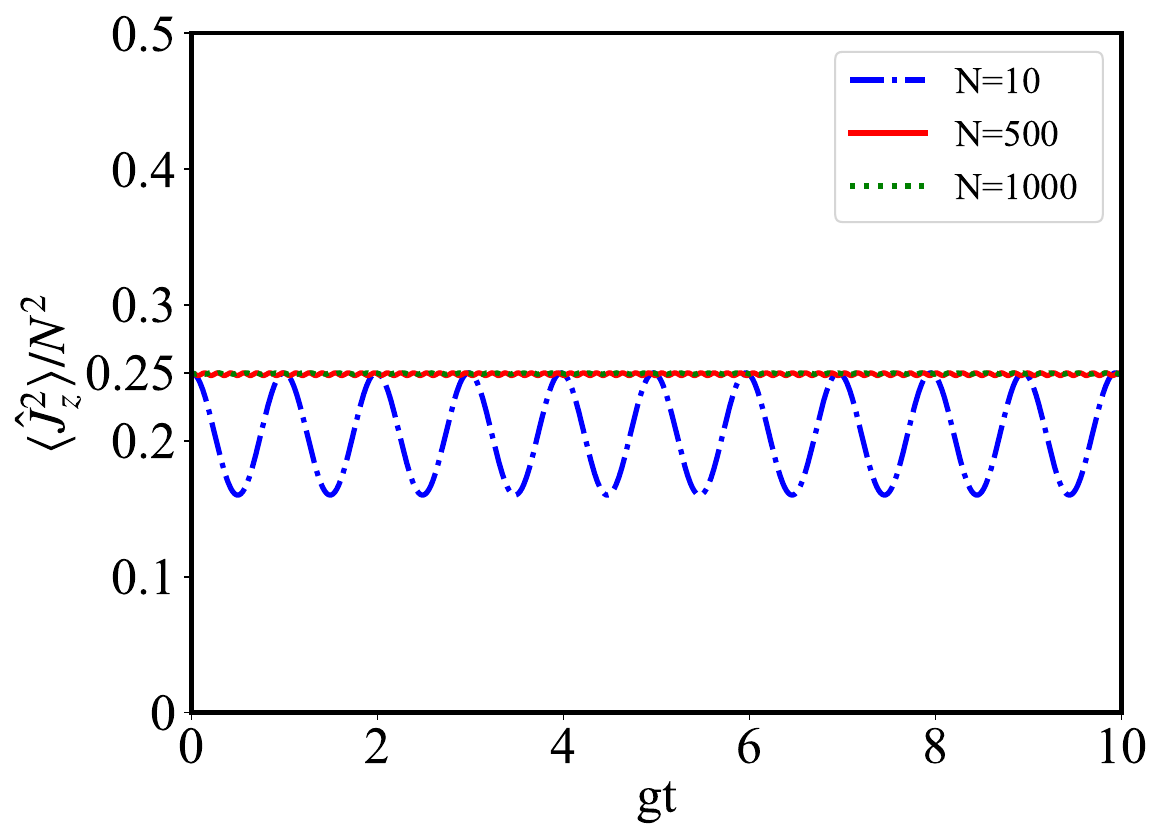}	
\caption{\label{fig11}The mean value of the operator $\langle \hat{J}_{z}^{2}\rangle/N^{2}$ over time $gt$ for different numbers of qubits when the initial state parameter $\tilde{\theta} = l\pi$ ($l$ is an odd number) for multiple qubits.}
\end{figure}

Moreover, when each qubit initially lacks coherence, i.e., when $\tilde{\theta} = k\pi$ ($k$ is an integer), the single-photon light field cannot trigger the generation of quantum entanglement between any two qubits among multiple identical qubits in the limit of the large particle number. Here we can explain this by the average value of the operator $\langle\hat{J}_{z}^{2}\rangle/N^{2}$.
As shown in Fig.~\ref{fig11}, when the initial parameter $\tilde{\theta}=l\pi$ ($l$ is an odd number), we plot the average value of the operator $\langle\hat{J}_{z}^{2}\rangle/N^{2}$ over time for different numbers of qubits, from which we can see that the average value of the operator $\langle\hat{J}_{z}^{2}\rangle/N^{2}$ converges to $1/4$ for a very large number of particles $N$. Since the initial state of the qubit system when the initial parameter $\tilde{\theta}=l\pi$ ($l$ is an odd number) is $|J,-J\rangle$, then from Eq.~(\ref{Eq96}) to Eq.~(\ref{Eq100}) we get $\langle\hat{J}_{+}\rangle/N=\langle[\hat{J}_{+},\hat{J}_{z}]_{+}\rangle/N^{2}=\langle\hat{J}_{+}^{2}\rangle/N^{2}=0$. In summary, in the limit of large particle number, when the initial parameter $\tilde{\theta}=l\pi$ ($l$ is an odd number), the reduced density matrix of any two qubits in the qubit system has all matrix elements as follows
\begin{eqnarray}
v_{\pm}&=&\frac{1}{2}\pm\frac{\langle\hat{J}_{z}\rangle}{N} ,\label{Eq101}\\
h_{\pm}&=&w=p=\mu=0 .\label{Eq102}
\end{eqnarray}
It is easy to know that when all the matrix elements of the density matrix of two qubits are as shown in the above equations, the quantum concurrence between two qubits is $0$, that is, there is no quantum entanglement between two qubits. Therefore, when $\tilde{\theta}=l\pi$ ($l$ is an odd number), the single-photon light field cannot cause quantum entanglement between any two qubits. In other words, in the limit of a large number of particles, the single-photon light field cannot trigger the generation of quantum entanglement between any two qubits among multiple identical qubits, all initially in the ground state. This conclusion is different from the two-qubit TC model, where the single-photon light field can trigger the generation of maximal quantum entanglement with a value of 1 between two qubits, both initially in the ground state.

\section{\label{Sec:4} Conclusion }
In conclusion, we investigated the triggering of quantum entanglement between two qubits or between any two qubits among multiple identical qubits by a single-photon number state light field. In the system of light interacting with two qubits, when the initial state of one qubit is determined and the other qubit is in a superposition of ground and excited states, we studied the effect of the initial excited state weight of the qubit in superposition on the quantum entanglement between the two qubits triggered by the single photon. We found that an excessively large excited state weight reduces the maximum quantum entanglement that can be achieved between the two qubits. Specifically, when both qubits are initially in the excited state, the single-photon light field cannot trigger quantum entanglement between the two qubits. However, when both qubits are initially in the ground state, the single-photon light field can trigger the generation of maximal quantum entanglement with a value of 1 between the two qubits. When the two qubits initially have the same coherence, regardless of whether they are identical qubits, a single photon can trigger the generation of non-Bell-type maximal quantum entanglement with a value of 1 between the two qubits when the initial coherence is maximum.

In the system of a single-photon number state light field interacting with multiple identical qubits, we also studied the effect of initial excited state weight and initial coherence of each qubit on the quantum entanglement between any two qubits triggered by the single-photon light field. We similarly found that a large initial excited state weight for each qubit in a superposition state reduces the maximum quantum entanglement between any two qubits triggered by the single photon. Consistent with the two-qubit TC model, in this system, a single photon cannot trigger quantum entanglement between any two qubits when all qubits are initially in the excited state. Contrary to the two-qubit TC model, a single-photon light field cannot trigger quantum entanglement between any two qubits among multiple qubits all initially in the ground state. We explained the reason for this phenomenon in the main text. Additionally, we found that in the limit of a large number of particles, the maximum quantum entanglement between any two qubits triggered by a single-photon light field varies almost independently of the number of qubits with the initial state parameter $\theta$, and the magnitude of the maximum quantum entanglement value also remains nearly constant regardless of the number of qubits. Furthermore, a single-photon light field can only trigger quantum entanglement between any two qubits when each qubit has coherence and the initial ground state weight is greater than the excited state weight.

\begin{acknowledgments}
W.J.L. was supported by the National Natural Science Foundation of China (Grants No. 12205092 and No. 12381240349), the Hunan Provincial Natural Science Foundation of China (Grant No. 2023JJ40208), and the Open fund project of the Key
Laboratory of Optoelectronic Control and Detection Technology of University of Hunan Province
(Grant No. 2022HSKFJJ038). 
\end{acknowledgments}  

\appendix

\section{\label{Appendix A}All matrix element representations of Eq.~(\ref{Eq2})}
Here, we list all the matrix elements of Eq.~(\ref{Eq2}) in the following \cite{Jiang_2024,fujii2004explicit,louisell1973quantum}
\begin{eqnarray}
\hat{U}_{11}&=&1+2\frac{\hat{A}(\hat{n}+1)-1}{\hat{C}(\hat{n}+1)}(\hat{n}+1), \label{EqA1}\\	\hat{U}_{44}&=&1+2\frac{\hat{A}(\hat{n}-1)-1}{\hat{C}(\hat{n}-1)}\hat{n}, \label{EqA2} \\
\hat{U}_{22}&=&\hat{U}_{33}=\frac{\hat{A}(\hat{n})+1}{2}, \label{EqA3}\\	\hat{U}_{23}&=&\hat{U}_{32}=\frac{\hat{A}(\hat{n})-1}{2}, \label{EqA4} \\
\hat{U}_{14}&=&2\frac{\hat{A}(\hat{n}+1)-1}{\hat{C}(\hat{n}+1)}\hat{a}^{2},  \label{EqA5}\\    	\hat{U}_{41}&=&2\frac{\hat{A}(\hat{n}-1)-1}{\hat{C}(\hat{n}-1)}\hat{a}^{\dagger2}, \label{EqA6}\\
\hat{U}_{12}&=&\hat{U}_{13}=-i\frac{\hat{B}(\hat{n}+1)}{\sqrt{\hat{C}(\hat{n}+1)}}\hat{a}, \label{EqA7}\\	\hat{U}_{21}&=&\hat{U}_{31}=-i\frac{\hat{B}(\hat{n})}{\sqrt{\hat{C}(\hat{n})}}\hat{a}^{\dagger}, \label{EqA8} \\
\hat{U}_{42}&=&\hat{U}_{43}=-i\frac{\hat{B}(\hat{n}-1)}{\sqrt{\hat{C}(\hat{n}-1)}}\hat{a}^{\dagger},	\label{EqA9}\\
\hat{U}_{24}&=&\hat{U}_{34}=-i\frac{\hat{B}(\hat{n})}{\sqrt{\hat{C}(\hat{n})}}\hat{a}, \label{EqA10} \\
\hat{A}(\hat{n})&=&\cos(gt\sqrt{\hat{C}(\hat{n})}),	\hat{B}(\hat{n})=\sin(gt\sqrt{\hat{C}(\hat{n})}), \nonumber \\
\hat{C}(\hat{n})&=&2(2\hat{n}+1), \nonumber \\
\hat{a}^{\dagger}f(\hat{n})&=&f(\hat{n}-1)\hat{a}^{\dagger},	\hat{a}f(\hat{n})=f(\hat{n}+1)\hat{a}. \nonumber
\end{eqnarray}
In all the above equations, $\hat{n}$ denotes the number operator of the light field.

\section{\label{Appendix B}Detailed calculation process of Eq.~(\ref{Eq94})}
Here, we give the exact computation of Eq.~(\ref{Eq94}). We insert the unit operator shown in Eq.~(\ref{Eq93}) in front of the initial state shown in Eq.~(\ref{Eq84}), so that the inner product of the eigenstates and initial state of the TC model is
\begin{eqnarray}
&&\langle\Phi_{J,M}(\{\lambda^{\sigma}\})|1\rangle\otimes|\text{\ensuremath{\tilde{\theta}}},0\rangle \nonumber \\
&=&\left\langle \widetilde{0}\right|\prod_{j=1}^{M}\hat{X}^{-}(\lambda_{j})|1\rangle\otimes|\text{\ensuremath{\tilde{\theta}}},0\rangle  \nonumber\\
&=&\left\langle J,-J\right|\otimes\left\langle 0\right|\prod_{j=1}^{M}\left(\hat{a}-\frac{\hat{J}_{-}}{\lambda_{j}^{\sigma}}\right)|1\rangle\otimes|\text{\ensuremath{\tilde{\theta}}},0\rangle   \nonumber\\
&=&\left\langle J,-J\right|\otimes\left\langle 0\right|\prod_{j=1}^{M}\left(\hat{a}-\frac{\hat{J}_{-}}{\lambda_{j}^{\sigma}}\right)\Bigg(\sum_{n=0}^{\infty}\left|n\right\rangle \left\langle n\right|\otimes \nonumber \\
&&\sum_{m=-J}^{J}\left|J,m\right\rangle \left\langle J,m\right|\Bigg)|1\rangle\otimes|\text{\ensuremath{\tilde{\theta}}},0\rangle   \nonumber\\
&=&\sum_{m=-J}^{J}d_{m}\left\langle J,-J\right|\otimes\left\langle 0\right|c(M, \{\lambda^{\sigma}\})\hat{a}\hat{J}_{-}^{M-1}|1\rangle\otimes\left|J,m\right\rangle    \nonumber\\
&=&\sum_{m=-J}^{J}d_{m}c(M, \{\lambda^{\sigma}\})\sqrt{\frac{(m+J)!(J-m+M-1)!}{(m+J-M+1)!(J-m)!}} \nonumber \\
&&\times\left\langle J,-J\right|J,m-M+1\rangle   \nonumber\\
&=&d_{M-J-1}c(M,\{\lambda^{\sigma}\})\sqrt{\frac{(M-1)!(2J)!}{(2J-M+1)!}}   \nonumber\\
&=&d_{M-J-1}C(M, \{\lambda^{\sigma}\}), \label{EqB1}
\end{eqnarray}
where 
\begin{eqnarray}
&&c(M, \{\lambda^{\sigma}\}) \nonumber\\
&=&(-1)^{M-1}\Bigg(\frac{1}{\lambda^{\sigma}_{2}\lambda^{\sigma}_{3}\cdots\lambda^{\sigma}_{M}}+\frac{1}{\lambda^{\sigma}_{1}\lambda^{\sigma}_{3}\cdots\lambda^{\sigma}_{M}}+ \nonumber\\
&&\cdots+\frac{1}{\lambda^{\sigma}_{1}\lambda^{\sigma}_{2}\cdots\lambda^{\sigma}_{M-2}\lambda^{\sigma}_{M}}+\frac{1}{\lambda^{\sigma}_{1}\lambda^{\sigma}_{2}\cdots\lambda^{\sigma}_{M-1}}\Bigg), \label{EqB2}
\end{eqnarray}
and $C(M, \{\lambda^{\sigma}\})=c(M, \{\lambda^{\sigma}\})\sqrt{\frac{(M-1)!(2J)!}{(2J-M+1)!}}$.

\section{\label{Appendix C}Detailed calculations for Eq.~(\ref{Eq96})-Eq.~(\ref{Eq100})}

In this section, we give the detailed calculations from Eq.~(\ref{Eq96}) to Eq.~(\ref{Eq100}). First, we start by giving the average of any operator O at time $t$ as
\begin{eqnarray}
\langle\hat{O}\rangle&=&\sum_{M=0}^{\infty}\sum_{E=0}^{\infty}\left\langle\Phi_{M}(t)\right|\hat{O} \left|\Phi_{E}(t)\right\rangle   \nonumber \\
&=&\sum_{M=0}^{\infty}\sum_{E=0}^{\infty}\Bigg(d_{M-J-1}^{*}\sum_{\sigma}^{K} \frac{C(M, \{\lambda^{*\sigma}\})}{N_{\sigma}^{2}}  e^{iE_{J,M}^{\sigma}t}\Bigg)\nonumber\\
&&\times\Bigg(d_{E-J-1}\sum_{\sigma}^{K} \frac{C(E, \{\lambda^{\sigma}\})}{N_{\sigma}^{2}}  e^{-iE_{J,E}^{\sigma}t}\Bigg)\nonumber\\
&&\times\langle\Phi_{J,M}(\{\lambda^{\sigma}\})|\hat{O}|\Phi_{J,E}(\{\lambda^{\sigma}\})\rangle . \label{EqC1}
\end{eqnarray}

In order to calculate the average value of operators $\hat{J}_{z}$, $\hat{J}_{z}^{2}$, $\hat{J}_{+}$, $(\hat{J}_{+}\hat{J}_{z}+\hat{J}_{z}\hat{J}_{+})$, and $\hat{J}_{+}^{2}$ at time $t$. We first calculate the result of multiplying these operators in different eigenstates of the TC model in the following way

\begin{eqnarray}
&&\langle\Phi_{J,M}(\{\lambda^{\sigma}\})|\hat{J}_{z}|\Phi_{J,E}(\{\lambda^{\sigma}\})\rangle \nonumber\\
&=&\text{\ensuremath{\langle J,-J|\otimes\langle0|\prod_{j=1}^{M}\left(\hat{a}-\frac{\hat{J}_{-}}{\text{\ensuremath{\lambda_{j}^{\sigma}}}}\right)\hat{J}_{z}\prod_{j=1}^{E}\left(\hat{a}^{\dagger}-\frac{\hat{J}_{+}}{\text{\ensuremath{\lambda_{j}^{\sigma}}}^{*}}\right)}}\nonumber\\
&&\times|J,-J\rangle\otimes|0\rangle \nonumber\\
&=&\langle J,-J|\otimes\langle0|(\hat{a}^{M}+x_{1}\hat{a}^{M-1}\hat{J}_{-}+\cdots+x_{M-1}\hat{a}\hat{J}_{-}^{M-1}\nonumber\\
&&+x_{M}\hat{J}_{-}^{M})\hat{J}_{z}(\hat{a}^{\dagger E}+x_{1}\hat{a}^{\dagger(E-1)}\hat{J}_{+}+\cdots+x_{E-1}\hat{a}^{\dagger}\hat{J}_{+}^{E-1}\nonumber\\
&&+x_{E}\hat{J}_{+}^{E})|J,-J\rangle\otimes|0\rangle \nonumber\\
&=&\langle\Phi_{J,M}(\{\lambda^{\sigma}\})|\hat{J}_{z}|\Phi_{J,E}(\{\lambda^{\sigma}\})\rangle\delta_{M,E} , \label{EqC2}\\
&&\langle\Phi_{J,M}(\{\lambda^{\sigma}\})|\hat{J}_{z}^{2}|\Phi_{J,E}(\{\lambda^{\sigma}\})\rangle\nonumber\\
&=&\langle\Phi_{J,M}(\{\lambda^{\sigma}\})|\hat{J}_{z}^{2}|\Phi_{J,E}(\{\lambda^{\sigma}\})\rangle\delta_{M,E}, \label{EqC3}\\
&&\langle\Phi_{J,M}(\{\lambda^{\sigma}\})|\hat{J}_{+}|\Phi_{J,E}(\{\lambda^{\sigma}\})\rangle \nonumber\\
&=&\text{\ensuremath{\langle J,-J|\otimes\langle0|\prod_{j=1}^{M}\left(\hat{a}-\frac{\hat{J}_{-}}{\text{\ensuremath{\lambda_{j}^{\sigma}}}}\right)\hat{J}_{+}\prod_{j=1}^{E}\left(\hat{a}^{\dagger}-\frac{\hat{J}_{+}}{\text{\ensuremath{\lambda_{j}^{\sigma}}}^{*}}\right)}}\nonumber\\
&&\times|J,-J\rangle\otimes|0\rangle  \nonumber\\
&=&\langle J,-J|\otimes\langle0|(\hat{a}^{M}+x_{1}\hat{a}^{M-1}\hat{J}_{-}+\cdots+x_{M-1}\hat{a}\hat{J}_{-}^{M-1}\nonumber\\
&&+x_{M}\hat{J}_{-}^{M})\hat{J}_{+}(\hat{a}^{\dagger E}+x_{1}\hat{a}^{\dagger(E-1)}\hat{J}_{+}+\cdots+x_{E-1}\hat{a}^{\dagger}\hat{J}_{+}^{E-1}\nonumber\\
&&+x_{E}\hat{J}_{+}^{E})|J,-J\rangle\otimes|0\rangle\nonumber\\
&=&\langle J,-J|\otimes\langle0|(\hat{a}^{M}+x_{1}\hat{a}^{M-1}\hat{J}_{-}+\cdots+x_{M-1}\hat{a}\hat{J}_{-}^{M-1}\nonumber\\
&&+x_{M}\hat{J}_{-}^{M})(\hat{a}^{\dagger E}\hat{J}_{+}+x_{1}\hat{a}^{\dagger(E-1)}\hat{J}_{+}^{2}+\cdots+x_{E-1}\hat{a}^{\dagger}\hat{J}_{+}^{E}\nonumber\\
&&+x_{E}\hat{J}_{+}^{E+1})|J,-J\rangle\otimes|0\rangle\nonumber\\
&=&\langle\Phi_{J,M}(\{\lambda^{\sigma}\})|\hat{J}_{+}|\Phi_{J,E}(\{\lambda^{\sigma}\})\rangle\delta_{M,E+1} , \label{EqC4} \\
&&\langle\Phi_{J,N}(\{\lambda^{\sigma}\})|(\hat{J}_{+}\hat{J}_{z}+\hat{J}_{z}\hat{J}_{+})|\Phi_{J,E}(\{\lambda^{\sigma}\})\rangle\nonumber\\
&=&\langle\Phi_{J,M}(\{\lambda^{\sigma}\})|\hat{J}_{+}|\Phi_{J,E}(\{\lambda^{\sigma}\})\rangle\delta_{M,E+1}, \label{EqC5}\\
&&\langle\Phi_{J,M}(\{\lambda^{\sigma}\})|\hat{J}_{+}^{2}|\Phi_{J,E}(\{\lambda^{\sigma}\})\rangle \nonumber\\
&=&\text{\ensuremath{\langle J,-J|\otimes\langle0|\prod_{j=1}^{M}\left(\hat{a}-\frac{\hat{J}_{-}}{\text{\ensuremath{\lambda_{j}^{\sigma}}}}\right)\hat{J}_{+}^{2}\prod_{j=1}^{E}\left(\hat{a}^{\dagger}-\frac{\hat{J}_{+}}{\text{\ensuremath{\lambda_{j}^{\sigma}}}^{*}}\right)}}\nonumber\\
&&\times|J,-J\rangle\otimes|0\rangle\nonumber\\
&=&\langle J,-J|\otimes\langle0|(\hat{a}^{M}+x_{1}\hat{a}^{M-1}\hat{J}_{-}+\cdots+x_{M-1}\hat{a}\hat{J}_{-}^{M-1}\nonumber\\
&&+x_{M}\hat{J}_{-}^{M})\hat{J}_{+}^{2}(\hat{a}^{\dagger E}+x_{1}\hat{a}^{\dagger(E-1)}\hat{J}_{+}+\cdots+x_{E-1}\hat{a}^{\dagger}\hat{J}_{+}^{E-1}\nonumber\\
&&+x_{E}\hat{J}_{+}^{E})|J,-J\rangle\otimes|0\rangle\nonumber\\
&=&\langle J,-J|\otimes\langle0|(\hat{a}^{M}+x_{1}\hat{a}^{M-1}\hat{J}_{-}+\cdots+x_{M-1}\hat{a}\hat{J}_{-}^{M-1}\nonumber\\
&&+x_{M}\hat{J}_{-}^{M})(\hat{a}^{\dagger E}\hat{J}_{+}^{2}+x_{1}\hat{a}^{\dagger(E-1)}\hat{J}_{+}^{3}+\cdots+x_{E-1}\hat{a}^{\dagger}\hat{J}_{+}^{E+1}\nonumber\\
&&+x_{E}\hat{J}_{+}^{E+2})|J,-J\rangle\otimes|0\rangle\nonumber\\
&=&\langle\Phi_{J,M}(\{\lambda^{\sigma}\})|\hat{J}_{+}^{2}|\Phi_{J,E}(\{\lambda^{\sigma}\})\rangle\delta_{M,E+2}. \label{EqC6}
\end{eqnarray}
In the above equations, we use coefficients $x_{i}$ ($i=1,2,\cdots,M$ or $E$). These coefficients are simply to make the calculation process understandable. Its does not affect our results, so we do not give specific expressions in here.

With the results above, we can get the average of all the operators in Eq.~(\ref{Eq96})-Eq.~(\ref{Eq100})
\begin{eqnarray}
&&\frac{\langle\hat{J}_{z}\rangle}{N}\nonumber\\
&=&\sum_{M=0}^{\infty}\sum_{E=0}^{\infty}\left\langle\Phi_{M}(t)\right|\hat{J}_{z} \left|\Phi_{E}(t)\right\rangle/N \nonumber\\
&=&\sum_{M=0}^{\infty}\sum_{E=0}^{\infty}\Bigg(d_{M-J-1}^{*}\sum_{\sigma}^{K} \frac{C(M, \{\lambda^{*\sigma}\})}{N_{\sigma}^{2}}  e^{iE_{J,M}^{\sigma}t}\Bigg)\nonumber\\
&&\times\Bigg(d_{E-J-1}\sum_{\sigma}^{K} \frac{C(E, \{\lambda^{\sigma}\})}{N_{\sigma}^{2}}  e^{-iE_{J,E}^{\sigma}t}\Bigg)\nonumber\\
&&\times\langle\Phi_{J,M}(\{\lambda^{\sigma}\})|\hat{J}_{z}|\Phi_{J,E}(\{\lambda^{\sigma}\})\rangle/N\nonumber\\
&=&\sum_{M=0}^{\infty}|d_{M-J-1}|^{2}\Bigg(\sum_{\sigma}^{K} \frac{C(M, \{\lambda^{*\sigma}\})}{N_{\sigma}^{2}}  e^{iE_{J,M}^{\sigma}t}\Bigg)\nonumber\\
&&\times\Bigg(\sum_{\sigma}^{K} \frac{C(M, \{\lambda^{\sigma}\})}{N_{\sigma}^{2}}  e^{-iE_{J,M}^{\sigma}t}\Bigg)\nonumber\\
&&\times\langle\Phi_{J,M}(\{\lambda^{\sigma}\})|\hat{J}_{z}|\Phi_{J,M}(\{\lambda^{\sigma}\})\rangle/N  , \label{EqC7}\\
&&\frac{\langle\hat{J}_{z}^{2}\rangle}{N^{2}}\nonumber\\
&=&\sum_{M=0}^{\infty}\sum_{E=0}^{\infty}\left\langle\Phi_{M}(t)\right|\hat{J}_{z}^{2} \left|\Phi_{E}(t)\right\rangle/N^{2} \nonumber\\
&=&\sum_{M=0}^{\infty}\sum_{E=0}^{\infty}\Bigg(d_{M-J-1}^{*}\sum_{\sigma}^{K} \frac{C(M, \{\lambda^{*\sigma}\})}{N_{\sigma}^{2}}  e^{iE_{J,M}^{\sigma}t}\Bigg)\nonumber\\
&&\times\Bigg(d_{E-J-1}\sum_{\sigma}^{K} \frac{C(E, \{\lambda^{\sigma}\})}{N_{\sigma}^{2}}  e^{-iE_{J,E}^{\sigma}t}\Bigg)\nonumber\\
&&\times\langle\Phi_{J,M}(\{\lambda^{\sigma}\})|\hat{J}_{z}^{2}|\Phi_{J,E}(\{\lambda^{\sigma}\})\rangle/N^{2}   \nonumber\\
&=&\sum_{M=0}^{\infty}|d_{M-J-1}|^{2}\Bigg(\sum_{\sigma}^{K} \frac{C(M, \{\lambda^{*\sigma}\})}{N_{\sigma}^{2}}  e^{iE_{J,M}^{\sigma}t}\Bigg)\nonumber\\
&&\times\Bigg(\sum_{\sigma}^{K} \frac{C(M, \{\lambda^{\sigma}\})}{N_{\sigma}^{2}}  e^{-iE_{J,M}^{\sigma}t}\Bigg)\nonumber\\
&&\times\langle\Phi_{J,M}(\{\lambda^{\sigma}\})|\hat{J}_{z}^{2}|\Phi_{J,M}(\{\lambda^{\sigma}\})\rangle/N^{2}    , \label{EqC8}\\
&&\frac{\langle\hat{J}_{+}\rangle}{N}\nonumber\\
&=&\sum_{M=0}^{\infty}\sum_{E=0}^{\infty}\left\langle\Phi_{M}(t)\right|\hat{J}_{+} \left|\Phi_{E}(t)\right\rangle/N \nonumber\\
&=&\sum_{M=0}^{\infty}\sum_{E=0}^{\infty}\Bigg(d_{M-J-1}^{*}\sum_{\sigma}^{K} \frac{C(M, \{\lambda^{*\sigma}\})}{N_{\sigma}^{2}}  e^{iE_{J,M}^{\sigma}t}\Bigg)\nonumber\\
&&\times\Bigg(d_{E-J-1}\sum_{\sigma}^{K} \frac{C(E, \{\lambda^{\sigma}\})}{N_{\sigma}^{2}}  e^{-iE_{J,E}^{\sigma}t}\Bigg)\nonumber\\
&&\times\langle\Phi_{J,M}(\{\lambda^{\sigma}\})|\hat{J}_{+}|\Phi_{J,E}(\{\lambda^{\sigma}\})\rangle/N\nonumber\\
&=&\sum_{M=0}^{\infty}d_{M-J-1}^{*}d_{M-J-2}\Bigg(\sum_{\sigma}^{K} \frac{C(M, \{\lambda^{*\sigma}\})}{N_{\sigma}^{2}}  e^{iE_{J,M}^{\sigma}t}\Bigg)\nonumber\\
&&\times\Bigg(\sum_{\sigma}^{K} \frac{C(M-1, \{\lambda^{\sigma}\})}{N_{\sigma}^{2}}  e^{-iE_{J,M-1}^{\sigma}t}\Bigg)\nonumber\\
&&\times\langle\Phi_{J,M}(\{\lambda^{\sigma}\})|\hat{J}_{+}|\Phi_{J,M-1}(\{\lambda^{\sigma}\})\rangle/N  , \label{EqC9}\\
&&\frac{\langle\hat{J}_{+}\hat{J}_{z}+\hat{J}_{z}\hat{J}_{+}\rangle}{N^{2}} \nonumber \\
&=&\sum_{M=0}^{\infty}\sum_{E=0}^{\infty}\left\langle\Phi_{M}(t)\right|(\hat{J}_{+}\hat{J}_{z}+\hat{J}_{z}\hat{J}_{+}) \left|\Phi_{E}(t)\right\rangle/N^{2} \nonumber\\
&=&\sum_{M=0}^{\infty}\sum_{E=0}^{\infty}\Bigg(d_{M-J-1}^{*}\sum_{\sigma}^{K} \frac{C(M, \{\lambda^{*\sigma}\})}{N_{\sigma}^{2}}  e^{iE_{J,M}^{\sigma}t}\Bigg)\nonumber\\
&&\times\Bigg(d_{E-J-1}\sum_{\sigma}^{K} \frac{C(E, \{\lambda^{\sigma}\})}{N_{\sigma}^{2}}  e^{-iE_{J,E}^{\sigma}t}\Bigg)\nonumber\\
&&\times\langle\Phi_{J,M}(\{\lambda^{\sigma}\})|(\hat{J}_{+}\hat{J}_{z}+\hat{J}_{z}\hat{J}_{+})|\Phi_{J,E}(\{\lambda^{\sigma}\})\rangle/N^{2}   \nonumber\\
&=&\sum_{M=0}^{\infty}d_{M-J-1}^{*}d_{M-J-2}\Bigg(\sum_{\sigma}^{K} \frac{C(M, \{\lambda^{*\sigma}\})}{N_{\sigma}^{2}}  e^{iE_{J,M}^{\sigma}t}\Bigg)\nonumber\\
&&\times\Bigg(\sum_{\sigma}^{K} \frac{C(M-1, \{\lambda^{\sigma}\})}{N_{\sigma}^{2}}  e^{-iE_{J,M-1}^{\sigma}t}\Bigg)\nonumber\\
&&\times\langle\Phi_{J,M}(\{\lambda^{\sigma}\})|(\hat{J}_{+}\hat{J}_{z}+\hat{J}_{z}\hat{J}_{+})|\Phi_{J,M-1}(\{\lambda^{\sigma}\})\rangle/N^{2}, \nonumber \\  \label{EqC10}\\
&&\frac{\langle\hat{J}_{+}^{2}\rangle}{N^{2}} \nonumber \\
&=&\sum_{M=0}^{\infty}\sum_{E=0}^{\infty}\left\langle\Phi_{M}(t)\right|\hat{J}_{+}^{2} \left|\Phi_{E}(t)\right\rangle/N^{2} \nonumber\\
&=&\sum_{M=0}^{\infty}\sum_{E=0}^{\infty}\Bigg(d_{M-J-1}^{*}\sum_{\sigma}^{K} \frac{C(M, \{\lambda^{*\sigma}\})}{N_{\sigma}^{2}}  e^{iE_{J,M}^{\sigma}t}\Bigg)\nonumber\\
&&\times\Bigg(d_{E-J-1}\sum_{\sigma}^{K} \frac{C(E, \{\lambda^{\sigma}\})}{N_{\sigma}^{2}}  e^{-iE_{J,E}^{\sigma}t}\Bigg)\nonumber\\
&&\times\langle\Phi_{J,M}(\{\lambda^{\sigma}\})|\hat{J}_{+}^{2}|\Phi_{J,E}(\{\lambda^{\sigma}\})\rangle/N^{2}   \nonumber\\
&=&\sum_{M=0}^{\infty}d_{M-J-1}^{*}d_{M-J-3}\Bigg(\sum_{\sigma}^{K} \frac{C(M, \{\lambda^{*\sigma}\})}{N_{\sigma}^{2}}  e^{iE_{J,M}^{\sigma}t}\Bigg)\nonumber\\
&&\times\Bigg(\sum_{\sigma}^{K} \frac{C(M-2, \{\lambda^{\sigma}\})}{N_{\sigma}^{2}}  e^{-iE_{J,M-2}^{\sigma}t}\Bigg)\nonumber\\
&&\times\langle\Phi_{J,M}(\{\lambda^{\sigma}\})|\hat{J}_{+}^{2}|\Phi_{J,M-2}(\{\lambda^{\sigma}\})\rangle/N^{2} . \label{EqC11}  
\end{eqnarray}
Since $d_{m}$ ($m=-J,-J+1,\cdots,J-1,J$) is the expansion coefficient of the coherent spin initial state of $N$ identical qubits over the Dicke state, $d_{M-J-1}=0$ when $M = 0$ or $M > 2J + 1$. So the above equation can be rewritten as
\begin{eqnarray}
\frac{\langle\hat{J}_{z}\rangle}{N}&=&\sum_{M=0}^{2J}|d_{M-J}|^{2}\Bigg(\sum_{\sigma}^{K} \frac{C(M+1, \{\lambda^{*\sigma}\})}{N_{\sigma}^{2}}  e^{iE_{J,M+1}^{\sigma}t}\Bigg)\nonumber\\
&&\times\Bigg(\sum_{\sigma}^{K} \frac{C(M+1, \{\lambda^{\sigma}\})}{N_{\sigma}^{2}}  e^{-iE_{J,M+1}^{\sigma}t}\Bigg)\nonumber\\
&&\times\langle\Phi_{J,M+1}(\{\lambda^{\sigma}\})|\hat{J}_{z}|\Phi_{J,M+1}(\{\lambda^{\sigma}\})\rangle/N \nonumber \\
&=&\sum_{M=0}^{2J}|d_{M-J}|^{2}F(\hat{J}_{z}, M,  t) , \label{EqC12} \\ 
\frac{\langle\hat{J}_{z}^{2}\rangle}{N^{2}}&=&\sum_{M=0}^{2J}|d_{M-J}|^{2}\Bigg(\sum_{\sigma}^{K} \frac{C(M+1, \{\lambda^{*\sigma}\})}{N_{\sigma}^{2}}  e^{iE_{J,M+1}^{\sigma}t}\Bigg)\nonumber\\
&&\times\Bigg(\sum_{\sigma}^{K} \frac{C(M+1, \{\lambda^{\sigma}\})}{N_{\sigma}^{2}}  e^{-iE_{J,M+1}^{\sigma}t}\Bigg)\nonumber\\
&&\times\langle\Phi_{J,M+1}(\{\lambda^{\sigma}\})|\hat{J}_{z}^{2}|\Phi_{J,M+1}(\{\lambda^{\sigma}\})\rangle/N^{2} \nonumber \\
&=&\sum_{M=0}^{2J}|d_{M-J}|^{2}F(\hat{J}_{z}^{2}, M,  t)  , \label{EqC13}\\ 
\frac{\langle\hat{J}_{+}\rangle}{N}&=&\sum_{M=0}^{\infty}d_{M-J-1}^{*}d_{M-J-2}\Bigg(\sum_{\sigma}^{K} \frac{C(M, \{\lambda^{*\sigma}\})}{N_{\sigma}^{2}} \nonumber\\
&&\times e^{iE_{J,M}^{\sigma}t}\Bigg)\Bigg(\sum_{\sigma}^{K} \frac{C(M-1, \{\lambda^{\sigma}\})}{N_{\sigma}^{2}}  e^{-iE_{J,M-1}^{\sigma}t}\Bigg)\nonumber\\
&&\times\langle\Phi_{J,M}(\{\lambda^{\sigma}\})|\hat{J}_{+}|\Phi_{J,M-1}(\{\lambda^{\sigma}\})\rangle/N  \nonumber\\
&=&\sum_{M=0}^{2J-1}d_{M-J+1}^{*}d_{M-J}\Bigg(\sum_{\sigma}^{K} \frac{C(M+2, \{\lambda^{*\sigma}\})}{N_{\sigma}^{2}}  \nonumber\\
&&\times e^{iE_{J,M+2}^{\sigma}t}\Bigg)\Bigg(\sum_{\sigma}^{K} \frac{C(M+1, \{\lambda^{\sigma}\})}{N_{\sigma}^{2}}  e^{-iE_{J,M+1}^{\sigma}t}\Bigg)\nonumber\\
&&\times\langle\Phi_{J,M+2}(\{\lambda^{\sigma}\})|\hat{J}_{+}|\Phi_{J,M+1}(\{\lambda^{\sigma}\})\rangle/N  \nonumber\\
&=&\sum_{M=0}^{2J-1}d_{M-J+1}^{*}d_{M-J}F(\hat{J}_{+},M,t), \label{EqC14} \\
&&\frac{\langle\hat{J}_{+}\hat{J}_{z}+\hat{J}_{z}\hat{J}_{+}\rangle}{N^{2}}\nonumber\\
&=&\sum_{M=0}^{\infty}d_{M-J-1}^{*}d_{M-J-2}\Bigg(\sum_{\sigma}^{K} \frac{C(M, \{\lambda^{*\sigma}\})}{N_{\sigma}^{2}}  \nonumber\\
&&\times e^{iE_{J,M}^{\sigma}t}\Bigg)\Bigg(\sum_{\sigma}^{K} \frac{C(M-1, \{\lambda^{\sigma}\})}{N_{\sigma}^{2}}  e^{-iE_{J,M-1}^{\sigma}t}\Bigg)\nonumber\\
&&\times\langle\Phi_{J,M}(\{\lambda^{\sigma}\})|(\hat{J}_{+}\hat{J}_{z}+\hat{J}_{z}\hat{J}_{+})\nonumber\\
&&\times|\Phi_{J,M-1}(\{\lambda^{\sigma}\})\rangle/N^{2}  \nonumber\\
&=&\sum_{M=0}^{2J-1}d_{M-J+1}^{*}d_{M-J}\Bigg(\sum_{\sigma}^{K} \frac{C(M+2, \{\lambda^{*\sigma}\})}{N_{\sigma}^{2}}  \nonumber\\
&&\times e^{iE_{J,M+2}^{\sigma}t}\Bigg)\Bigg(\sum_{\sigma}^{K} \frac{C(M+1, \{\lambda^{\sigma}\})}{N_{\sigma}^{2}}  e^{-iE_{J,M+1}^{\sigma}t}\Bigg)\nonumber\\
&&\times\langle\Phi_{J,M+2}(\{\lambda^{\sigma}\})|(\hat{J}_{+}\hat{J}_{z}+\hat{J}_{z}\hat{J}_{+})\nonumber\\
&&\times|\Phi_{J,M+1}(\{\lambda^{\sigma}\})\rangle/N^{2} \nonumber\\  \nonumber\\
&=&\sum_{M=0}^{2J-1}d_{M-J+1}^{*}d_{M-J}F(\hat{J}_{+}\hat{J}_{z}+\hat{J}_{z}\hat{J}_{+},M,t), \label{EqC15} \\
&&\frac{\langle\hat{J}_{+}^{2}\rangle}{N^{2}}\nonumber\\
&=&\sum_{M=0}^{2J-2}d_{M-J+2}^{*}d_{M-J}\Bigg(\sum_{\sigma}^{K} \frac{C(M+3, \{\lambda^{*\sigma}\})}{N_{\sigma}^{2}} \nonumber\\ 
&&\times e^{iE_{J,M+3}^{\sigma}t}\Bigg)\Bigg(\sum_{\sigma}^{K} \frac{C(M+1, \{\lambda^{\sigma}\})}{N_{\sigma}^{2}}  e^{-iE_{J,M+1}^{\sigma}t}\Bigg)\nonumber\\
&&\times\langle\Phi_{J,M+3}(\{\lambda^{\sigma}\})|\hat{J}_{+}^{2}|\Phi_{J,M+1}(\{\lambda^{\sigma}\})\rangle/N^{2}  \nonumber\\
&=&\sum_{M=0}^{2J-2}d_{M-J+2}^{*}d_{M-J}F(\hat{J}_{+}^{2},M,t). \label{EqC16}
\end{eqnarray}

From the above we can see that the average value of these operators at time $t$ is actually determined by some functions $F(\hat{J}_{z}, M,  t)$, $F(\hat{J}_{z}^{2}, M,  t)$, $F(\hat{J}_{+},M,t)$, $F(\hat{J}_{+}\hat{J}_{z}+\hat{J}_{z}\hat{J}_{+},M,t)$, $F(\hat{J}_{+}^{2},M,t)$, and the probability distribution $d_{m}$.

\nocite{*}

\bibliography{GQEbyAP}

\end{document}